\documentstyle[preprint,eqsecnum,aps]{revtex}
\tighten
\begin{document}
\draft
\tolerance=10000
\hfuzz=5pt
\preprint{HD-TVP-97/05}
\title{Spectral density functions and their sum rules in an effective
chiral field theory}
\author{S. P. Klevansky}
\address{Institut f\"{u}r Theoretische Physik, Universit\"{a}t Heidelberg,\\
Philosophenweg 19, D-69120, Heidelberg, Germany.}
\author{R. H. Lemmer \cite{RHL}}
\address{Max-Planck-Institut f\"ur Kernphysik, Postfach 10 39 80,\\
D-69029 Heidelberg, Germany}
\date{\today}
\maketitle
\begin{abstract}
The validity of Weinberg's two sum rules for massless QCD, as well as the
six additional sum rules introduced into  chiral perturbation theory
by Gasser and Leutwyler, are investigated for the extended
Nambu-Jona-Lasinio chiral model that includes vector and axial vector
degrees of freedom.
A detailed analysis of the vector, axial vector and coupled pion plus 
longitudinal axial vector modes is given.
 We show that, under Pauli-Villars
regularization of the meson polarization 
amplitudes that determine the spectral density
functions, all of the sum rules involving inverse moments higher than
zero  are automatically obeyed by the model spectral densities. By contrast,
the zero moment sum rules acquire a non-vanishing right hand side that is proportional to
the quark condensate density of the non-perturbative groundstate.
We use selected sum rules in conjunction with
other calculations to obtain explicit expressions for the scale-independent
coupling constants  $\bar l_i$ of chiral perturbation theory in the combination
$\bar l_i+\ln(m^2_\pi/\mu^2)$,
to evaluate the Das-Guralnik-Mathur-Low-Young current algebra expression for the electromagnetic mass difference of the pion,
and to compute the pion electrical polarizability $\alpha_E$.
The sum rule calculations
set an upper limit of
$\alpha_E\leq \alpha/(8\pi^2m_\pi f^2_\pi) \approx 6\times10^{-4}$fm$^3$ on
this quantity,
that is independent of the quark mass up to ${\cal O}(m^2_\pi/m^2)$,
and {\it only} depends on the fundamental constants:
pion mass $m_\pi$ and pion decay constant $f_\pi$, in addition to fine structure
constant $\alpha\approx 1/137$.
\end{abstract}
\vskip 0.2in
No. of manuscript pages: 100;  No. of tables: 7;  No. of figures: 21.

\clearpage
\renewcommand{\baselinestretch}{1.3}
\footnotesize \normalsize


\section{Introduction}

Our present knowledge of the low energy hadronic spectra and the properties
thereof is based on the premise that chiral symmetry is a good
approximate symmetry of the underlying quantum chromodynamics (QCD)
Lagrangian.     This has found verification time and again in understanding
the nature of the low-lying octet of pseudoscalar mesons in particular.
Calculational methods have
been developed primarily in two complementary directions. On the one hand,
the current algebra tools of the 1960's \cite{nar89}
plus the two sum rules
of Weinberg \cite{wein67,ber75}
attempt to address the properties of the QCD Lagrangian {\it per se}
in the perturbative regime. On the other, the chiral
perturbation theory (CHPT) approach \cite{wein79,gl84,leutwyler91},
which is based on an effective Lagrangian
that only contains mesonic degrees of freedom,
addresses the low-energy non-perturbative regime of QCD.
The latter approach introduces terms into this
Lagrangian in an expansion in momenta.   The unknown coefficients of these
terms in each order of the expansion, which in the two flavor case amount to
ten in total, have to be fixed by experimentally measurable quantities.

An effective model approach of describing the low energy sector {\it ab
initio} in terms of quark degrees of freedom has also been studied for
some years now in the so-called Nambu--Jona-Lasinio (NJL) model
\cite{njl61,weise90,vogl91,klev92,hk94,bij94,brz94}. In its
original two flavor formulation, the minimal version of this model
contains scalar
and pseudoscalar fields in a chirally symmetric combination,
and the model only has three parameters:
the (strong) coupling $G_1$, an averaged current quark mass
$\hat m = (m_u^0 + m_d^0)/2$, and a cutoff $\Lambda$ that sets the scale of
the theory.
Mesons are constructed as collective excitations within this model, and the
Goldstone theorem is fulfilled in the chiral limit.    The spontaneous
breakdown of chiral symmetry leads to a dynamically generated quark mass.
In addition, it can easily be demonstrated within the framework of this
model that the Goldberger-Treiman and Gell-Mann--Oakes--Renner relations are
 fulfilled, and that
 the quark axial current is conserved.   In addition, 
Dashen's theorem \cite{dash69}  that prevents a chiral neutral pion from
acquiring a
mass  shift in the presence of an electromagnetic field, also
continues to be obeyed \cite{dstl95}.
These properties, which are a consequence of chiral symmetry,  continue to hold
in an
extended version of the NJL  model or ENJL model, as we will abbreviate it
 \cite{weise90,brz94}.
In this extended model, the vector and axial
vector degrees of freedom are explicitly included, and an additional
(strong)
coupling strength $G_2$ is required in order to couple these modes into the
Lagrangian.    It is 
imperative in
 demonstrating that
the expected properties of chiral symmetry  hold,
  that a suitable
regularization scheme   be employed 
under which the relevant Ward identities continue to
hold \cite{dstl95,dstl94}.

The primary purpose of this paper is to investigate the sum rules
of both the
Weinberg type \cite{wein67} and also those introduced by
Gasser and Leutwyler \cite{gl84}.
These sum rules involve, {\it inter alia  } the vector as well as axial
vector spectral densities, and thus it appears necessary in regarding a 
quark model, to have such degrees of freedom explicitly supported.    Hence our
interest in using the ENJL model for this purpose.    However, the 
application of this model to this problem is fraught with 
difficulties.   On the one hand, the identification of the vector and
axial vector meson modes with the physical $\rho$ and $a_1$ mesons is
especially problematic, and we will return to comment on their analysis
somewhat later on in this introduction.  We turn first rather to the issue
of the spectral functions themselves.   We note that either version of 
 the NJL model
constitutes an effective
field theory that does not confine the  explicit quark degrees of
freedom that it contains. Consequently one expects the strength
distributions of the meson spectra, which contain
interacting quark-antiquark pairs, to behave unphysically in the high energy
domain.
This domain is naturally important if the sum rules are to work.  However, in
spite of these
deficiencies, one can show
explicitly \cite{dkl96} that under the 
Pauli-Villars (PV) regularization of the
amplitudes, Weinberg's first sum rule
\begin{eqnarray}
\int_0^\infty\frac{ds}{s}\big(\rho^{V}_1(s)-\rho^{A}_1(s)\big)=f^2_\pi\,,
\label{e:introw1}
\end{eqnarray}
is obeyed exactly, while  the second one satisfies a modified equation,
\begin{eqnarray}
\int_0^\infty
ds\big(\rho^{V}_1(s)-\rho^{A}_1(s)\big)=-m\langle\bar\psi\psi\rangle\ne 0\,.
\label{e:introw2}
\end{eqnarray}
Both statements hold to leading order in the number of colors. Here $f_\pi$
is the pion decay constant, and $m$ and
$\langle\bar\psi\psi\rangle$ are the quark mass and scalar quark condensate in
the ground state.
These conclusions also hold for the minimal version of the NJL model.
This is perhaps all the more surprizing, since  the minimal version
does not support either vector mode as a
fundamental excitation, so that it contains no
resonance structures of any kind in these channels.

 Besides the two Weinberg sum rules,  we are able to demonstrate that
 the six additional sum rules introduced by
Gasser and Leutwyler are also
obeyed by the spectral densities of the ENJL model.
However, these results come at a price: as pointed out in  \cite {iz80},
 for example, the Pauli-Villars
regularization procedure effectively introduces additional spinor fields that
may correspond to
indefinite metric sectors of the Hilbert space.
In the present case, the use of the PV procedure
leads to  negative and therefore unphysical spectral densities
at large momentum transfers.
However, this occurs in such a fashion as to give an overall result that is in
accordance with the sum rule.
We thus interpret the, perhaps unexpected,
 continued validity  of the
sum rules under these circumstances
 as a consequence of the underlying chiral symmetry
and its faithful representation in the  Pauli-Villars regularization method.

The results obtained from the ENJL model fall naturally under scrutiny with
regard to
their accordance or lack thereof with chiral perturbation theory. The
retention of the sum rules in both approaches suggests that there could be
considerable overlap in these results.
In fact, the ENJL model provides a dynamical basis for calculating the
empirical 
 Gasser-Leutwyler
coupling constants of chiral perturbation theory.  We evaluate these
constants in both the NJL and  ENJL models, and express them as far as possible
in terms of the
physically measurable parameters $f_\pi$ and $g_A$, the quark axial
form factor.
One finds that one additional physically non-measurable quantity,
the (constituent) quark mass $m$, also
occurs in these expressions, giving them this single  parameter dependence.
 However, since the calculations of all quantities within the ENJL model
 have only been
performed to leading order ${\cal O}(N_c)$ in the number of colors,
one has to  identify the calculated
coefficients with the Gasser-Leutwyler constants with some care.
The reason for this is that the physically meaningful scale-independent
 constants $\bar l_i$ of Gasser and Leutwyler
are not well-defined in the chiral limit, where they diverge logarithmically
with vanishing
pion mass. This comes about due to the appearance of
chiral logarithms in CHPT that are to be anticipated on general
grounds \cite{lp73}. In either version of the  NJL model,
such chiral logarithms  appear as a part of the ${\cal}(1/N_c)$
non-chiral corrections
that necessarily include mesons in intermediate states
\cite{kh95,dstl95}. Thus we must identify the coefficients obtained from the
model
calculations with the $renormalized$ $l^r_i$'s of Gasser and Leutwyler,
that have a well-defined chiral limit, and not directly with
the $\bar l_i$ as is commonly done
 (see \cite {mk94} and further
references cited therein\footnote{We note that the
good agreement claimed in previous calculations can be traced back to the
fact that the chiral logarithm is a slowly varying function.}).
We do this here. In this way, predictions can
be obtained for the combinations
\begin{eqnarray}
 \bar l_i+\ln(m^2_\pi/\mu^2)\sim l^r_i.
\label{e:combination}
\end{eqnarray}
 Here the mass $\mu$ is a scale parameter introduced via the
dimensional regularization of the chiral perturbation theory amplitudes
\cite{gl84}.
 One can explicitly validate this approach \cite{kh95}: the coefficients
that have been evaluated to ${\cal O}(N_c)$ in either version of the NJL model are finite in the
chiral limit, and acquire  chiral logarithms (at the scale
 $\mu=2m\approx m_\sigma$ of the $\sigma$ meson mass)
 upon incorporating the meson loops.

The
results obtained in this manner
are in reasonable agreement with the known empirical values of the $\bar
l$'s, after including the logarithm as shown.
As already noted, this agreement,
however, depends explicitly on the calculated value of
the
dynamically generated  quark mass
as obtained from the gap equation.  An important
exception to this remark is the
result for the  electric polarizability,  $\alpha_E $, of charged pions.
This coefficient can be obtained directly from
Holstein's sum rule \cite{holstein90} in conjunction with the
ENJL expressions for the spectral densities and the pion radius.  One
obtains
the simple form
\begin{eqnarray}
\alpha_E= \frac{\alpha}{m_\pi}\frac{1}{48\pi^2f^2_\pi}(\bar l^{ENJL}_6
-\bar l^{ENJL}_5)
=\frac{g_A^2}{8\pi^2 f^2_\pi}\frac{\alpha}{m_\pi}
\leq \frac{1}{8\pi^2 f^2_\pi}\frac{\alpha}{m_\pi}
= 6.05\times 10^{-4}\,{\rm fm^3}
\end{eqnarray}
where  $\alpha = e^2/4\pi\approx 1/137$,
and the other symbols have their usual meanings.
This is an extremely interesting  result for several reasons.
 Firstly it coincides exactly in form with the
CHPT expression for the polarizability in terms of the $\bar l_i$
to one-loop order\cite{burgi96}. Secondly, the
terms coming from the dynamically generated 
quark mass $m$ cancel exactly in the difference
$\bar l^{ENJL}_6-\bar l^{ENJL}_5$,
leading to a result that {\it only contains physical constants}.
Thirdly, only the axial meson degree of freedom contributes to the
polarizability through $g_A$; there is no contribution from
the $\rho$ meson.
Fourthly, since
 $g^2_A\leq 1$, this establishes a {\it parameter free} prediction for the
upper limit on the ENJL value of $\alpha_E$, at least
to ${\cal O}(\alpha N_c)$.
Using the estimated value $g_A= 0.75$, one has the numerical result, $\alpha_E =
3.40\times 10^{-4}$fm$^3$.
  This answer can  be compared directly
with the prediction \cite{holstein90} of chiral perturbation theory to one-loop
order,
 $\alpha^{chpt}_E=2.68\times 10^{-4}$fm$^3$,
that uses the empirical values of the $\bar l$'s.    Experimentally, 
the value of $\alpha_E$ is still very uncertain. 
Current analysis of the available data gives values of $\alpha_E$ ranging
between $2.2\pm 1.6$ to $20 \pm 12 \times 10^{-4}$ fm$^3$ 
\cite{jb90,bab92,antipov83,aiberg86,pp94}. 

One can also evaluate the
current algebra result of  Das {\it et al.}
 \cite{low67} for the electromagnetic splitting of the
charged to neutral pion masses using the
 ENJL spectral densities.
 The Das {\it et al.} formula links the 
  splitting  $\Delta m^2_\pi = m_{\pi^\pm}^2
 - m_{\pi^0}^2$ to a modified sum rule
 for the spectral
 density difference $\rho^V_1-\rho^A_1$, integrated over all momentum
 transfers of the virtual photons that are responsible for the
 electromagnetic self-energy of the pion.
  The resulting ENJL expression for this mass squared difference  can be
written
  in the compact form \cite{dkl96}
\begin{eqnarray}
\Delta m^2_\pi=3ie^2\int\frac{d^4 q
}{(2\pi)^4}\frac{1}{q^2}F_P(q^2)F_V(q^2)F_A(q^2)
\label{e:introdeltam}
\end{eqnarray}
where the $F's$ are pion and quark form factors that are defined in Section II
of the
main text. 
The special assumption that the form factors
in Eq.~(\ref{e:introdeltam}) describe  point-like mesons without internal structure,
leads to the result originally derived by Das {\it et al.}.
Further evaluation of this expression is
postponed to Section IV.
There we also give the set of gauge invariant pion polarization
diagrams to ${\cal O}(\alpha N_c)$ that corresponds to this result.
The electromagnetic gauge invariance of the mass difference expression 
 is thereby made explicit.

As has been alluded to in an earlier part of the introduction, the
determination and interpretation of the vector and axial vector 
excitation modes that are supported by the ENJL model is not
straightforward, but remain essential ingredients for the construction
of the spectral densities.   We have therefore investigated these modes
anew.
The  particular innate difficulty in these calculations arises from
the
fact that both the $\rho$ meson and the axial vector meson $a_1$
have empirical masses that lie {\it above} the  $4m^2$ threshold for free
quark pair creation,
a non-physical
process from the point of view of confinement. The existence of this
 threshold means that the complex plane
for the momentum transfer variable squared $s$
 has a unitarity cut along $4m^2 \leq s\leq \infty$.
Accordingly the
 ENJL model masses have to be sought as the poles of the relevant meson
 propagators  that have been continued through
this cut
onto the contiguous or ``second'' Riemann sheet.

Taken at face value, these facts seem to make a model of this nature of
dubious value for the vector modes.   In fact, the
 situation is far more complicated than this.
A careful analysis of the vector polarization
shows that this function has {\it two} isolated poles on the second sheet for our
parameter choice:
one  on the real axis
below $4m^2$,   and a second complex pole in the lower half plane of that
sheet.
The real pole corresponds  to a  $\rho$ mode containing {\it virtually}
bound  $\bar q q$ pairs\footnote{The occurrence of this state in the ENJL model
was
first pointed out in \cite{tkm91}.
Increasing the coupling strength $G_2$ of the
vector plus axial vector terms in the Lagrangian causes this mode to move
through the
branch cut
onto the physical sheet, to become a true bound state for
the model $\rho$ meson.}.
The $\rho$ spectral density function  receives contributions
from both of these poles.
A single pole approximation to describe the resonance
 behavior of the vector strength in terms of a  single mass plus a  width
 parameter
is thus incomplete. While this density still displays a
single-peak
structure that can be placed around the observed physical $ \rho$ mass
 for an appropriate
choice of parameters, the extent to which the virtual state strength
intrudes\footnote{The enhancement of the neutron-proton scattering
cross-section in the singlet channel over that in the
triplet channel due to the presence of a
virtually bound singlet state in the former, is a case in point
\cite{rgs53}.}
into the energy domain of the peak is parameter-dependent. Thus
one cannot unambiguously associate the position of the peak
with the mass squared
of the $\rho$ meson as being given by either the virtual bound state pole,
or the real part of the complex pole.
  We give full details of this problem in Section III.
As such, the interpretation of the $\rho$ meson, like the $\pi$ meson,
differs vastly from the conventional view of being simply a bound state of
two constituents.   Note in passing that, simply
taken on its own, the $\rho$ meson is of considerable
interest currently, given its special role in the interpretations of the
observed excess in the lepton spectra in heavy-ion collisions in the
intermediate mass range \cite{lepton97}.

By contrast, the axial vector polarization displays a single complex pole
in the
lower half plane of the second sheet, and the axial vector spectral density has
a single peak structure that can be associated with this pole.  The positions
of peaks in both the vector and the axial vector
spectral functions are basically controlled by the vector coupling
strength $G_2$.  
While the value of this coupling can be fixed via $g_A$ at
 $G_2=(1-g_A)/(8f^2_\pi) \approx 3.61$GeV$^{-2}$
 [from Eq.~(\ref{e:axialg})], plus the known value of $f_\pi\approx93$MeV and
an assumed value for $g_A=0.75$, the actual peak positions are
fairly robust
against moderate changes in $G_2$. Taking the resulting
peaks to define the $\rho(770)$ and $a_1(1260)$ meson masses,
one obtains $m_\rho=713$MeV, $m_{a_1}=1027$MeV for the vector masses of the
ENJL model,
to be compared with the observed masses of $768.1\pm 0.6$MeV and $1230\pm
40$MeV \cite{data96}.
 It is interesting to observe that the predicted
ratio $m_{a_1}/m_\rho =1.44\approx \sqrt 2$ for the above value of $G_2$ is
remarkably close to the
original Weinberg estimate \cite{wein67} of exactly $\sqrt2$.

The values for
the model masses are satisfactory from an experimental point of view. One
thus has to try to reconcile these results with the fact that neither
spectral function contains its dominant physical decay channel,
$\rho\rightarrow
\pi\pi$ or $a_1\rightarrow \rho\pi$ respectively,
but rather describes the (hypothetical)
 $\rho\rightarrow \bar qq$  or $a_1\rightarrow\bar qq$ decays allowed by the
non-confining ENJL Lagrangian.
This is a long-recognised deficiency \cite{weise90}.
On the other hand, the model spectral densities
continue to saturate the first sum rule, Eq.~(\ref{e:introw1}).
From these facts alone,
one can  argue with some justification that
the positions of the peaks in these distributions are more significant than the
particular manner in which they decay, and therefore do lead to meaningful
masses.
 One should recall in this regard that, in Weinberg's original discussion
of the problem, the vector meson mass ratio was determined by using delta
function distributions
for the vector densities that ignored all decay channels, but with strengths
that allowed the sum rules to be satisfied.   More generally, the fact that
both Weinberg's and Gasser and Leutwyler's sum rules are satisfied by the
model spectral density functions in spite of their undesirable decay
properties,
suggests that results based on integrals over these densities do
nevertheless have physical significance.

An analysis of the ENJL model commences with the construction of the
irreducible polarizations that are now coupled in the isovector pseudoscalar
and
longitudinal axial channels.  This is done in Section II. We are not the first
authors to carry out such an analysis
(see e.g. \cite{weise90,brz94}). Hence this section is in the nature of a
summary
of those aspects of the
problem that are necessary for an understanding of the sum rules, and
it also serves to set our  notation.
{}From a knowledge of the irreducible polarization functions, the new meson
propagators in the
extended model can be constructed.   This is done in such a fashion that
the generalization from the NJL model becomes obvious, and so that the
NJL results are simply recovered on setting $G_2$ to zero.    Central
to the derivations is that, as far as possible, quantities are expressed
in terms of physical variables, the pion decay constants $f_\pi$ and the
pion quark coupling strength, $g_{\pi qq}$, in the ENJL model.    One
observes
that these quantities undergo a simple multiplicative scaling
$f_\pi=\sqrt{g_A}f_p$ and $g_{\pi qq}=g_p/\sqrt{g_A}$ 
via the quark axial form factor $g_A$  with
respect to their NJL values $f_p$ and $g_p$ in such a way as to preserve
the
chiral theorems, like the Goldberger-Treiman relation.  In fact, apart from such
scaling, the main effect of including the vector and axial vector degrees of freedom
is to drastically redistribute the associated NJL spectral densities into two peaks
that serve to determine the masses of these modes in the model.
Conservation of the axial vector current can also be demonstrated
explicitly.   Here one finds that the ENJL model leads to non-trivial
vector and axial
form factors $F_V(q^2)$ and $G_A(q^2)$ for the quark currents
such that $F_V(0)=1$, but $G_A(0)=g_A\le 1$.

There is of couse an overriding $caveat$ on all of these statements: the issue
of regularization. In addition to being non-confining, the NJL model
and its ENJL extension  are both  non-renormalizable.  Thus a cutoff
$\Lambda$ in their mass spectra is necessary in order to
obtain finite answers. One anticipates that the magnitude of
 this cutoff will be
similar
 for either version of the model, but this parameter has in any
event to be of order of the chiral symmetry-breaking scale
$\Lambda_{\chi}\sim 1$GeV of QCD. Otherwise the NJL Lagrangian
could not reasonably fulfill the role of a  low energy effective model theory.
However the problem is more complicated than this. One
has to also ensure that the regularization scheme employed respects the chiral
Ward
identities satisfied by those amplitudes that express the consequences of the
chiral symmetry of this model Lagrangian. The Pauli-Villars regularization
scheme \cite{iz80}
fulfulls these
requirements
\cite{dstl95,dstl94}: in particular, it preserves those chiral Ward identities
between
regulated amplitudes that are necessary for a
demonstration of the sum rules\footnote{This is not to say
that the  Pauli-Villars  regularization scheme will  guarantee
all Ward identities. A discussion of the related questions that arise in the case of the
PV regularizarion of the $\sigma$-model, for example, can be found in \cite{gl69}.}.
We take up the PV regularization of the polarization amplitudes in some detail
in
Appendix A. There we show that the PV scheme can be implemented in
conjunction with either unsubtracted, or once-subtracted dispersion
relations that lead, not surprizingly, to a very different asymptotic behavior
of the polarization amplitude in question. This feature has a direct bearing on
the ability 
of the associated spectral density, which is given by the imaginary part of the
polarization, to satisfy the sum rules.


 The spectral density functions carrying specific quantum numbers are next
introduced in Section III
with the aid of the appropriate current-current correlators and  the
K\"allen-Lehman
spectral representation.  We establish the analytic properties of the
polarization functions. Then we analyse the analytic behavior of their
corresponding spectral
density functions under PV-regularization, and show by means of contour
integration
techniques that all eight sum rules, the two of Weinberg, and the six of
Gasser and Leutwyler, are obeyed by ENJL spectral densities regulated in this
manner.
The only exceptions that occur are for the zero-moment sum rules, like in
Eq.~(\ref{e:introw2}),
where a finite value instead of zero appears on the right hand side. We use
these
results in conjunction with the non-chiral expansions for the physical
pion mass and decay constant of Section II to identify  several of the
coupling
parameters of chiral perturbation theory.
In parallel with these developments,
we also explore the analytic properties of polarization amplitudes that have
been
obtained from once-subtracted dispersion relations prior to regularization.
Such amplitudes are  shown to give rise to Landau ghost poles at space-like
$q^2$
in the  $\pi$, $\rho$, and $a_1$ channels. Some of the consequences
of nevertheless admitting such unphysical propagators are briefly discussed.

In Section IV,  we discuss the use of ENJL spectral densities in conjunction
with sum rules
to compute two electromagnetic properties
of the pion: the electric polarizability coefficient of the charged pion
from Holstein's sum rule \cite{holstein90},
and the $\pi^\pm-\pi^0$ mass splitting from the current algebra sum rule of
Das, Guralnik, Mathur, Low and Young \cite{low67}. We also provide an
alternative
derivation of the mass splitting in terms of Feynman diagrams that leads to
exactly the same result.   An explicit demonstration of how Dashen's
theorem is satisfied by the ENJL model electromagnetic self-energies for the
neutral pion is also given.
A summary of results and conclusions is presented in Section V, while
Appendix A
contains additional calculational details of a more technical nature that do
not
appear in the main text.

\section{ Review of formalism}
Since the properties of Nambu-Jona-Lasinio type Lagrangians have already been the
subject of extensive analysis in the recent literature
\cite{njl61,weise90,vogl91,klev92,hk94,bij94,brz94}, we use this section to briefly summarize only those
aspects that will be required as background for the specific developments to
follow in Sections III and IV.

\subsection{Two-point functions}
In this section, we discuss the two-point
correlation functions from which the spectral densities we wish to study
are obtained.  Since the calculational procedure is the same for all
spectral densities, let us denote by $J^a(x)$ a generic operator with 
$SU(2)$ isospin and Lorentz indices summarized by the index $a$. 
Then the associated time-ordered
two-point correlators, or polarization
functions, are given by
\begin{eqnarray}
\int d^4x\;e^{iq\cdot x}\langle 0|T\{J^a(x)J^b(0)\}|0\rangle
\label{e:JJGEN}
\end{eqnarray}
in momentum space, where $T$ is the usual time-ordering operator.
These currents are assumed to be bilinear forms of the
two-flavor quark fields $\psi(x)$ and their Dirac adjoints, i.e.
\begin{eqnarray}
J^a(x)=\bar\psi(x)\Gamma^a\psi(x),
\end{eqnarray}
where the vertex $\Gamma^a$ characterizes the nature of the excitation.
Using the notation of
Ref.~\cite{gl84}, the bilinear forms $P,S,V,A$ for isospin currents $J^a$ are 
\begin{eqnarray}
\begin{array}{lll}
P^a(x)=\bar\psi(x)i\gamma_5\tau^a\psi(x)&0^{-+}&\quad \pi\\
S^a(x)= \bar\psi(x)\tau^a\psi(x)&0^{++}&\quad a_0\\ 
V^a_{\mu}(x)=\bar\psi(x)\gamma_\mu {\tau^a\over 2}\psi(x)&1^{--}&\quad \rho\\
A^a_{\mu}(x)= \bar\psi(x)\gamma_\mu\gamma_5 {\tau^a\over 2}\psi(x)&1^{++}&
\quad a_1 
\end{array}
\label{e:isovec}
\end{eqnarray}
with $\tau^a$, $a=1\;2\;3$ being the $SU(2)$ Pauli isospin matrices.
The spin, parity and charge conjugation
quantum numbers $J^{PC}$
of the excitations they lead to relative to the invariant vacuum, as well as
representative low-lying meson states that carry these quantum numbers, are also indicated.
In addition to these isospin currents, it is useful to consider the isoscalar densities
\begin{eqnarray}
\begin{array}{lllll}
P^0(x)=\bar \psi(x)i\gamma_5\psi(x)&0^{-+}&\eta&\\
S^0(x)=\bar\psi(x)\psi(x)&0^{++}&f_0&or\;\sigma
\label{e:S0P0}
\end{array}
\end{eqnarray}
The factor of
$1/2$ that is included in Eq.~(\ref{e:isovec}) in the definition of the
vector and axial vector currents is purely convention.
If we further assume that the dynamics
of these quark currents are governed by a two-flavor $U(1)\times SU_L(2)
\times SU_R(2)$ chirally symmetric Lagrangian of the Nambu-Jona-Lasinio type,
then both the vector and the axial vector (Noether) currents will be conserved. If this is the case,
then the
Lorentz and and isospin structure of the expression (\ref{e:JJGEN})
is known
ahead of time. If we write
\begin{eqnarray}
\int d^4x\;e^{iq\cdot x}\langle 0|T\{J^a_\mu(x)J^b_\nu(0)\}|0\rangle
= -{i\over 4}  \tilde\Pi^{JJ}_{\mu\nu;ab}(q^2),
\label{e:JJ}
\end{eqnarray}
where $J^a_\mu(x)$ is given by either $V^a_\mu$ or $A^a_\mu$, then both correlation functions become purely transverse tensors in their
Lorentz indices and diagonal in their isospin indices, i.e.
\begin{eqnarray}
\tilde\Pi^{JJ}_{\mu\nu;ab}(q^2)=\tilde\Pi^{JJ}(q^2)T_{\mu\nu}\delta_{ab},
\label{e:conserved}
\end{eqnarray}
where $T_{\mu\nu}=(g_{\mu\nu}-\hat{q}_\mu\hat{q}_{\nu})$ with
$\hat{q}_\mu=q_\mu/\sqrt{q^2}$.
In a similar fashion, the scalar and pseudoscalar two-point functions have the
structure
\begin{eqnarray}
\int d^4x\;e^{iq\cdot x}\langle 0|T\{J^a(x)J^b(0)\}|0\rangle
= -i \tilde\Pi^{JJ}(q^2)\delta_{ab},
\label{e:JJSP}
\end{eqnarray}
where $J^a(x)$ is given by either $S^a$ or $P^a$ for the isovector currents, or
by $S^0$ or $P^0$ for the isoscalar densities. In the latter case, there is
no  isospin
factor. Note that in Eqs.~(\ref{e:JJ}) to (\ref{e:JJSP}),
we have introduced the full polarization functions $\tilde\Pi^{JJ}(q^2)$. In order
to achieve a uniformity of definition for these, we have introduced a factor
$1/4$ in Eq.~(\ref{e:JJ}) to compensate for the factor $1/2$ carried in
Eq.~(\ref{e:isovec}) in the current definition.

It is important to bear in mind that the transverse
form given
by Eq.~(\ref{e:conserved}) holds in the first place for the exact
 vector or axial
vector polarization in a chirally symmetric theory, and therefore must also 
hold for any approximate calculation of
either function. Therefore any approximate treatment has to be implemented in
 such a manner that current
conservation is maintained at each level of approximation: otherwise 
the physical
consequences of the symmetries contained in the Lagrangian could 
become distorted. 
A case in point that is discussed again below is the mixing of the
 $longitudinal$ axial vector
and pseudoscalar pion modes already at the lowest order of approximation in the
minimal NJL model that is necessary to ensure axial
current conservation in the chiral limit \cite{mk94}.

\subsection{Effective chiral Lagrangian}

In order to calculate the two-point correlation functions, we need to
specify the Lagrangian $\cal L$ governing their dynamics. For this purpose,
we identify $\cal L$ with the two-flavor Nambu-Jona-Lasinio Lagrangian,
 that has
been extended to include vector and axial degrees of freedom in a chirally
invariant fashion \cite{weise90,bij94,brz94}. We write
\begin{equation}
{\cal L}_{\rm ENJL} = \bar{\psi} \big[{\rm i} {\partial{\mkern -10.mu}{/}}
- \hat{m}\big] \psi + G_{1} \Big[ (\bar{\psi} \psi)^2 +
  (\bar{\psi} {\rm i} \gamma_5 \mbox{\boldmath$\tau$} \psi)^2 \Big]
 - G_{2} \Big[ (\bar{\psi} \gamma_{\mu} \mbox{\boldmath$\tau$} \psi)^2 +
  (\bar{\psi} \gamma_{\mu} \gamma_5 \mbox{\boldmath$\tau$} \psi)^2 \Big]
\label{e:lagra}
\end{equation}
where the isovector and axial vector terms have to appear with a common
coupling constant $G_2$ in order to retain chiral invariance\footnote{One could
equally well include additional chirally invariant
terms in this Lagrangian, like the separately chirally invariant
isoscalar vector and the axial vector terms (carrying the quantum
numbers of the $\omega$ and $f_1$ mesons).
However, since the important
collective low-lying excitations contained in $ {\cal L}_{\rm NJL}$ are the
pions which are also isovector, the isovector vector mesons are the relevant
degrees of freedom for our purposes.}. In this expression, the $\psi's$ are
iso-doublet, color-triplet Dirac quark fields and $\hat m$ is the averaged
current quark mass $\hat m = (m^0_u + m^0_d)/2$.
We also note that ${\cal L}_{\rm ENJL}$
breaks the $U_A(1)$ symmetry in maximal fashion due to the 't Hooft term
\cite{th76} that it
contains \cite{klev92} and which models the QCD instanton interaction.
The $G's$ should be assumed to
scale like $N_C^{-1}$ if the correct QCD scaling properties for the quark
and meson masses are to be maintained.

In the following, it will be convenient to refer to results that are obtained
with $G_2$ either present or absent as pertaining to the ENJL or NJL model respectively.
Then, apart from the pionic and sigma-like excitations  with
$J^{PC}=0^{-+}$ and $0^{++}$
that are collective modes of the NJL Lagrangian built out of coherent
superpositions of strongly interacting $\bar qq$ pairs of binding energy $2m$ and
zero respectively \cite{njl61}, the ENJL Lagrangian also supports additional collective
modes with $J^{PC}=1^{--}$ and $1^{++}$ that coincide with
the quantum numbers of the $\rho$ and $a_1$ meson.
Although we will continue to refer to these modes as ``meson'' states of the
corresponding name for brevity, to what
extent this identification with the actual physical mesons of the same quantum
numbers is appropriate, is a matter for later discussion.

\subsection{Schwinger-Dyson amplitudes to leading order
${\cal O}(N_c)$ via diagram summation: the RPA approximation}
Due to the specific structure of the Lagrangian in Eq.~(\ref{e:lagra}), the
Schwinger-Dyson [SD] integral equations for the full polarization
functions can be solved explicitly \cite{njl61} in terms of the
lowest order single quark loop irreducible polarization diagrams.
These individual loops are denoted as $\Pi^{JJ^\prime}$ together with
the appropriate additional spinor and flavor structure. They may be decomposed
according to their Lorentz structure as
\begin{mathletters}
\begin{eqnarray}
&&\Pi^{VV}_{\mu\nu;ab}(q^2)=\Pi^{VV}_T(q^2)T_{\mu\nu}\delta_{ab}
\label{e:VV}
\\          
&&\Pi^{AA}_{\mu\nu;ab}(q^2)=[\Pi^{AA}_T(q^2) T_{\mu\nu}+\Pi^{AA}_L(q^2) L_{\mu\nu}]\delta_{ab}
\label{e:AA}
\\
&&\Pi^{AP}_{\mu;ab}(q^2)=\Pi^{AP}(q^2) \hat{q}_\mu \delta_{ab},\qquad \hat{q}_\mu=\frac{q_\mu}{\sqrt{q^2}}
\label{e:AP}
\\
&&\Pi^{PP}_{ab}(q^2)=\Pi^{PP}(q^2)\delta_{ab}
\label{e:PP}
\\
&&\Pi^{SS}_{ab}(q^2)=\Pi^{SS}(q^2)\delta_{ab}.
\label{e:SS}
\end{eqnarray}
\end{mathletters}
The two remaining polarizations $\Pi^{P_0P_0}(q^2)$ and $\Pi^{S_0S_0}(q^2)$ generated by the
density operators $P^0$ and $S^0$ are scalars in both Lorentz and isospin space.
Here $L_{\mu\nu}=\hat{q}_\mu\hat{q}_\nu$ is a longitudinal tensor such that
$T_{\mu\nu}+L_{\mu\nu}=g_{\mu\nu}$.  As has been indicated explicitly, the
irreducible polarizations are all functions of $q^2$.
The full polarization functions that were introduced in Eqs.~(\ref{e:JJ}) to
(\ref{e:JJSP}) and are denoted by $\tilde\Pi^{JJ^\prime}$ have the
same invariant structure as the irreducible one-loop amplitudes that generate them.
Note that the vertices that these one-loop diagrams connect are indicated by the following
superscripts, 
\begin{eqnarray}
V\rightarrow\gamma_\mu\tau_a,\quad A\rightarrow \gamma_\mu\gamma_5\tau_a, 
\quad P\rightarrow i\gamma_5\tau_a, 
\quad S\rightarrow \tau_a,\quad P_0\rightarrow i\gamma_5,
\quad S_0\rightarrow 1.
\end{eqnarray}

All such one-loop diagrams involve traces over all quark variables, which 
make them proportional
to $N_c$. Thus their products with $G_1$ or $G_2$ are independent of $N_c$.
 Hence
the RPA approximation for the polarization amplitudes, which involves a
selective summation of diagrams
to all orders in 
the (strong) interaction parameters $G_1$ and $G_2$, will only contain
interaction effects
to leading order ${\cal O}(1)$ in the expansion in powers of the inverse 
color variable $1/N_c$ if, as here, we
replace the exact irreducible polarization by its one-loop approximation.

We can now write down separate SD integral equations for the pure
vector ($\rho$), pure transverse axial vector,
 and coupled longitudinal axial vector-pseudoscalar ($a_L,\pi$)
 channels. We can do this for either the scattering
amplitudes in these channels or the polarization diagrams. If we do the
latter, then one sees from the structure of ${\cal L}_{\rm NJL}$ that the full
vector polarization is simply given by the geometric progression of one-loop
irreducible amplitudes connected via the interaction in this channel,
\begin{eqnarray}
\tilde\Pi^{VV}_T=  \Pi^{VV}_T-2G_2\Pi^{VV}_T 
\tilde\Pi^{VV}_T=\frac{\Pi^{VV}_T}{1+2G_2\Pi^{VV}_T},
\label{e:vecsolu}
\end{eqnarray}
and is automatically transverse and strictly ${\rm O}(N_c)$ since the
denominator is ${\rm O}(1)$ in the number of colors. We refer to this in the following as the
random phase approximation, or RPA because of its similarity with the
diagram summation technique used in discussing collective modes in
nuclear physics. The axial vector and
pseudoscalar
polarization functions are by contrast coupled by the amplitude $\Pi^{AP}_{\mu;ab}$
that allows for transitions between the longitudinal part of the axial
vector and the pseudoscalar
pion modes. This feature was first observed in the early field theoretic formulations
of Weinberg's model \cite{gg69} where it led to a finite pion field
renormalization. The situation here is very similar although there are no
elementary meson fields present in the Lagrangian. Taking this off-diagonal
polarization amplitude into account,
one has
\begin{eqnarray}
\tilde\Pi^{AA}_{\mu\nu}=\Pi^{AA}_{\mu\nu}+
\Pi^{AA}_{\mu\rho}(-2G_2)g^{\rho\rho^{\prime}}\tilde\Pi^{AA}_{\rho^{\prime}\nu}
+\Pi^{AP}_\mu(2G_1)\tilde\Pi^{PA}_\nu,
\label{e:axvecsolu1}
\end{eqnarray}
after dropping the now redundant isospin labels.
Note that $\Pi^{AP}_\mu\tilde\Pi^{PA}_\nu=\Pi^{AP}\tilde\Pi^{PA}\hat{ q}_\mu \hat{q}_\nu$
is longitudinal. Hence one can split the solution of the above equation into parts
that are purely transverse, and parts that are purely longitudinal:
\begin{eqnarray}
\tilde\Pi^{AA}_T=  \Pi^{AA}_T-2G_2\Pi^{AA}_T \tilde\Pi^{AA}_T=\frac{\Pi^{AA}_T}{1+2G_2\Pi^{AA}_T}
\label{e:AT}
\end{eqnarray}
and
\begin{eqnarray}
\tilde\Pi^{AA}_L=\Pi^{AA}_L-2G_2\Pi^{AA}_L\tilde\Pi^{AA}_L
+2G_1\Pi^{AP}\tilde\Pi^{PA}.
\label{e:AL}
\end{eqnarray}

The pseudoscalar polarization function can likewise be found as the solution
of the pair of coupled SD equations,
\begin{eqnarray}
\tilde\Pi^{PP}&=& \Pi^{PP}+\Pi^{PP}(2G_1)\tilde\Pi^{PP}+\Pi^{PA}(-2G_2)
\tilde\Pi^{AP},
\label{e:PPS}
\end{eqnarray}
while
\begin{eqnarray}
\tilde\Pi^{PA}\hat{q}_\mu&=&\Pi^{PA}\hat{q}_\mu+\Pi^{PA}\hat{q}^\nu(-2G_2)\tilde\Pi^{AA}_{\nu\mu}
+\Pi^{PP}(2G_1)\tilde\Pi^{PA}\hat{q}_\mu
\nonumber
\end{eqnarray}
or
\begin{eqnarray}
\tilde\Pi^{PA}=\Pi^{PA}+\Pi^{PA}(-2G_2)\tilde\Pi^{AA}
+\Pi^{PP}(2G_1)\tilde\Pi^{PA},
\label{e:PAC}
\end{eqnarray}
since $\Pi^{PA\,;\nu}\tilde\Pi^{AA}_{\nu\mu}=\Pi^{PA}\tilde\Pi^{AA}_L\hat{q}_\mu$
projects out onto the longitudinal part of the axial polarization.
Thus we end up with a single equation for the transverse
axial polarization, but three coupled equations (\ref{e:AL}-\ref{e:PAC}) for the pseudoscalar polarization and the
longitudinal part of the axial polarization. These latter equations can be
written as a single matrix equation in the {$P,A$} space as
\begin{eqnarray}
\bf\tilde\Pi=\Pi+\Pi K \tilde\Pi,
\end{eqnarray}
where the boldface symbols have the matrix structure,
\begin{eqnarray}
\bf{\Pi}=
\left(\begin{array}{cc}
\Pi^{PP}&\Pi^{PA}\\
\Pi^{AP}&\Pi^{AA}_L
\end{array}\right)\;,
\qquad
\bf{K}=
\left(\begin{array}{cc}
2G_1 & 0 \\
0 & -2G_2
\end{array}\right).
\label{e:mtrx}
\end{eqnarray}
The $\bf\tilde\Pi$ has the same matrix structure as $\bf\Pi$.
While the solution to this matrix equation may be found without difficulty \cite {vogl91},
it is more instructive to proceed as follows. We split up the matrix of interaction
strengths
as
\begin{eqnarray}
\bf{K}=\bf{K_1+K_2}=
\left(\begin{array}{cc}
2G_1 & 0 \\
0 & 0
\end{array}\right)
 +
\left(\begin{array}{cc}
0 & 0 \\
0 & -2G_2
\end{array}\right),
\end{eqnarray}
and introduce the auxillary matrix of polarization functions $\bf\hat{\Pi}$,
distinguished by carrying a hat, that are generated by the $axial$ vector part
of the interaction acting in isolation. This auxillary matrix satisfies the equation
\begin{eqnarray}
\bf\hat{\Pi}= \Pi+\Pi K_2\hat{\Pi},
\end{eqnarray}
with solution
\begin{eqnarray}
\bf\hat{\Pi}=(1-\Pi K_2)^{-1}\Pi.
\label{e:matrixsolu}
\end{eqnarray}
Subtracting the matrix equations satisfied by $\bf\tilde\Pi$ and
$\bf\hat{\Pi}$ respectively and using the solution for
$\bf\hat{\Pi}$ given above, one finds that
\begin{eqnarray}
\bf\tilde\Pi=\hat{\Pi}+\hat{\Pi}K_1\tilde\Pi,
\end{eqnarray}
with solution
\begin{eqnarray}
\bf\tilde\Pi=(1-\hat{\Pi} K_1)^{-1}\hat{\Pi}.
\label{e:tildesolu}
\end{eqnarray}
This result shows that the role of the longitudinal axial mode can be completely
absorbed in the determination of the auxillary polarization matrix $\hat{\Pi}$.
The effect of this is to renormalize the one-loop polarization functions as follows,
as may be seen by evaluating Eq.~(\ref{e:matrixsolu}):
\begin{eqnarray}
\hat{\Pi}^{PP}&=&\Pi^{PP}-2G_2\frac{\Pi^{PA}\Pi^{AP}}{1+2G_2\Pi^{AA}_L}
\\
\hat{\Pi}^{AP}&=&\frac{\Pi^{AP}}{1+2G_2\Pi^{AA}_L}
\\
\hat{\Pi}^{AA}_L&=&\frac{\Pi^{AA}_L}{1+2G_2\Pi^{AA}_L},
\end{eqnarray}
where the argument $q^2$ has been suppressed for brevity.
Physically this renormalization comes about
 due to the partial summation to infinity
of one-loop diagrams connected by the longitudinal piece of the axial vector
interaction.
Together with the explicit solution for the transverse axial polarization
given by Eq.~(\ref{e:AT}) above, Eqs.~(\ref{e:matrixsolu}) and (\ref{e:tildesolu})
provide us with a complete description of the polarization matrix.
It will turn out that the longitudinal part of $\tilde\Pi^{AA}_{\mu\nu}$ 
vanishes $identically$ in the presence of the pseudoscalar 
interaction term contained in $\cal{L}_{\rm NJL}$ (in the chiral limit).
This is not true of $\hat{\Pi}^{AA}_{\mu\nu}$ that continues to have a longitudinal
component.

Of the remaining polarization functions, only the isoscalar one,
$\Pi^{S_0S_0}$
is modified by the interactions contained in ${\cal L}_{\rm NJL}$. Using the
RPA again, one finds
\begin{eqnarray}
\tilde\Pi^{S_0S_0}(q^2)=\frac{\Pi^{S_0S_0}(q^2)}{1-2G_1\Pi^{S_0S_0}(q^2)}.
\label{e:polsigmabar}
\end{eqnarray}

All that is required now are the explicit forms
of the matrix elements of the polarization at the one loop level to complete
the solution for the various polarization functions. These are given below.

\subsection{Calculation of the one-loop polarization functions}
The one-loop diagrams corresponding to the isospin currents of
Eq.~(\ref{e:isovec}) are given by
\begin{mathletters}
\begin{eqnarray}
\\
&&\Pi^{VV}_{\mu\nu;ab}=
i\int\frac{d^4p}{(2\pi)^4}Tr[\gamma_\mu\tau_a S(p+q)\gamma_\nu\tau_b S(p)]
\label{e:vectorpol}
\\
&&\Pi^{AA}_{\mu\nu;ab}=
i\int\frac{d^4p}{(2\pi)^4}Tr[\gamma_\mu\gamma_5\tau_a S(p+q)\gamma_\nu\gamma_5\tau_b S(p)]
\label{e:axialvectorpol}
\\
&&\Pi^{AP}_{\mu;ab}=i\int\frac{d^4p}{(2\pi)^4}Tr[\gamma_\mu\gamma_5\tau_a S(p+q)i\gamma_5\tau_b S(p)]
\label{e:iap}
\\
&&\Pi^{PP}_{ab}=i\int\frac{d^4p}{(2\pi)^4}Tr[i\gamma_5\tau_a S(p+q)i\gamma_5\tau_b S(p)]
\label{e:ipp}
\\
&&\Pi^{SS}_{ab}=i\int\frac{d^4p}{(2\pi)^4}Tr[\tau_a S(p+q)\tau_b S(p)]
\label{e:s1s1},
\end{eqnarray}
\end{mathletters}
where the trace $Tr$ operation runs over all intrinsic variables, and where
$S(p)=(\not\!p-m+i\epsilon)^{-1}$ is the quark propagator.
The corresponding isoscalar diagrams are
\begin{mathletters}
\begin{eqnarray}
&\Pi^{P_0P_0}&=i\int\frac{d^4p}{(2\pi)^4}Tr[i\gamma_5 S(p+q)i\gamma_5 S(p)]
\\
&\Pi^{S_0S_0}&=i\int\frac{d^4p}{(2\pi)^4}Tr[S(p+q)S(p)].
\end{eqnarray}
\end{mathletters}
We first consider the chiral limit, $\hat{m}=0$, and start
 with the scalar and pseudoscalar polarization diagrams,
as well as the ($\pi,a_L$) coupled diagram.
It has been shown \cite{vogl91,klev92,hk94} that
all these diagrams
may be written in terms of the mean field (or Hartree) expressions
 for the pion
weak decay constant $f_p$, the $\pi q q$ coupling constant $g_p$ and
pion form factor $F_P$ in the chiral limit. One has\footnote{The expressions 
for the scalar and pseudoscalar
amplitudes also contain 
surface terms \cite{will93} that arise because of the momentum
variable shifts that have to be introduced to simplify them using the 
quadratically divergent gap
equation. The surface terms are removed after Pauli-Villars regularization
as discussed in more detail in Ref.~\cite{dstl95}.}
\begin{mathletters}
\begin{eqnarray}
&&\Pi^{PP}=(2G_1)^{-1}+\frac{f^2_p(q^2)}{m^2}q^2 =\Pi^{P_0P_0}
\label{e:HPP}
\\
&&\Pi^{AP}=2if^2_p(q^2)\sqrt{\frac{q^2}{m^2}}=(\Pi^{PA})^*
\label{e:HAP}
\\
&&\Pi^{S_0S_0}=(2G_1)^{-1}+\frac{f^2_p(q^2)}{m^2}(q^2-4m^2)
=\Pi^{SS},
\label{e:HS0S0}
\end{eqnarray}
\end{mathletters}
after introducing the abbreviation
\begin{eqnarray}
f^2_p(q^2)=f^2_pF_P(q^2).
\label{e:fpirun}
\end{eqnarray}
Here
\begin{mathletters}
\begin{eqnarray}
f^2_p=-4iN_cm^2I(0)
\label{e:fpimean}
\\
g^2_p=[-4iN_cI(0)]^{-1}
\label{e:gpimean}
\\
F_P(q^2)=I(q^2)/I(0),
\label{e:ffpimean}
\end{eqnarray}
\end{mathletters}
where $I(q^2)$ is the common loop integral
\begin{eqnarray}
I(q^2)= \int\frac{d^4p}{(2\pi)^4}\frac{1}{[(p+q)^2-m^2][p^2-m^2]}.
\label{e:eye}
\end{eqnarray}
A suitable regularization of this divergent integral is discussed in Appendix A.
Notice that the mean field pion decay constant and pion-quark coupling
satisfy the Goldberger-Treiman relation $f_pg_p=m$ at the quark level, where the quark mass $m$ arises
from spontaneous chiral symmetry
breaking in the groundstate of ${\cal L}_{\rm NJL}$. This mass is self-consistently determined by the single quark self-energy $\Sigma$
as $m=\Sigma(m)$, the so-called ``gap'' equation \cite{njl61}. In the Hartree mean-field approximation, one finds for
either the NJL model or its ENJL extension that
\begin{eqnarray}
m\approx \Sigma_H&=&2G_1 \int\frac{d^4p}{(2\pi)^4}Tr[iS(p)]
\nonumber
\\
&=&16G_1N_cm\int\frac{d^4p}{(2\pi)^4} \frac{i}{p^2-m^2+i\epsilon},
\label{e:gap}
\end{eqnarray}
since only the scalar interaction in ${\cal L}_{\rm NJL}$ is operative in determining
$\Sigma_H$. This equation
leads to the relations $\Pi^{PP}(0)=\Pi^{S_0S_0}(4m^2)=(2G_1)^{-1}$
that have been used to simplify
$\Pi^{PP}$ and $\Pi^{S_0S_0}$ in Eqs.~(\ref{e:HPP}) and (\ref{e:HS0S0}).

We now turn to the vector and axial vector polarization diagrams of Eqs.~(\ref{e:vectorpol}) and (\ref{e:axialvectorpol}).
Commuting $\gamma_5 S(p)=S(-p)\gamma_5$ under the trace in the latter
expression leads to
\begin{eqnarray}
\Pi^{AA}_{\mu\nu;ab}(q^2)-\Pi^{VV}_{\mu\nu;ab}(q^2)=
4f^2_pF_P(q^2)g_{\mu\nu}\delta_{ab}.
\label{e:poldiff}
\end{eqnarray}
One also notes that $\Pi^{VV}_{\mu\nu;ab}$ is purely transverse, as may be established
by contracting with $q^{\mu}$ and using the identity $\not\!\!q=S^{-1}(p+q)-
S^{-1}(p)$. Then
\begin{eqnarray}
q^\mu\Pi^{VV}_{\mu\nu;ab}=i\int\frac{d^4p}{(2\pi)^4}
Tr\left[S(p)\gamma_\nu-S(p+q)\gamma_\nu\right]\delta_{ab}=0,
\label{e:veclong}
\end{eqnarray}
{\it provided} that the necessary shift in the momentum integration variable
in the divergent integral on the right is
allowed. If so, one can conclude that
\begin{eqnarray}
\Pi^{AA}_T(q^2)-\Pi^{VV}_T(q^2)=\Pi^{AA}_L(q^2)=4f^2_p(q^2),
\label{e:ward1}
\end{eqnarray}
by inserting the axial and vector polarizations in terms of
their transverse and longitudinal
parts from Eqs.~(\ref{e:AA}) and (\ref{e:VV}), projecting out with the
help of the relations $T^{\mu\nu}T_{\mu\nu}\;=\;T^{\mu\nu}g_{\mu\nu}=3$ and
$L^{\mu\nu}L_{\mu\nu}\;=\;1$, and using Eq.~(\ref{e:fpirun}).

The relations (\ref{e:ward1}), which are in the nature of Ward identities,
are crucial for proving the Weinberg sum rules \cite{wein67} as well as establishing
the relation between the Feynman diagram structure and the Das {\it et al.}
\cite{low67} formulation for the electromagnetic pion mass splitting in the
ENJL model. Since all these quantities involve formally divergent 
integrals at the one-loop level, their regularization has to be
implemented in such a fashion that Eqs.~(\ref{e:poldiff}), (\ref{e:veclong})
and Eq.~(\ref{e:ward1})
on which
these relations are based, remain valid between the
regulated quantities as well. We discuss this later. The explicit form of
$\Pi^{VV}_T(q^2)$ after regularization is given in the appendix. It is shown
there that $\Pi^{VV}_T(0)=0$. Thus one sees from Eq.~(\ref{e:ward1})
that
\begin{eqnarray}
\Pi^{AA}_T(0)=\Pi^{AA}_L(0)=4f^2_p.
\label{e:ward3}
\end{eqnarray}

The ingredients needed for constructing the polarization
matrix $\bf\Pi$ are now available: the pseudoscalar and mixed polarization
diagrams appear in Eqs.~(\ref{e:HPP}, \ref{e:HAP}), and $\Pi^{AA}_L$ is
already known from Eq.~(\ref{e:ward1}).
The resulting one-loop polarization in ($P,A$) space is 
given by the hermitian matrix
\begin{eqnarray}
\bf{\Pi}=
\left(\begin{array}{cc}
(2G_1)^{-1}+f^2_p(q^2){q^2\over m^2} & -2if^2_p(q^2)\sqrt{{q^2\over m^2}}\\
2if^2_p(q^2)\sqrt{{q^2\over m^2}} & 4f^2_p(q^2)
\end{array}\right).
\end{eqnarray}
After renormalization by the longitudinal axial mode 
according to Eq.~(\ref{e:matrixsolu}), this becomes
\begin{eqnarray}
\bf\hat{\Pi}=
\left(\begin{array}{cc}
(2G_1)^{-1}+f^2_\pi (q^2){q^2\over m^2} & -2if^2_\pi (q^2)\sqrt{{q^2\over m^2}}\\
2if^2_\pi (q^2)\sqrt{{q^2\over m^2}} & 4f^2_\pi (q^2)
\end{array}\right).
\end{eqnarray}
Notice that the $sole$ effect of the
longitudinal mode 
is to replace the factor $f_p(q^2)$ of Eq.~(\ref{e:fpirun}) by
\begin{eqnarray}
f_\pi(q^2)=f_p(q^2)[1+8G_2f^2_p(q^2)]^{-{1\over 2}};
\quad f^2_p(q^2)=f^2_pF_P(q^2).
\label{e:fpisc}
\end{eqnarray}
It is shown later [Eq.~(\ref{e:fpren})] that the chiral limit
$q^2\rightarrow 0$ of this expression
gives the pion weak decay constant $f_\pi$ for the ENJL model.
Since
\begin{eqnarray}
({\bf1-\hat{\Pi} K_1})^{-1}= {\cal D}^{-1}
\left(\begin{array}{lr}
1 & 0\\
4iG_1f^2_\pi (q^2)\sqrt{\frac{q^2}{m^2}} & {\cal D}
\end{array}\right),                                                  
\label{e:inverse}
\end{eqnarray}
where the determinant ${\cal D}$ is
\begin{eqnarray}
{\cal{D}}=Det({\bf1-\hat{\Pi} K_1})=1-2G_1\hat{\Pi}^{PP}(q^2),
\end{eqnarray}
the matrix elements of $\bf\tilde\Pi=\bf(1-\hat{\Pi} K_1)^{-1}\hat{\Pi}$
now follow as
\begin{eqnarray}
\tilde\Pi^{PP}&=&
\frac{\hat{\Pi}^{PP}(q^2)}{1-2G_1\hat{\Pi}^{PP}(q^2)}
\label{e:ppren}
\\
\tilde\Pi^{AP}&=&\frac{2if^2_\pi(q^2)\sqrt{q^2\over m^2}}
{1-2G_1\hat{\Pi}^{PP}(q^2)}
=-\frac{i}{G_1}\sqrt{\frac{m^2}{q^2}}
\\
\tilde\Pi^{PA}&=&(\tilde\Pi^{AP})^*
\\
\tilde\Pi^{AA}_L&=&0,
\label{e:fullong}
\end{eqnarray}
where
\begin{eqnarray}
\hat{\Pi}^{PP}(q^2)=(2G_1)^{-1}+\frac{f^2_\pi(q^2)}{m^2}q^2
\label{e:polren}
\end{eqnarray}
is the renormalized version of the 
pseudoscalar polarization of (\ref{e:HPP}) .
Thus $\bf\tilde\Pi$ has a zero in the lower right hand corner:
the full longitudinal polarization $vanishes$ as it must, to ensure axial
current conservation when the
longitudinal axial vector ($a_L$) mode mixing with $\pi$ 
is properly taken into account.
Note that this conclusion also remains true for the minimal
NJL model, where $G_2=0$ \cite{mk94}.

\subsection{Propagators}
Together with Eqs.~(\ref{e:vecsolu}), (\ref{e:AT}) and (\ref{e:polsigmabar}),
the above suite of equations completes the solutions for the
transverse vector and axial vector, scalar, and the coupled pseudoscalar-longitudinal
axial vector polarization functions for the ENJL model in the RPA approximation.
We now turn to the
task of constructing the associated propagators for these various modes.
\subsubsection{The $\rho$ and transverse $a_1$ mode propagators}
The scattering amplitude of a quark-antiquark pair arising from the
exchange of either vector meson mode has the form
\begin{eqnarray}
(\Gamma^{\mu;a}_J)_{12}\left[{1 \over i}D^{JJ}_{\mu\nu, ab}(q^2)\right]
(\Gamma^{\nu;b}_J)_{34},
\end{eqnarray}
where
\begin{eqnarray}
\Gamma^{\mu;a}_J=\gamma^\mu \tau^a \quad 
{\rm or}\quad \gamma^\mu\gamma_5 \tau^a,
\end{eqnarray}
for the vector or axial vector channels respectively.
The $D^{JJ}_{\mu\nu;ab}=D^{JJ}_{\mu\nu}\delta_{ab}$ defines
 the corresponding meson propagator.
In the RPA approximation,
it is built up from the interaction lines
$-2iG_2$ and $2iG_1$ that connect 0, 1, 2,.... polarization bubbles to the
initial and
final vertices.
In the vector channel, the sequence of contributions for $-iD^{VV}_{\mu\nu}$ reads
\begin{eqnarray}
-iD^{VV}_{\mu\nu}&=&(-2iG_2)g_{\mu\nu}+(-2iG_2)\Big\{\Pi^{VV}_T T_{\mu\nu}
+ (-2G_2) (\Pi^{VV}_T)^2T_{\mu\rho}T^{\rho}_{\nu}+ \cdots\Big\}(-2iG_2)
\nonumber
\\
&=&\frac{(-2iG_2)}{1+2G_2\Pi^{VV}_T}T_{\mu\nu}+(-2iG_2)L_{\mu\nu},
\label{e:vecpro}
\end{eqnarray}
since the irreducible vector polarization is purely transverse. Hence the transverse
and longitudinal parts of the rho propagator are
\begin{eqnarray}
-iD^{VV}_T=\frac{(-2iG_2)}{1+2G_2\Pi^{VV}_T},\qquad -iD^{VV}_L=(-2iG_2).
\label{e:vecprot}
\end{eqnarray}
The corresponding expression for the axial vector propagator is complicated by
the presence of the coupling amplitudes $\Pi^{AP}_\mu$. In place of
Eq.~(\ref{e:vecpro}), one has the series
\begin{eqnarray}
-iD^{AA}_{\mu\nu}=(-2iG_2)g_{\mu\nu}+(-2iG_2){1\over i}
\Big\{\Pi^{AA}_{\mu\nu}
+ \Pi^{AA}_{\mu\rho}(-2G_2)g^{\rho\rho^\prime}\Pi^{AA}_{\rho^\prime\nu}
+ \cdots
\nonumber
\\
+\Pi^{AP}_\mu(2G_1)\Pi^{PA}_\nu+\cdots\Big\}(-2iG_2).
\end{eqnarray}
The series in the curly brackets is just the series for the full axial
polarization $\tilde\Pi^{AA}_{\mu\nu}$ given in Eq.~(\ref{e:axvecsolu1}).
However, as
this is again purely transverse, see Eq.~(\ref{e:fullong}), the results for
the transverse and longitudinal parts of $-iD^{AA}_{\mu\nu}$ are, according to
Eq.~(\ref{e:AT}), identical
with those of $-iD^{VV}_{\mu\nu}$ with $\Pi^{VV}_T$ replaced by $\Pi^{AA}_T$.
Thus
\begin{eqnarray}
-iD^{AA}_T=\frac{(-2iG_2)}{1+2G_2\Pi^{AA}_T},\qquad -iD^{AA}_L=(-2iG_2).
\label{e:axvecpro}
\end{eqnarray}
The longitudinal part of this propagator also has non-vanishing off-diagonal
elements,
as we discuss below.

\subsubsection{The coupled $\pi-a_L$ mode propagators}
As we have seen, the pion and longitudinal axial modes $\pi$ and $a_L$
are
coupled through the off-diagonal elements in $\tilde\Pi$. The corresponding SD equation for the resulting propagator is thus
a matrix equation involving the full polarization matrix $\tilde\Pi$,
\begin{eqnarray}
-i{\bf D}&=&i{\bf K}+i{\bf K}{1\over i}{\bf\Pi}\, i{\bf K} +
i{\bf K}{1\over i}{\bf\Pi}\, i{\bf K}{1\over i}{\bf\Pi}\, i{\bf K}+\cdots
\nonumber
\\
&=&i{\bf K}+i{\bf K}{1\over i}\Big\{{\bf\Pi}
+{\bf K\,\Pi}+{\bf K\,\Pi\, K\,\Pi}+\cdots\Big\}i{\bf K}
\nonumber
\\
&=& i{\bf K}+i{\bf K}{1\over i}{\bf\tilde\Pi}\,i{\bf K},
\end{eqnarray}
which in view of Eq.~(\ref{e:tildesolu}) has the solution 
\begin{eqnarray}
-i{\bf D}=i{\bf K}+i{\bf K}({\bf 1-\hat{\Pi}\,K_1})^{-1}{\bf\hat{\Pi}\,K}.
\end{eqnarray}
Insert the known value of the inverse matrix to obtain
\begin{eqnarray}
-i{\bf D}=-iD^{PP}(q^2)\left(\begin{array}{cc}
A&iB\\
-iB&C
\end{array}\right), 
\end{eqnarray}
where $A,\,B,\,C$ are the real matrix elements 
\begin{eqnarray}
A&=&1 \\
\label{e:A}
B&=&4G_2f^2_\pi(q^2)\sqrt{\frac{q^2}{m^2}} \\
\label{e:B}
C&=&2G_2f^2_\pi(q^2)\frac{q^2}{m^2},
\label{e:C}
\end{eqnarray}
and where the renormalized propagator for a chiral pion,
\begin{eqnarray}
-iD^{PP}(q^2)=\frac{2iG_1}{1-2G_1\hat
{\Pi}^{PP}(q^2)}=\frac{-im^2}{q^2f^2_\pi(q^2)}=
 \frac{-ig^2_{\pi qq}(q^2)}{q^2F_P(q^2)},
\label{e:DDren}
\end{eqnarray}
has been factored out. Here 
\begin{eqnarray}
g_{\pi qq}(q^2)=g_p[1+8G_2f^2_p(q^2)]^{1\over 2}.
\label{e:grun}
\end{eqnarray}
As shown below, the chiral limit of this function
gives the $renormalized$ $\pi qq$ coupling constant $g_{\pi qq}(0)= g_{\pi qq}$.
The auxillary constants
$f_p$ and $g_p$ have thus disappeared from the problem, to be replaced
by their renormalized values $f_\pi$ and $ g_{\pi qq}$. Although this will be
justified later by examining the conserved axial current, one already sees that
these renormalized values also satisfy the GT relation $f_\pi g_{\pi qq}=m$.

The matrix ${\bf D}$ enters into the $\bar qq$ scattering amplitude
and has to be diagonalized
in order to identify the independent $\pi$ and longitudinal axial eigenmodes and
their coupling constants to the quarks.
To this end, we introduce the real symmetric matrix 
\begin{eqnarray}
M= \left(\begin{array}{cc}
A&B\\
B&C
\end{array}\right)
\end{eqnarray}
that has real eigenvalues
\begin{eqnarray}
\lambda_{\pi,a_L}(q^2)={1\over 2}(A+C)\pm {1\over 2}\sqrt{(A-C)^2+4B^2},
\label{e:eigen}
\end{eqnarray}
where the signs correspond to the identification 
of the $\pi$ and $a_L$
eigenmodes
according to the expected limits
$\lambda_\pi\rightarrow 1,\:\lambda_{a_L}\rightarrow 0$ as $G_2\rightarrow 0$.
The orthogonal transformation matrix that renders $\bf M$ diagonal is 
\begin{eqnarray}
{\bf U}=
\left(\begin{array}{cc}
\cos\theta&-\sin\theta\\
\sin\theta&\cos\theta
\end{array}\right)\,, \qquad
\tan 2\theta&=&\frac{2B}{(A-C)}.
\label{e:theta}
\end{eqnarray}
The scattering amplitude can then be written as follows
\begin{eqnarray}
T&=&(i\gamma_5\tau\, ,-i\not\!\hat{q}\gamma_5\tau){\bf M}
\left(\begin{array}{c}
i\gamma_5\tau\\
i\not\!\hat{q}\gamma_5\tau
\end{array}\right) D^{PP}
\nonumber
\\
&=&(i\gamma_5\tau\, ,-i\not\!\hat{q}\gamma_5\tau){\bf U}
\Big\{{\bf U}^T{\bf M}{\bf U}\Big\}{\bf U}^T
\left(\begin{array}{c}
i\gamma_5\tau\\
i\not\!\hat{q}\gamma_5\tau
\end{array}\right) D^{PP}
\nonumber
\\
&=&T_\pi+T_{a_L},
\end{eqnarray}
where
\begin{eqnarray}
T_\pi&=&\left[(\cos\theta -\not\!\hat{q}\sin\theta)i\gamma_5\tau\right]
\otimes\left[(\cos\theta+\not\!\hat{q}\sin\theta)i\gamma_5\tau\right]
\lambda_\pi D^{PP}(q^2) 
\label{e:pimode}
\end{eqnarray}
and
\begin{eqnarray}
T_{a_L}&=&\left[(\sin\theta+\not\!\hat{q}\cos\theta)i\gamma_5\tau\right]
\otimes\left[(\sin\theta-\not\!\hat{q}\cos\theta)i\gamma_5\tau\right]
\lambda_{a_L} D^{PP}(q^2).
\label{e:amode}
\end{eqnarray}
The separate contributions of the propagating pion and longitudinal axial
eigenmodes are now self-evident.
However, since the eigenvalue $\lambda_{a_L}\sim q^2$ as $q^2\rightarrow 0$
according to Eq.~(\ref{e:eigen}), the product $\lambda_{a_L}D^{PP}(q^2)$ does not have
a pole at $q^2=0$. Thus the longitudinal axial scattering amplitude does not represent
a propagating physical particle carrying mass. Nevertheless  this amplitude is an essential
adjunct to the propagating pion eigenmode, which does represent a physical
particle, as far as conservation of the isospin current is concerned,
as well as the interaction of the pion with electromagnetic fields.
These comments will be elucidated in the later sections.
By contrast, from the structure of $D^{PP}(q^2)$,
 it is evident that the pion propagator
part of $T$ does have a pole at $q^2=0$:
the Goldstone mode has been preserved by the RPA. However,
as already noted before (see e.g. \cite{weise90}),
 the coupling constant of the quarks to the
(Goldstone) pion as well as its weak decay constant are renormalized, and an
additional pseudovector
coupling constant, $g_{pv}$, is induced by the $\pi-a_L$ coupling in the
ENJL Lagrangian. One confirms this by examining the residue of $T_\pi$
at the pole $q^2\rightarrow 0$. In this limit, one finds from Eqs.~(\ref{e:eigen}) and
(\ref{e:theta}) that
\begin{eqnarray}
\lambda_\pi&\rightarrow& 1
\\
\cos\theta&\rightarrow& 1
\\
{\sin\theta\over \sqrt{q^2}}&\rightarrow& 4G_2{f^2_\pi\over m}.
\label{e:thetasm}
\end{eqnarray}
Then $T_\pi$ can be recast as \cite{weise90} 
\begin{eqnarray}
T_\pi\rightarrow [(g_{\pi qq}-\not\!q {g_{pv}\over 2m})i\gamma_5\tau]
\otimes[(g_{\pi qq}+\not\!q {g_{pv}\over 2m})i\gamma_5\tau]{1\over q^2}
\label{e:tpi}
\end{eqnarray}
in the vicinity of the pion pole, that allows one to identify the pseudoscalar
and induced pseudovector pion-quark coupling
constants $g_{\pi qq}$ and $g_{pv}$ as
\begin{eqnarray}
g_{\pi qq}=\lim_{q^2\to 0}g_{\pi qq}(q^2)\sqrt{\lambda_\pi(q^2)}\cos\theta
=g_p[1+8G_2f^2_p]^{1\over 2}=\frac{g_p}{\sqrt{g_A}}
\label{e:gren}  
\end{eqnarray}
and
\begin{eqnarray}
g_{pv}= \lim_{q^2\to 0}2m\,g_{\pi qq}(q^2)\sqrt{\lambda(q^2)}\;{\sin\theta\over \sqrt{q^2}}
=g_{\pi qq}(8G_2f^2_{\pi})=g_{\pi qq}(1-g_A).
\label{e:gpv}
\end{eqnarray}
The quark axial form factor $g_A$ that is identified in Eq.~(\ref{e:axialg}) to follow, has
been introduced to simplify the final expressions for these two coupling constants.

\subsection{Chiral symmetry invariance to ${\cal O}(N_c)$ in the  RPA}
Apart from satisfying Goldstone's theorem in the RPA, the ENJL
model also obeys the remaining chiral theorems as realized by the 
Goldberger-Treiman relation and axial current
conservation. The Goldberger-Treiman relation can be established by
 calculating
the one pion to vacuum matrix element of the axial 
current that defines the pion weak decay constant $f_\pi$. In the quark
model,
this matrix element is
determined by connecting the vertex for the incoming pion, in this case
$i(g_{\pi qq}-\not\!\!{q}\;{g_{pv}}/{2m})i\gamma_5\tau_b$,
via a quark loop to the axial current vertex $\gamma_\mu\gamma_5\tau_a/2$.
Since the induced pseudovector term in the $\pi qq$ interaction only
couples to the longitudinal part of the axial polarization,
the resulting single quark loop diagram, can be re-expressed in terms
of the $\Pi^{AP}$ and $\Pi^{AA}_L$.
The details are summarized below.   One has
\begin{eqnarray}
\langle 0|A_{\mu;a}|\pi_b(q)\rangle=if_\pi q_\mu\delta_{ab}
={1\over 2}\Pi^{AP}_{\mu;ab}(0)g_{\pi qq}
-{1\over 2}\Pi^{AA}_{\mu\nu;ab}(0) q^\nu\frac{ig_{pv}}{2m}.
\label{e:axmat}
\end{eqnarray}
But
\begin{eqnarray}
\Pi^{AP}_{\mu;ab}(q^2)=2i\frac{f^2_p(q^2)}{m}q_\mu\delta_{ab}\quad {\rm and}\quad
\Pi^{AA}_{\mu\nu;ab}(q^2) q^\nu=4f^2_p(q^2)q_{\mu}\delta_{ab},
\label{e:aux}
\end{eqnarray}
in view of Eqs.~(\ref{e:HAP}) and (\ref{e:ward1}).
Hence the pion decay constant satisfies the relation 
\begin{eqnarray}                                 
f_\pi={f^2_p\over m}(g_{\pi qq}-g_{pv}).
\label{e:fpiax}
\end{eqnarray}
However
\begin{eqnarray}
(g_{\pi qq}-g_{pv})= g_{\pi qq}g_A
\end{eqnarray}
follows from Eqs.~(\ref{e:gren}) and (\ref{e:gpv}). Inserting this difference
into the expression for $f_\pi$, one finds
\begin{eqnarray}
f_\pi=\sqrt{g_A}f_p=f_p[1+8G_2f^2_p]^{-{1\over 2}}.
\label{e:fpren}
\end{eqnarray}
Thus $f_\pi$ and $g_{\pi qq}$ are again
connected by a Goldberger-Treiman relation,
\begin{eqnarray}
f_\pi g_{\pi qq}=\sqrt{g_A}f_p \frac{g_p}{\sqrt{g_A}}=m.
\label{e:gtenjl}
\end{eqnarray}

We also establish the axial current conservation for the Lagrangian
given by Eq.~(\ref{e:lagra}) and identify the quark isovector axial form
factor. The expected modification of the quark current due to the exchange
of transverse axial $a_1$ plus the $\pi$ and longitudinal
$a_L$ axial
eigenmodes is depicted in Fig.~\ref{f:isospin current}. Since the current operator represented by
the sum of these diagrams is to eventually be sandwiched between
initial and final quark momentum states $p$ and $p^\prime$ where $q=p^\prime-
p$ is the momentum entering the vertex, we can freely use the
identity $\not\!\!q\gamma_5=2m\gamma_5$. Translating the Feynman diagrams in
Fig.~\ref{f:isospin current} with the help of Eqs.~(\ref{e:aux}),  
(\ref{e:pimode}), (\ref{e:amode}) and (\ref{e:axvecpro}) in
terms of the polarization loops and propagators they represent, one finds,
 after
some calculation, that  
\begin{eqnarray}
J^a_{\mu 5} =\gamma_\mu\gamma_5{\tau^a\over 2} -\frac{2m\gamma_5}{q^2}&q_\mu&
\frac{f^2_p(q^2)}{f^2_\pi(q^2)}
\Big\{\lambda_\pi
\big(\cos\theta-\frac{2m}{\sqrt{q^2}}\sin\theta\big)^2
+\lambda_a\big(\sin\theta+\frac{2m}{\sqrt{q^2}}\cos\theta\big)^2\Big\}
{\tau^a\over 2}
\nonumber
\\
&-&\frac{2G_2\Pi^{AA}_T}{1+2G_2\Pi^{AA}_T}T_{\mu\nu}\gamma^{\nu}
\gamma_5{\tau^a\over 2}.
\label{e:ja5}
\end{eqnarray}
The contents of the curly bracket is given by
\begin{eqnarray}
\Big\{\cdots\Big\}=A+\frac{4m^2}{q^2}C-\frac{4m}{\sqrt{q^2}}B
=1-8G_2f^2_\pi(q^2) =\frac{1}{1+8G_2f^2_p(q^2)},
\end{eqnarray}
see Eq.~(\ref{e:curly}) of Appendix A.  This just cancels out the factor
$f^2_p(q^2)/f^2_\pi(q^2)$ in view of Eq.~(\ref{e:fpisc}) and we are left with
\begin{eqnarray}
J^a_{\mu 5}& =&\Big (\gamma_\mu\gamma_5 -\frac{2m\gamma_5}{q^2}q_\mu
-\frac{2G_2\Pi^{AA}_T}{1+2G_2\Pi^{AA}_T}T_{\mu\nu}\gamma^{\nu}\gamma_5\Big){\tau^a\over 2}
\nonumber
\\
&=& \Big (g_{\mu\nu} -\frac {q_\mu q_\nu}{q^2}
-\frac{2G_2\Pi^{AA}_T}{1+2G_2\Pi^{AA}_T}
T_{\mu\nu}\Big)\gamma^{\nu}\gamma_5{\tau^a\over 2}
\nonumber
\\
&=&\Big(\frac{1}{1+2G_2\Pi^{AA}_T}\Big)
T_{\mu\nu}\gamma^{\nu}\gamma_5{\tau^a\over 2}
\nonumber\\
&=&G_A(q^2)T_{\mu\nu}\gamma^{\nu}\gamma_5{\tau^a\over 2}.
\label{e:axcurrent}
\end{eqnarray}
Thus the axial current is purely transverse and consequently conserved,
$q^\mu J^a_{\mu 5}=0$. However the quarks have
acquired an axial form factor
\begin{eqnarray}
 G_A(q^2)=\frac{1}{1+2G_2\Pi^{AA}_T(q^2)}
\label{e:axformfac}
\end{eqnarray}
associated with the $\pi+a_L$ and $a_1$ exchanges that contribute
to the current.  At zero momentum transfer this gives the renormalized
axial coupling as
\begin{eqnarray}
g_A=G_A(0)=[1+2G_2\Pi^{AA}_T(0)]^{-1}=[1+8G_2f^2_p]^{-1}=[1-8G_2f^2_\pi],
\label{e:axialg}
\end{eqnarray}
in view of Eq.~(\ref{e:ward3}) followed by Eq.~(\ref{e:fpren}).
It will prove useful to introduce a renormalized form for the axial form factor 
\begin{eqnarray}
F_A(q^2)=G_A(q^2)/G_A(0)=g_A^{-1}G_A(q^2)
\label{e:axialformfac}
\end{eqnarray}
that has the property
 that $F_A(0)=1$. One notes the essential role of the axial $a_1$  degree of freedom in obtaining
all of these results. In particular, the presence of the diagonalized version of the
longitudinal axial meson propagator in Eq.~(\ref{e:ja5}) is crucial for axial
current conservation. One also observes that
the quark's axial coupling to the current
has been reduced from unity down to $g_A<1$ by the associated meson clouds.
The same remark is true of the weak decay constant that is also reduced
below its mean field value to $f^2_\pi=g_A f^2_p$, as shown in
Eq.~(\ref{e:fpren}). Thus one has to identify the renormalized
$f_\pi$ and not
$f_p$ with the physical value of $\approx 93\,$MeV.

Vector current conservation
follows in the same way, except that here $\rho^0$ exchange replaces
the $\pi$ and axial modes. Otherwise
the calculation is the same as above. However, the ``bare'' current and meson
exchange current are separately conserved in this case.
Translating Fig.~\ref{f:isospin current} once more, one finds
\begin{eqnarray}
J^a_\mu=\Big (g_{\mu\nu} 
-\frac{2G_2\Pi^{VV}_T}{1+2G_2\Pi^{VV}_T}
T_{\mu\nu}\Big)\gamma^{\nu}{\tau^a\over 2}
=\Big(\frac{1}{1+2G_2\Pi^{VV}_T}\Big)T_{\mu\nu}\gamma^\nu\frac{\tau^a}2
=F_V(q^2)T_{\mu\nu}\gamma^{\nu}{\tau^a\over 2},
\nonumber\\
\label{e:vecformfac}
\end{eqnarray}
after invoking the identity $\not\!q=0$ that is appropriate when calculating
$\langle p^\prime|J^a_\mu|p\rangle$. Thus $J^a_\mu$ is also conserved, and
the vector form factor of the quark is
\begin{eqnarray}
F_V(q^2)=\frac{1}{1+2G_2\Pi^{VV}_T(q^2)}.
\label{e:vectorformfac}
\end{eqnarray}
In this case, there is no renormalization at zero momentum transfer, $F_V(0)=1$, since
$\Pi^{VV}_T(0)=0$.

\subsection{Explicit symmetry-breaking and chiral perturbation theory}
The results obtained so far all refer to the chiral limit of vanishing current quark mass.
We close this section by briefly indicating their extension to the case
$\hat{m}\neq0$ that is required for the discussion of some of the sum rules in Section III.
As a byproduct,
this extension will also allow us to make contact and compare with the results of
chiral perturbation theory (CHPT) that has been extensively used as a theory
for studying hadron properties in the non-perturbative regime of QCD
\cite{leutwyler91,burgi96,urech95}.

When a non-zero current quark mass is included, the gap
equation for the new quark mass that replaces the $m$ of Eq.~(\ref{e:gap})
is 
\begin{eqnarray}
m^*-\hat{m} =\Sigma(m^*)\approx \Sigma_H(m^*),
\label{e:ncgap}
\end{eqnarray}
that has a revised solution $m^*\neq m$. The changed gap equation alters the
structure of the  pseudoscalar
polarization from that given by Eq.~(\ref{e:polren}) to
\begin{eqnarray}
\Pi^{PP}_{nc}
=\big\{\frac{m-\hat{m}}{2G_1m}+\frac{f^{2}_\pi
(q^2)}{m^{2}}q^2\big\}_{m=m^*},
\label{e:polrenstar}
\end{eqnarray}
evaluated at the new mass $m^*$.   The new index $nc$ explicitly denotes
the use of
 non-chiral functions.
Carrying out the same set of matrix multiplications as those leading to
Eqs.~(\ref{e:ppren})-(\ref{e:fullong}), it is found that the non-chiral
expressions for $\tilde\Pi^{PP}$ and $\tilde\Pi^{AP}$ have exactly
the same form as their chiral counterparts, but
with $\hat{\Pi}^{PP}$ replaced by 
$\Pi^{PP}_{nc}$. 
Furthermore, the longitudinal polarization now no longer vanishes
since the associated axial current is not conserved.
The final expressions for the non-chiral case are thus
\begin{eqnarray}
&&\tilde\Pi^{PP}_{nc}=
\frac{\Pi^{PP}_{nc}(q^2)}{1-2G_1\Pi^{PP}_{nc}(q^2)}
\label{e:polstar}
\\
&&\tilde\Pi^{AP}_{nc}= \frac{2if^2_\pi(q^2)\sqrt{q^2\over m^2}}{1-2G_1\Pi^{PP}_{nc} (q^2)}
\\
&&\tilde\Pi^{PA}_{nc}=\frac{-2if^2_\pi(q^2)\sqrt{q^2\over m^2}}{1-2G_1\Pi^{PP}_{nc} (q^2)}
\\
&&\tilde\Pi^{AA}_{L;nc}
=\frac{4f^2_\pi(q^2)}{1-2G_1\Pi^{PP}_{nc}(q^2)}\;{\hat{m}\over m},
\label{e:star}
\end{eqnarray}
that do not permit further simplification. All amplitudes now have to be
evaluated for a quark mass $m=m^*$ given by the root of
Eq.~(\ref{e:ncgap}).

The propagator matrix representing the mixing of the pion and longitudinal axial
vector modes has the same form as before,
\begin{eqnarray}
-i{\bf D}_{nc}=-iD^{PP}_{nc}(q^2)\left(\begin{array}{cc}
A&iB\\
-iB&C
\end{array}\right) 
\end{eqnarray}
with $A$ and $B$ unchanged except for being evaluated at $m=m^*$, but $C$
contains an additional term that is proportional to $\hat{m}$,
\begin{eqnarray}
A&=&1 \\
B&=&4G_2f^2_\pi(q^2)\sqrt{\frac{q^2}{m^{*2}}} \\
C&=&2G_2\big[\frac{f^2_\pi (q^2)}{m^{*2}}q^2-(2G_1)^{-1}
{\hat{m}\over m^*}[1-8G_2f^2_\pi (q^2)]\big].
\label{e:abcnc}
\end{eqnarray}
The propagator for a non-chiral pion,
\begin{eqnarray}
-iD^{PP}_{nc}(q^2)=\frac{2iG_1}{1-2G_1\Pi^{PP}_{nc}(q^2)},
\label{e:DDnc}
\end{eqnarray}
has been extracted.
The pion pole has thus been shifted away
from the Goldstone mode at $q^2=0$
to some finite value at $q^2=m^2_\pi$  that gives the mass of the ENJL pion.
To lowest order in the current quark mass, this pole lies at 
\begin{eqnarray}
m^2_\pi=\frac{m\hat{m}}{2G_1f^2_\pi}+O(\hat{m}^2).
\label{e:enjlms}
\end{eqnarray}
The corresponding pole for the pion mass of the minimal NJL model lies at
\begin{eqnarray}
m^2_p=\frac{m\hat{m}}{2G_1f^2_p}+O(\hat{m}^2).
\label{e:njlms}
\end{eqnarray}
The quark mass in these expressions now
refers to the mass obtained from the gap equation in the chiral limit,
$\hat{m}=0$.  Using the mean field result,
 $m= -2G_1\langle\bar\psi\psi\rangle$,
that connects $m$ to the condensate density in either the NJL or the ENJL
version of the model in the chiral limit, one can recast
these equations as
\begin{eqnarray}
m^2_\pi f^2_\pi=m^2_pf^2_p=
\frac{m\hat{m}}{2G_1}=-\hat{m}\langle\bar\psi\psi\rangle+O(\hat{m}^2)
\label{e:gmor}
\end{eqnarray}
to confirm that the 
Gell-Mann Oakes Renner current algebra relation \cite{gmor68} is satisfied by
either version of the NJL model \cite{vogl91,klev92}.

The diagonalization of the propagator matrix proceeds as before, and one regains
Eqs.~(\ref{e:pimode}) and (\ref{e:amode}) with the rotation angle calculated
using the revised values of $A\,,B\,,C$ and with $D^{PP}$ replaced by
$D^{PP}_{nc}$. The non-chiral analogs of the pseudoscalar and induced pseudovector coupling constants
of Eqs.~(\ref{e:gren}) and (\ref{e:gpv}) are related to the residue at the
pion pole of $D^{PP}_{nc}(q^2)$.
If we use capital letters to distinguish the non-chiral coupling
constants from the chiral ones, then
\begin{eqnarray}
G_{\pi qq}=
\lim_{q^2\to m^2_\pi}\big[\partial\Pi^{PP}_{nc}/\partial q^2\big]^{-{1\over 2}}
\sqrt{\lambda_\pi(q^2)}\,\cos\theta
\end{eqnarray}
and
\begin{eqnarray}
G_{pv}=\lim_{q^2\to m^2_\pi}2m^*
\big[\partial\Pi^{PP}_{nc}/\partial q^2\big]^{-{1\over 2}}
\sqrt{\lambda_\pi(q^2)}\;\frac{\sin\theta}{\sqrt{q^2}}.
\end{eqnarray}
The modified quark mass $m^*$ enters into all these expressions
in a very complicated way. However, since the current quark
mass ($\hat{m}\sim 3-5\,$ MeV)
is expected to be of order of $(2-3)\%$ of the quark mass that is generated
dynamically via the chiral gap equation, a perturbative approach suggests itself
\cite{mk94} that is analogous to the procedure used in CHPT. This approach seeks
to express the non-chiral amplitudes in terms of their chiral counterparts to
first non-vanishing order in the current quark mass $\hat{m}$. Since the
relevant physical values of $q^2\sim m^2_\pi$ are always of $O(\hat{m}$), which in turn
are of the same order as the shift $\delta m=(m^* -m)$ away from the chiral mass,
all amplitudes have to be expanded simultaneously in $q^2$ $and$ $\delta m$
about zero.  Basic to these expansions are the following results:

\vspace{0.5cm}
\noindent (i) {\it The mass shift.} The mass shift $\delta m$
to $O(\hat{m})$ is obtained by solving the non-chiral gap equation (\ref{e:ncgap})
to this order. One finds \cite{mk94} $\delta m\approx {\hat{m}}/{8G_1f^2_p}$,
a  result that can be rewritten in either of the two equivalent forms
\begin{eqnarray}
\delta m=\frac{m_p^2}{4m},\quad \quad {\rm or}\quad \quad
\delta m=\frac{g_Am^2_\pi}{4m},
\label{e:deltam}
\end{eqnarray}
that lie between 20 and 13 MeV depending on the parameters used,
see  Table~\ref{table1}.
The physical 
origin of the mass shift $m\rightarrow m^*$ that is induced
  can be understood directly in terms of an additional 
self-energy change due to the $\sigma$ mode being exchanged between the quark
and an external scalar field $S_0=\bar\psi\psi$,
 that is associated with the current quark mass
in the Lagrangian.  The diagrams entering the gap equation that determine 
$m^*$ to ${\cal O}(\hat m)$ are shown in Fig.~\ref{f:currentmass}.
The  additional quark self-energy that is due to the  non-vanishing current
 quark mass is given 
diagramatically by Fig.~\ref{f:currentmass} (b) and (c):
translating these Feynman diagrams, one obtains 
\begin{eqnarray}
-i\delta m&=&-i\hat{m}+(-i\hat{m})[-i\Pi^{S_0S_0}(0)][-iD^{S_0S_0}(0)]
\nonumber
\\
&=&\frac{-i\hat{m}}{1- 2G_1\Pi^{S_0S_0}(0)},
\end{eqnarray}
since the $\sigma$ propagator is given by 
\begin{eqnarray}
-iD^{S_0S_0}(q^2)=\frac{2iG_1}{1-2G_1\Pi^{S_0S_0}(q^2)}
=\frac{-im^2}{(q^2-4m^2)f^2_p(q^2)},
\end{eqnarray}
in terms of the scalar polarization of Eq.~(\ref{e:HS0S0}).
The $\sigma$ exchange has 
generated an effective scalar form factor $[1-2G_1\Pi^{S_0S_0}(0)]^{-1}$ that
modifies the bare
current quark mass vertex in an analogous way to what was found for the vector and axial
vector current vertices, and its structure as a geometric series in $G_1$
confirms that the calculation of $\delta m$ is still non-perturbative in the
strong coupling constant. Inserting the
zero momentum transfer value of the combination 
$1-2G_1\Pi^{S_0S_0}(0)=8G_1f^2_p$ from  Eq.~(\ref{e:HS0S0}),
one regains Eq.~(\ref{e:deltam}).

\vspace{0.5cm}
\noindent(ii) {\it Non-chiral corrections to the pion-quark coupling constants.} The revision of the pion
polarization to order $O(\hat{m}^2)$ is obtained by expanding
$\Pi^{PP}_{nc}$ to lowest in $q^2$ and $\delta m$. To do so,
 first calculate the
variation of the basic integral $I(q^2)$ in Eq.~(\ref{e:eye}) with $q^2$ and $m^2$. We temporarily write $I(q^2,m^{*2})$
in order to emphasize the two independent variables involved
in this function, and obtain
\begin{eqnarray}
I(q^2,m^{*2})=I(0,m^2)+\frac{i}{(4\pi)^2}\frac{q^2}{6m^2}
-\frac{i}{(4\pi)^2}\frac{\delta m^2}{m^2}+\cdots,
\end{eqnarray}
with the help of the relations
\begin{eqnarray}
\frac{\partial I(0,m^2)}{\partial q^2}= \frac{i}{(4\pi)^2}\frac{1}{6m^2}\,;
\qquad
\frac{\partial I(0,m^2)}{\partial m^2}=-\frac{i}{(4\pi)^2}\frac{1}{m^2},
\nonumber
\end{eqnarray}
that are easily established.
Here $m^{*2}=m^2+\delta m^2+O(\hat{m}^2)$, where $\delta m^2=2m\delta m 
={1\over 2}m^2_p$,
from Eq.~(\ref{e:deltam}).
Hence $f^2_p(q^2)=-4iN_cm^{*2}I(q^2,m^{*2})$ in Eq.~(\ref{e:fpirun}) has the expansion
\begin{eqnarray}
f^2_p(q^2,m^{*2})= f^2_p\big[1+\frac{q^2}{8\pi^2f^2_p}-
\frac{3m^2_p}{8\pi^2f^2_p}+
\frac{m^2_p}{2m^2}+O(\hat{m}^2)\big],
\end{eqnarray}
from which one concludes that $f^2_\pi(q^2)$ of Eq.~(\ref{e:fpisc}) behaves like
\begin{eqnarray}
f^2_\pi(q^2,m^{*2})=f^2_\pi\big[1+\frac{g^2_A}{8\pi^2f^2_\pi}q^2+\frac{g^2_A}{2m^2}
m^2_\pi-\frac{3g^3_A}{8\pi^2f^2_\pi}m^2_\pi+O(\hat{m}^2)\big].
\label{e:fpiexp}
\end{eqnarray}
We have set $N_c=3$ in both of these expansions, and used the relations
$m^2_p=g_Am^2_\pi$ and $f^2_p=f^2_\pi/g_A$ to eliminate the NJL values of the
pion mass and decay constants in favor of those pertaining to the chiral ENJL
case.
Then one finds that the pseudoscalar polarization has the expansion
\begin{eqnarray}
\Pi^{PP}_{nc}(q^2,m^{*2})&=&\frac{m^*-\hat{m}}{2G_1m^*}+q^2\frac{f^2_\pi}{m^2}\big[
1+\frac{g^2_A}{8\pi^2f^2_\pi}q^2-\frac{3g^3_A}{8\pi^2f^2_\pi}m^2_\pi-
\frac{g_A(1-g_A)}{2m^2}m^2_\pi+O(\hat{m}^2)\big]. \nonumber \\
&&
\end{eqnarray}
Accordingly, the derivative of this expression at the ENJL pion mass $m_\pi$ that enters
the determination of the
pseudoscalar and pseudovector coupling constants is
\begin{eqnarray}
\big[\partial \Pi^{PP}_{nc}(q^2,m^{*2})/
\partial q^2\big]_{q^2\rightarrow m^2_\pi}&=&
g^{-2}_{\pi qq}
\big[1-g^2_A(3g_A-2)
\frac{m^2_\pi}{8\pi^2f^2_\pi}-g_A(1-g_A)\frac{m^2_\pi}{2m^2}+
O(\hat{m}^2)\big]. \nonumber \\
&&
\end{eqnarray}
Now,
the final determination of the coupling constants, and thus also the non-chiral corrections
to the pion weak decay constant, also requires the expansion of the root
$\lambda_\pi(m_\pi^2)$ as well as $\sin\theta$ and $\cos\theta$ to lowest order in $m^2_\pi$.
To obtain these, we need the
expanded forms of the coefficients $A,B$ and $C$ of Eqs.~(\ref{e:abcnc}).
One then finds  that
\begin{eqnarray}
&&\lambda_\pi(m^2_\pi)\approx 1+(4G_2f^2_\pi)^2\frac{m^2_\pi}{m^2}\,,\quad
\sqrt{\lambda_\pi(m^2_\pi)}\cos\theta \approx 1
\nonumber
\\
&&(2m^*/m_\pi)\sqrt{\lambda_\pi(m^2_\pi)}\sin\theta
\approx 8G_2f^2_\pi(m^2_\pi,m^{*2})
\label{e:expans} 
\end{eqnarray}
to lowest order in $\hat{m}$,
after dropping terms of $O(\hat{m}^2)$. Here,
the function $f^2_\pi(m^2_\pi,m^{*2})$ is
given by the expansion in Eq.~(\ref{e:fpiexp}) evaluated at $q^2=m^2_\pi$.
 Hence the coupling constants take the form
\begin{eqnarray}
G_{\pi qq}=g_{\pi qq}\big[1+g^2_A(3g_A-2)\frac{m^2_\pi}{16\pi^2f^2_\pi}
+g_A(1-g_A)\frac{m^2_\pi}{4m^2}+O(\hat{m}^2)\big]
\end{eqnarray}
and
\begin{eqnarray}
G_{pv}=g_{\pi qq}(1-g_A)\big[1-3g^3_A\frac{m^2_\pi}{16\pi^2f^2_\pi}
+g_A(1+g_A)\frac{m^2_\pi}{4m^2}+O(\hat{m}^2)\big].
\end{eqnarray}
As a check, notice that this last relation correctly reduces to (\ref{e:gpv}) in the chiral limit.

\vspace{0.5cm}
\noindent(iii) {\it Non-chiral corrections to the pion mass and weak decay constant.}
The values of the physical pion mass and weak decay constant
$M_\pi$ and $F_\pi$, expressed 
 as expansions in powers of $\hat{m}$, can now be found. The
weak decay constant is obtained by evaluating the right hand side of Eq.~(\ref{e:axmat})
at $q^2=m^2_\pi$ instead of zero, while replacing $g_{\pi qq}-g_{pv}$
by $G_{\pi qq}-G_{pv}$ and $m$ by $m^*$. One obtains
\begin{eqnarray}
F_\pi=f_\pi\big[1+\frac{m^2_\pi}{16\pi^2f^2_\pi}
(\underbrace{4\pi^2\frac{f^2_\pi}{m^2}g_A^2-3g_A^3}_{\sim l^r_4})
+ O(\hat{m}^2)\big],
\label{e:physfpi}
\end{eqnarray}
to the indicated order. The corrections to the ENJL pion mass of
Eq.~(\ref{e:enjlms}) follow upon expanding the defining equation
for the value of the pole in the
non-chiral propagator $D^{PP}_{nc}(q^2)$ to higher powers
than the first in $\hat{m}$:
\begin{eqnarray}
M^2_\pi&=&\frac{m^*\hat{m}}{2G_1f^2_\pi(M^2_\pi,m^{*2})}
\nonumber\\
&=&m^2_\pi\big[1-\frac{m^2_\pi}{32\pi^2f^2_\pi}
\big(\underbrace{8\pi^2\frac{f^2_\pi}{m^2}g_A(2g_A-1)-4g^2_A(3g_A-1)}_{\sim l^r_3}\big)+O(\hat{m}^2)\big]
\label{e:physmass}
\end{eqnarray}
after using Eq.~(\ref{e:enjlms}) to eliminate the combination
$m\hat{m}/(2G_1f^2_\pi)=m^2_\pi$.

\vspace{0.5cm}
\noindent(iv) {\it The Gasser-Leutwyler coupling constants of chiral 
perturbation theory.}
The coefficients of the $m^2_\pi$ terms in expansions like
Eqs.~(\ref{e:physfpi}) and (\ref{e:physmass}) that have been underscored
with braces can be used to identify the ENJL model predictions for two 
of the coupling constants,
$\bar l_3$ and $\bar l_4$
of CHPT that were introduced by Gasser and Leutwyler \cite{gl84}.
However, 
as they stand,
these equations actually determine the renormalized
 coupling constants $l^r_i$, and not
the scale-invariant versions $\bar l_i$ of Gasser and Leutwyler,
 which are the
physical parameters to be extracted from experiment.
The price one pays for introducing the scale invariance in
$\bar l_i$ is that the latter coupling constants then
all diverge logarithmically like $-\ln m^2_\pi$ in the chiral limit. Such
``chiral logarithms'', that are expected on general grounds \cite{lp73},
only appear in the NJL model or its ENJL extension when non-chiral
${\cal O}(1/N_c)$ corrections to the mean field results are included.
 As we have
 not incorporated such 
corrections into our mean field
treatment of the problem, we must identify the underscored
quantities not with $\bar l_i$, but with the combination
\begin{eqnarray}
\bar l_i+\ln( {m^2_\pi}/{\mu^2}) \sim l_i^r,
\label{e:lbar} 
\end{eqnarray}
which is well-defined in the chiral limit. Here $\mu$ is the as yet arbitrary scale parameter
of CHPT\footnote{The arbitrariness in the choice of the scale $\mu$
is reflected in the slow logarithmic dependence on it and the overall smallness
of the logarithmic correction in any event.   This explains the good
agreement of previous calculations with the empirical values of the 
$\bar l_i's$, see \cite{mk94}.}.
This way of including the chiral logarithms has been verified
explicitly in the calculation of the pion radius parameter $\bar l_6$ in \cite{kh95}, and for
the pion electromagnetic mass difference in \cite{dstl95}. The same conclusion holds
for the remaining coupling constants.
We thus have
\begin{eqnarray}
\bar l^{ENJL}_3+\ln ({m^2_\pi}/{\mu^2})&=&
\big\{8\pi^2\frac{f^2_\pi}{m^2}(2g_A-1)-4g_A(3g_A-1)\big\}g_A
\label{e:l3}
\\
\bar l^{ENJL}_4+\ln ({m^2_\pi}/{\mu^2})&=&\big
\{4\pi^2\frac{f^2_\pi}{m^2}-3g_A\big\}g_A^2.
\label{e:l4}
\end{eqnarray}
These expressions reduce to the values given previously 
\cite{mk94} (in a somewhat different form)
for the NJL model ($g_A\rightarrow 1$, $f^2_\pi
\rightarrow f^2_p$ and $m^2_\pi\rightarrow
m^2_p$):
\begin{eqnarray}
\bar l^{NJL}_3+\ln ({m^2_p}/{\mu^2})&=&8\pi^2\frac{f^2_p}{m^2}-8
\\
\bar l^{NJL}_4+\ln ({m^2_p}/{\mu^2})&=&4\pi^2\frac{f^2_p}{m^2}-3.
\end{eqnarray}
In Eqs.(\ref{e:l3}) and (\ref{e:l4}), the combinations on the left are
specified as functions of two physically measurable quantities, the pion
decay constant $f_\pi$ as well as the axial form factor $g_A$.   The only
unknown factor is the quark mass $m$.     Numerical values for these 
quantities will be discussed in the next section, after the parameters of
the model have been fixed.

\subsection{Parameters.}
The determination of the parameters appearing in the effective Lagrangian 
follows the
standard procedure of attempting to reproduce some of the experimental
 properties of
the mesons and quarks that can be calculated from the model.
In the case of the ENJL model with the chiral symmetry-breaking  current quark
mass included, one has the two interaction strengths,
$G_1$ and $G_2$,
an averaged current quark mass $\hat m$,  and a
regulating cutoff $\Lambda$, to determine, i.e. four parameters in all.
On the other hand, it is clear that the $\bar l_i$'s derived above to
${\cal O}(\hat m^2)$  only refer to the solution for $m$ from the gap equation
(\ref{e:gap}) in the absence of a current quark mass.
This is the general situation. We may thus ignore the effect of $\hat m$ on
 the
quark mass, and instead regard $\hat m$ as interchangeable
 with the pion mass
$m_\pi$ in view of the GMOR relation (\ref{e:gmor})
rather than as an additional parameter.
This then leaves three parameters to fix: $G_1\,,G_2$ and $\Lambda$.
In order to determine these, we use the physical
value $F_\pi\approx 93$MeV of the pion  decay
constant\footnote{A more accurate value is $F_\pi=92.4$MeV,
see Ref.~\cite{hol90}.}  to fix $f_\pi$, coupled with the
inferred value of the condensate density per light quark flavor of
${\langle\bar uu\rangle}\approx {\langle\bar dd\rangle}\sim
(-200$MeV$)^3$ to $(-300$MeV$)^3$, and a typical value for the quark
axial form factor $g_A$ to set these parameters. Choosing the value of $g_A$
is, according to Eq.~(\ref{e:axialg}), equivalent to choosing a value
for the interaction strength $G_2$ multiplying the vector fields,
 once $f_\pi$ is
fixed.
We take  $g_A=0.75$, since this
 leads to a correct estimate of the nucleon axial form
factor $g^N_A =(5/3)g_A\approx 1.26$ using the naive quark model
formula \cite{weise90,mr74}. Then $G_2$ is fixed at $G_2=(1-g_A)/(8f^2_\pi)=
3.61$GeV$^{-2}$.
The ratio $f^2_\pi/m^2$ follows  
 from Eq.~(\ref{e:fpimean}) plus $f^2_\pi=g_Af^2_p$.
We use the Pauli-Villars (PV) regularization scheme for this calculation. 
The PV regulated version of Eq.~(\ref{e:fpimean}) is given by 
Eq.~(\ref{e:regpicoupling}) in the appendix.
One has
\begin{eqnarray}
\frac{f^2_{\pi}}{m^2}&=&-\frac{3g_A}{4\pi^2}\sum_{a=0}C_a\ln\,\frac{M^2_a}{m^2}
=\frac{3g_A}{4\pi^2}L(x)
\nonumber\\
L(x)&=&2\ln(1+x)-\ln(1+2x),\quad x=\frac{\Lambda^2}{m^2},
\label{e:ratio}
\end{eqnarray}
for $N_c=3$, after inserting the standard choice \cite{iz80} $C_a=(1,1,-2)$ and $M^2_a=m^2+\alpha_a\Lambda^2$
with $\alpha=(0,2,1)$ for the PV regularization\footnote{As an aside we comment
that, when regulated in $O(4)$ momentum space, the corresponding
expression for $L(x)$  is
\begin{eqnarray}
L_{O(4)}(x)=\ln(1+x)-x/(1+x)
\nonumber
\end{eqnarray}
Here $x=\Lambda^2_{O(4)}/m^2$  and $\Lambda_{O(4)}$ is a new cutoff.
The scaling $\Lambda^2_{O(4)}\rightarrow (2\,\ln2)\,\Lambda^2$ essentially
maps this function onto $L(x)$, i.e. $L_{O(4)}[(2\,\ln2)\,x]\approx L(x)$
to a very high degree of approximation, see Fig.~13 of
Ref.~ \cite{dstl94}. The results to follow are thus essentially unchanged
in the $O(4)$ regularization scheme,
 apart from requiring a larger cutoff. However, for the reasons already 
outlined in the
Introduction, we prefer to use the PV regularization.}.
Thus $f^2_\pi/m^2$,
 and therefore also the combination $\bar l_i+\ln(m^2_\pi/\mu^2)$ are only functions
of the one remaining dimensionless combination $x=\Lambda^2/m^2$, i.e.
the
ratio 
of the square of the  regularization scale to the spontaneously
generated quark mass at that scale.   Thus one may write
\begin{eqnarray}
\bar l^{ENJL}_3(x)+\ln(m^2_\pi/\mu^2)&=&6g^2_A(2g_A-1)L(x)-4g^2_A(3g_A-1)
\label{e:l3x}
\\
\bar l^{ENJL}_4(x)+\ln(m^2_\pi/\mu^2)&=&\big(3L(x)-3\big)g^3_A,
\label{e:l4x}
\end{eqnarray}
which reduce to the especially simple expressions 
\begin{eqnarray}
\bar l^{NJL}_3(x) +\ln(m^2_p/\mu^2)&=&6L(x)-8
\\
\bar l^{NJL}_4(x) +\ln(m^2_p/\mu^2)&=&3L(x)-3
\label{e:l4nx}
\end{eqnarray}
for the NJL values of these coefficients.

For fixed $f_\pi$ and $g_A$, Eq.~(\ref{e:ratio}) determines how $m$ must vary with the
ratio $x$. The corresponding value of the strong coupling $G_1$ that is
required to reproduce this value of $m$ follows from the gap equation
that $m$ must satisfy.  Using the PV regulated version of the gap equation
given in the appendix as Eq.~(\ref{e:densconnect})  for $N_c=3$, one finds
\begin{eqnarray}
G_1^{-1}=\frac{3m^2}{\pi^2}\big[2x\ln\big\{
\frac{1+2x}{1+x}\big\}-L(x)\big].
\label{e:G1PV}
\end{eqnarray}
Knowing $G_1$ and $m$, the  quark condensate density
per light quark
flavor is then given by
\begin{eqnarray}
-\langle \bar uu\rangle=-\langle \bar dd\rangle=\frac{m}{4G_1}+O(\hat{m}).
\label{e:cond}
\end{eqnarray}

The variation with $x=\Lambda^2/m^2$ of the quark mass $m$ and
condensate density is shown in Fig.~\ref{f:nchrlm} for the ENJL model. The input
parameters are as stated above: $f_\pi=93$MeV and $g_A=0.75$.
If we  insist that  $f_p$ must  take on the value of the physical pion decay constant
of $93$MeV for the minimal NJL model case also, then
both these
quantities simply scale down with  $\sqrt{g_A}$:
$m_{NJL}=\sqrt {g_A}\,m_{ENJL}$ and $\langle\bar qq\rangle^{1/3}_{NJL}
=\sqrt {g_A}\,\langle\bar qq\rangle^{1/3}_{ENJL}$, so that
$\Lambda_{NJL}=\sqrt{g_A}\,\Lambda_{ENJL}$,
and $\{G_1\}_{NJL}=
g^{-1}_A\,\{G_1\}_{ENJL}$. It is interesting that
the $\rho$ meson degree of freedom takes no part in this scaling.
One notes that both $m$ and the quark condensate are slow functions of $x$:
a fourfold change from $x=4$ to $x=16$ can
accomodate an uncertainty
in the empirical
value of the condensate density that lies between
 $-(300$MeV$)^3\le \langle\bar uu\rangle\le
-(250$MeV$)^3$.
A comparison of the parameters and the 
numerical results for the two 
cases is given in Table~\ref{table1} for $x=16$.

In Tables~\ref{table2a} and \ref{table2b},
 we have compiled the
analytic expressions and predicted numerical values for the  
Gasser-Leutwyler scale-independent coupling parameters 
as extracted from Eqs.~(\ref{e:l3x}),
(\ref{e:l4x}) for $\bar l_3$ and $\bar l_4$  for both the NJL and ENJL
models, and compared these values with
experiment.    In addition, we have included the values of $\bar l_5$,
$\bar l_6$ and $l_7$ (for which there is no barred version) and which are
calculated from Eqs.~(\ref{e:l5})
and (\ref{e:l6}) as well as Eq.~(\ref{e:l7}) in the sections to come.

 The predicted values for these parameters have been calculated
using the same input data as for Table~\ref{table1}.
The logarithmic contribution is shown separately in this table.
This contribution has been calculated
at the scale $\mu=2m_\sigma\approx 2m = 485$MeV or $528$MeV
respectively in these two cases, that is suggested by the structure of the pion radius
calculation \cite{kh95}. Gasser and Leutwyler fix 
$\mu=M_{\eta}=549$MeV at the $\eta$ meson mass.  From this table, one notes (i) that
either version of the model gives very similar predictions for the coupling
constants of chiral perturbation theory, and (ii) that the
agreement with the empirical data is quite acceptable, the more so since no attempt
was made to fit any of the $l's$ directly.   The only exception to this is 
$\bar l_3$, which is more than twice as large as the suggested empirical
value which, however, carries a large error.

\section{Spectral density functions and sum rules}
In this section, we construct the spectral density functions
 that give the meson mode (mass)$^2$ strength distributions in the various
 channels,
 and discuss their sum rule properties.
Six such spectral functions can be defined and there are eight sum rules
linking them that we discuss.
We first introduce the spectral functions, and we then evaluate them in the
ensuing subsections within the ENJL model.

We start with the isovector vector and axial vector polarization 
given by Eq.~(\ref{e:JJ}) and consider their spectral densities $\rho^J$. We write

\begin{mathletters}
\begin{eqnarray}
&&{1\over 2\pi}\int{d^4 x}\,e^{iq\cdot x}\langle 0|V^a_\mu(x) V^b_\nu(0)|0\rangle =
\rho^V_1(q^2)(-T_{\mu\nu})\delta^{ab}
\\
&&{1\over 2\pi}\int{d^4 x}\,e^{iq\cdot x}\langle 0|A^a_\mu(x) A^b_\nu(0)|0\rangle =
\rho^A_1(q^2)(-T_{\mu\nu})\delta^{ab}+\rho^A_0(q^2)L_{\mu\nu}\delta^{ab}.
\end{eqnarray}
\end{mathletters}
The subscripts $0$ and $1$ indicate the spin projection carried by these spectral
densities.  Since the normalized transverse and longitudinal projectors
$T_{\mu\nu}$ and $L_{\mu\nu}$ where $T_{\mu\nu}+L_{\mu\nu}=g_{\mu\nu}$ have been used,
these identifications of the vector and axial vector
densities agree with those introduced in Weinberg's original paper on sum rules,
and equal $q^2$ times the vector densities of Gasser and Leutwyler \cite{gl84}.

In a seminal paper \cite{wein67},
 Weinberg proved two QCD sum rules that now bear
his name, and which are
obeyed by the inverse
moments of these spectral densities in the chiral limit
in which both the vector and axial vector currents are conserved:
\begin{mathletters}
\begin{eqnarray}
\int_0^\infty\frac{ds}{s}\big(\rho^V_1(s)-\rho^A_1(s)\big)=f^2_\pi+O(\hat{m})
\label{e:w1}
\\
\int_0^\infty ds\big(\rho^V_1(s)-\rho^A_1(s)\big)=O(\hat{m}).
\label{e:w2}
\end{eqnarray}
\end{mathletters}
The first sum rule is exactly fulfilled in the chiral limit,
 where, as our notation suggests,
$f^2_\pi$ represents the value of the {\it physical} weak decay constant
$F^2_\pi=f^2_\pi+O(\hat{m})$ in the limit of vanishing pion mass.
The second sum rule contains additional QCD
corrections \cite{nar89,ber75}.
These relations assume (see Ref.~\cite{nar89} for a recent overview)
that the associated
polarization functions satisfy unsubtracted dispersion relations. Then, 
with the aid of the K\"{a}llen-Lehmann spectral representation
\cite{iz80}, one can write
\begin{eqnarray}
{1\over 4}\tilde\Pi^{JJ}_T(s)
=-\int_0^\infty dt\,\frac{\rho^J(t)}{t-s}
=\frac{1}{4\pi}\int_0^\infty dt\,\frac{Im[\tilde\Pi^{JJ}_T(t)]}{t-s},
\label{e:disptilde}
\end{eqnarray}
with $J=V,A$, after setting $s=q^2$.
The spectral density distributions for the vector and axial vector polarizations
are then
\begin{eqnarray}
\rho^J_1(s)=-{1\over 4\pi}\,Im[\tilde\Pi^{JJ}_T(s)]\,,
\qquad {\rm and } \qquad \rho^A_0(s)={1\over 4\pi}\,Im[\tilde\Pi^{AA}_L(s)].
\end{eqnarray}

In addition to the Weinberg sum rules for the 
vector and axial vector densities,
further sum rules have been derived by Gasser and Leutwyler for higher inverse
moments of these densities,
as well as for the moments of the scalar and pseudoscalar densities, that are
specific to expansions in the
current quark mass $\hat{m}$ in the context of chiral perturbation theory \cite{gl84}. To introduce these sum rules,
we use the same definitions as these authors in the scalar - pseudoscalar sector
and write
\begin{mathletters}
\begin{eqnarray}
&&{1\over 2\pi}\int{d^4 x}\,e^{iq\cdot x}\langle 0|S^0(x) S^0(0)|0\rangle =
\rho^{S}(q^2)
\\
&&{1\over 2\pi}\int{d^4 x}\,e^{iq\cdot x}\langle 0|P^a(x) P^b(0)|0\rangle =
\rho^{P}(q^2)\delta^{ab}
\\
&&{1\over 2\pi}\int{d^4 x}\,e^{iq\cdot x}\langle 0|S^a(x) S^b(0)|0\rangle =
\tilde\rho^{S}(q^2)\delta^{ab}
\label{e:3.5c}
\\
&&{1\over 2\pi}\int{d^4 x}\,e^{iq\cdot x}\langle 0|P^0(x) P^0(0)|0\rangle =
\tilde\rho^{P}(q^2)
\label{e:3.5d}
\end{eqnarray}
\end{mathletters}
with
\begin{eqnarray}
\rho^J(s)={1\over \pi}Im[\tilde\Pi^{JJ}(s)]
\end{eqnarray}
for the scalar and
pseudoscalar correlators: here $J=S,P$ identifies the character of the relevant 
density operator.
The tilde over the scalar isovector and pseudoscalar isoscalar spectral
densities in Eqs. (\ref{e:3.5c}) and (\ref{e:3.5d})
 serves to distinguish them from the
isoscalar and isovector densities of the same spatial character in the first pair of equations.
The density operators appearing in these correlation functions have been defined
in Eqs.~(\ref{e:isovec}) and (\ref{e:S0P0}).  Then Gasser and Leutwyler
show by standard techniques \cite{nar89} that
\begin{mathletters}
\begin{eqnarray}
\int_0^\infty\frac{ds}{s^2}\big(\rho^V_1(s)-\rho^A_1(s)\big)
={1\over {48\pi^2}}(\bar l_5-1)+O(\hat{m})
\label{e:gl-2}
\\
\int_0^\infty\frac{ds}{s^2}\rho^A_0(s)=\frac{f^2_\pi}{m^2_\pi}+O(\hat{m}),
\label{e:gla0}
\end{eqnarray}
\end{mathletters}
to the indicated order in $\hat{m}$.
Two further sum rules that require a non-vanishing current quark mass are
\begin{mathletters}
\begin{eqnarray}
\int_0^\infty\frac{ds}{s}\hat{m}^2\big(\rho^S(s)-\rho^P(s)\big)
&=&-f^2_\pi m^2_\pi+O(\hat{m}^2)
\label{e:gl3}
\\
\int_0^\infty\frac{ds}{s}\hat{m}^2\big(\tilde\rho^S(s)-\tilde\rho^P(s)\big)&=&
-2m^4_\pi l_7+O(\hat{m}^3).
\label{e:gl4}
\end{eqnarray}
\end{mathletters}
As noted previously, $f^2_\pi$ and $m^2_\pi$ refer to the lowest non-vanishing
order expressions for the pion decay constant and pion mass, considered as a function of $\hat{m}$;
$\bar l_5$ and $l_7$ are simply two further parameters introduced by CHPT.
These authors also show that
\begin{mathletters}
\begin{eqnarray}
\int_0^\infty ds\big(\rho^S(s)-\rho^P(s)\big)&=&O(\hat{m})\\
\int_0^\infty ds\big(\tilde\rho^S(s)-\tilde\rho^P(s)\big)&=&O(\hat{m}),
\label{e:glsp}
\end{eqnarray}
\end{mathletters}
that mimic Weinberg's second sum rule for the scalar minus pseudoscalar
 densities.

\subsection{Vector densities and their properties}
Recently it was shown \cite{dkl96} that Weinberg's first sum
rule is obeyed exactly, and the second sum rule obeyed in modified form,
by the vector and axial vector densities generated by
$both$ the NJL model and its ENJL extension, provided that these
densities were calculated from polarization functions that were
PV-regulated.
Here we expand on these, perhaps unexpected,
results and also extend the discussion to cover the 
remaining sum rules stated above.

We first examine the vector and axial vector spectral densities for the NJL
model.    These correspond to densities at the one-loop level of
approximation.
 Their difference comes
directly from the imaginary part of the Ward identity relation
in Eq.~(\ref{e:ward1}) after Pauli-Villars regularization. The result 
 is
\begin{eqnarray}
\rho^{(0)V}_1(s)-\rho^{(0)A}_1(s)
=\frac{N_c m^2}{4\pi^2}
\big\{\sqrt{1-{4m^2\over s}}\,\theta(s-4m^2)\big\}_{PV}
\sim-\frac{N_cm^2}{\pi^2}\frac{\Lambda^4}{s^2}\,,\quad s\rightarrow \infty,
\end{eqnarray}
which is derived in detail
in the appendix as Eq.~(\ref{e:axvecdens}), and where the PV notation is
defined.
Without the PV-regulating instruction in place, this difference reaches the constant
value $N_c m^2/(4\pi^2)$ at large $s$, and neither 
sum rule would converge.
 With it, this difference is sufficiently convergent
 to give finite answers for both sum rules. By direct
integration, one in fact finds that
\begin{eqnarray}
\int_0^\infty\frac{ds}{s}\big(\rho^{(0)V}_1(s)-
\rho^{(0)A}_1(s)\big)=f^2_p\,,
\quad (\rm NJL),
\label{e:weinmin1}
\end{eqnarray}
while
\begin{eqnarray}
\int_0^\infty ds\big(\rho^{(0)V}_1(s)-
\rho^{(0)A}_1(s)\big)=-m\langle\bar\psi\psi\rangle\ne 0\,.
\quad (\rm NJL).
\label{e:weinmin2}
\end{eqnarray}
Details of this calculation can be found in
Eqs.~(\ref{e:weinbergmin1}) and (\ref{e:weinbergmin2}).
Thus the PV-regulated vector spectral densities
of the minimal NJL model continue to obey the sum rules even though these
one-loop densities do not contain any vector
or axial vector resonances structures of any kind.
One sees that the first Weinberg sum rule
is obeyed exactly, with $f^2_\pi$ replaced by its mean field equivalent
$ f^2_p$ appropriate to the NJL model.
In particular, one realises that the PV regularization of the NJL
spectral densities has introduced
$precisely$
the correct amount of additional negative or ``unphysical''
 spectral density to satisfy
the first Weinberg sum rule exactly\footnote{We also remark that the same sum rules
for the NJL densities are obtained if the the polarization functions
$\Pi^{VV}_T(s)$
and $\Pi^{AA}_T(s)$ that generate them are regulated via an $O(4)$ cutoff in
momentum space. However the problem then is that the `Ward identity'',
Eq.~(\ref{e:ward1}), that lies at the root of both sum rules does not
necessarily
hold without introducing additional assumptions to deal with the
non-uniqueness of regulating quadratic divergences under $O(4)$. This
difficulty is discussed further in Refs.~\cite{will93} and \cite{dstl95}.
We do not employ the $O(4)$ regularization method in this paper.}.
 The second sum rule is obeyed in
a modified form 
that is connected  with the existence of the
 quark condensate in the NJL ground state.  Only for the case that the 
condensate density vanishes, does the second Weinberg relation hold exactly.
The value of the right hand side of Eq.(\ref{e:weinmin2}) is typically
(250MeV)$^4$ in the broken phase.

The situation for the ENJL generated spectral densities is quite different.
Now quark-quark interactions are taken into account that effectively screen
out the undesirable high energy behavior of the one-loop densities, as is
shown below.   This in turn means 
that two possibilities are available: (i) to PV-regulate both the real and
imaginary parts of the irreducible polarization functions, which implies
that these satisfy an unsubtracted dispersion relation involving
PV-regulated spectral densities.    Polarizations and associated quantities
that are calculated in this scheme are denoted as ``barred'' functions 
in the following. (ii) Alternately, one can regulate only the real part of
the irreducible polarization functions, leaving the imaginary part and thus
the spectral densities unregulated.   Such a polarization function satisfies
a once subtracted dispersion relation containing unregulated spectral 
densities.    These quantities are denoted as ``unbarred'' in what follows.
Details of the schemes are given in the appendix.
There it is shown explicitly that,
 from the point of
view of dispersion relations, the distinction between the barred and
unbarred amplitudes can be traced back to using either the PV-regulated or
unregulated one-loop
spectral densities in
the dispersion relation calculations given in
Eqs.~(\ref{e:vectorunsub}) and (\ref{e:vector2sub}).

We now examine the ENJL-generated spectral densities as regulated in the
 barred scheme and show in more detail
that the same form of results as was found for the NJL model in 
Eqs.(\ref{e:weinmin1}) and (\ref{e:weinmin2}) also hold in the ENJL case.
The vector and axial vector densities are obtained from the
corresponding polarization functions that include
all interactions in the RPA approximation, as recorded in
Eqs.~(\ref{e:vecsolu}) and (\ref{e:AT}).
Then
\begin{eqnarray}
\rho^V_1(s)=-{1\over 4\pi}Im\Big(\frac{\bar \Pi^{VV}_T(s)}
{1+2G_2\bar \Pi^{VV}_T(s)}\Big)
\\
\rho^A_1(s)=-{1\over 4\pi}Im\Big(\frac{\bar \Pi^{AA}_T(s)}
{1+2G_2\bar \Pi^{AA}_T(s)}\Big).
\end{eqnarray}
The barred polarization functions
are  given  in 
 Eqs.~(\ref{e:vecpolbar}) and
(\ref{e:axialpolreg}) of the appendix.
Carrying out the indicated operations, one
finds that
\begin{eqnarray}
\rho^V_1(s)&=&{\rho}^{(0)V}_1(s)|\bar F_V(s)|^2
\label{e:rhov}
\\
\rho^A_1(s)&=& {\rho}^{(0)A}_1(s)|\bar G_A(s)|^2,
\label{e:rhoa}
\end{eqnarray}
where $\bar F_V$ and $\bar G_A$ are defined in Eqs.~(\ref{e:vectorformfac}) and
(\ref{e:axformfac}) with $\Pi$ replaced by its regulated version $\bar \Pi$.
Thus the densities of the
ENJL model are simply those of the minimal NJL model, modified by
the modulus squared of the corresponding quark current form factor.

Plots of the ENJL vector and axial vector densities are shown in
Fig.~\ref{f:strengths} using the typical set of parameters given in the
second line of
Table~\ref{table1} that are appropriate in the chiral limit (we refer 
in the following to this set of parameters as the
standard set).
The different threshold momentum dependence of the spectral densities in 
Fig.~\ref{f:strengths} comes about via the behavior of   
the factors $\rho_1^{(0)V}$ or $\rho_1^{(0)A}$.  These factors behave like
 $p^{2L+1}$ = $p$ or $p^3$,
where $p=(1/2)\sqrt{s-4m^2}\;$ is the decay momentum 
 for the decay of the spin one bosons with
$J^{PC}
=1^{--}$ or $1^{++}$ respectively into a $\bar qq$ pair.   The 
$L$ dependence comes about
since the parity
and
charge conjugation quantum numbers of the $\bar qq$ system are given by 
$P=-(-1)^L$ and
$C=(-1)^{L+S}$. The opposite
parity and charge conjugation in the two channels then 
requires that the spins of
the $\bar qq$ pair be aligned to $S=1$ in both cases, but also to  carry
internal angular momentum
$L=1$ in the second case.
 One sees that the presence of the quark form factors
drastically redistributes both spectral densities, causing a
pile-up of strength at low momentum transfers $s=q^2$ by suppressing
the high energy tails of $\rho^{(0)V}_1$ and $\rho^{(0)A}_1$.
However, at large values of $s$, both spectral densities become negative due to
the asymptotic behavior of the regulated one loop densities $\rho_1^{(0)V}$
and $\rho_1^{(0)A}$.  
This is shown explicitly in Fig.~\ref{f:neg} for the vector density of both
the ENJL case and the (one loop) NJL case. The strong redistribution of the
vector strength due to the inclusion of the vector mode is especially clear
in this figure.   In this case, the 
redistribution of strength takes place in such a manner
that the integral over $\rho^V_1(s)/s$ always vanishes.

 In Fig.~\ref{f:diffstr}, we have
shown the difference $\rho_1^{(V)} - \rho_1^{(A)}$, 
plotted out to values of $s$
that extend beyond both PV-induced thresholds, to
illustrate the extent of modification introduced by the vector and axial
vector degrees of freedom.   Again one sees that there is a pile-up of 
strength at low momentum transfers.

If one were to follow the second, unbarred regularization scheme, then 
the PV-regulated one-loop NJL
spectral densities of
Eqs.~(\ref{e:vecdens}) and (\ref{e:axvecdensexp}) are replaced 
by their unregulated
versions
\begin{eqnarray}
\rho^{(0)V}_1(s)\rightarrow\frac{N_cs}{24\pi^2}\big\{\sqrt{1-{4m^2\over s}}
(1+{1\over 2}{4m^2\over s})\theta(s-4m^2)\big\}
\label{e:vecdenstxt}
\end{eqnarray}
and
\begin{eqnarray}
\rho^{(0)A}_1(s)\rightarrow\frac{N_cs}{24\pi^2}
\big\{\sqrt[3]{1-{4m^2\over s}}
\theta(s-4m^2)\big\}
\label{e:axvecdenstxt}
\end{eqnarray}
in Eqs.~(\ref{e:rhov}) and (\ref{e:rhoa}).
This effectively removes the bars 
on $F_V(s)$ and $G_A(s)$.
Since
\begin{eqnarray}
\rho^V_1(s)\sim \rho^A_1(s)\sim \frac{1}{s(\ln\,|s|)^2}, \quad\quad
|s|\rightarrow\infty,
\end{eqnarray}
from Table~\ref{table4}, one then obtains spectral densities for the 
ENJL model that, on the one hand, always remain positive definite, instead
of behaving like $-1/s$, but which on the other hand, are
essentially indistinguishable from those given in Fig.~\ref{f:strengths}
over the resonance regions on the scale of this figure.
Later we discuss some consequences of making this choice of regularization
instead.

If one associates the peaks in either the regulated or unregulated
spectral densities with the masses squared $m^2_\rho$ and $m^2_{a_1}$
of the $\rho$ and $a_1$  modes of the ENJL model, then for our standard set
of parameters, one finds $m_\rho=0.713$GeV and $m_{a_1}=1.027$GeV from
Fig.~\ref{f:strengths}. This is in fact in reasonable qualitative agreement with the
measured values \cite{data96} of these meson masses, which lie at $0.768$GeV and $1.230$GeV respectively.
However, since both spectral densities peak above the
$\bar q q$ decay threshold at $4m^2$, which is unphysical from the point of view of
confinement, the identification of these peaks with the meson mode
masses can be questioned.
On the other hand, we show below that Weinberg's
first sum rule is satisfied exactly by the associated vector minus axial vector
densities that exhibit these peak values, so that the integrated strength
is $exactly$ reproduced in accordance
with the tenets of chiral symmetry, on which this sum rule is based.
In fact, in Weinberg's original
treatment \cite{wein67}, the meson masses were determined by saturating this sum rule
with delta function 
strength distributions for the vector and axial vector spectral densities
that simply ignored the presence of any decay channels as a first approximation.
Since the ENJL spectral densities also saturate the first sum rule, it is
thus  reasonable to use the peak values of these
spectral densities (that have been broadened by replacing the physical
$\rho\rightarrow \pi\pi$ and $a_1\rightarrow \rho\pi$ 
decay channels by the unphysical $\bar qq$
decay channel in each case) to identify the associated
meson masses in the ENJL model,
with, as we have
seen, not dissimilar predictions for these masses.

The masses of these modes should on the general basis of propagator theory
also be related to the poles of the corresponding propagators.
Below we investigate
the analytic structure of the form factors $F_V(s)$ and $F_A(s)$, or equivalently the
transverse vector and axial vector propagators $D^{VV}_T(s)$ and $D^{AA}_T(s)$
given in Eqs.~(\ref{e:vecprot}) and (\ref{e:axvecpro}).

\subsubsection{Pole structure of the $\rho$ and $a_1$ propagators}

One knows on general
grounds \cite{iz80}
that any complex poles in the scattering amplitudes or propagators $D^{VV}_T(s)$ and $D^{AA}_T(s)$
that correspond to physical resonances
must be located on their second
(or ``unphysical'') Riemann sheets of the cut $s$-plane,
for reasons of causality. From Eqs.~(\ref{e:vecprot})
and (\ref{e:axvecpro}), one sees that the poles of these functions as
a function of $s$ are governed by the roots of
the combinations 
\begin{eqnarray}
1+2G_2\Pi^{VV}_T(s)\quad{\rm and}\quad 1+2G_2\Pi^{AA}_T(s).
\end{eqnarray}
We will find that for a range of coupling strengths
 $G_1$ and $G_2$, including those in our standard
parameter set of Table~\ref{table1}, 
the axial vector mode has a single pole in the lower half plane of
the second Riemann sheet, while the vector mode has two poles on that sheet:
a real root
below $4m^2$, corresponding to a virtual bound state, and a second complex root in the
lower half plane. The positions of the two $\rho$ poles and the $a_1$ pole 
 are given numerically in Table~\ref{table3}
for the standard ENJL parameter set.   Details of this analysis now follow.

We discuss the vector
mode first. According to Eqs.~(\ref{e:vecpolbar}) and (\ref{e:vecpolbar0}), the polarization function for this channel is given by
\begin{eqnarray}
\bar\Pi^{VV}_T(s)
=-\frac{2}{3}\frac{f^2_p}{m^2}s+\frac{N_cs}{6\pi^2}
\big\{{1\over 3}+(3-f)(\sqrt{f}\coth^{-1}\sqrt{f}-1)\big\}_{PV},
\label{e:vecpolbartxt}
\end{eqnarray}
where $f=1-4m^2/s$.  
This form of the regulated vector polarization function satisfies an unsubtracted dispersion relation
exactly of the form given in Eq.~(\ref{e:disptilde}),
if
the PV-regularization instruction is left in place on the second factor. If this instruction is
removed (we indicate this by removing the bar) then $\Pi^{VV}_T(s)/s$ satisfies
a once-subtracted dispersion relation.
A detailed derivation of these statements
is found in the appendix.

The analytic
properties of $\Pi^{VV}_T(s)$ are determined by the behavior of the function
\begin{eqnarray}
J(z)=\sqrt{f}\coth^{-1}\sqrt{f}\,,\quad f=1-\frac{1}{z},
\nonumber
\end{eqnarray}
that is
defined in Eq.~(\ref{e:JC}), with $z=s/4m^2$. Considered as a function of a complex
variable $z$, this function is single-valued in the cut $z$-plane
with the cut extending from $z=1$ to $\infty$ along the real axis. The schedule
of forms
for $J(z)$ and its analytic continuation $\tilde J(z)$ onto
the  contiguous Riemann sheet connected through
this cut is given in Eqs.~(\ref{e:J1c}) through (\ref{e:Jtildc}).
These expression show that both $J(x)$ and $\tilde J(x)$ are real in
the interval $0\leq x\leq 1$ on the real axis.
Plots of the real part of $J(z)$
and $\tilde J(z)$ as a function of the real variable $z=s/4m^2$, are shown
in Fig.~\ref{f:ReJ}.

The corresponding behavior of the real part of $\bar\Pi^{VV}_T(s)$ is shown 
in Fig.~\ref{f:branch}
as a function of the real variable $z=s/4m^2$ on the first and second sheets
of the cut $s$-plane.  For $s\geq 4m^2$ this function has to be continued
 onto the
the upper lip of the cut in accord with the Feynman prescription \cite{iz80}.
The values of the polarization along the upper lip of the cut on the
first sheet join smoothly with their analytic continuation onto the 
lower lip of the cut on the
second sheet. At the branch point, these values bifurcate into two branches
of the function that assume different values along the sector
 $0\le z\le 1$ of the two sheets.
The values on the first
and second sheets are reached by allowing the
variable $z$ to pass either infinitesimally above or infinitesimally
 below the branch point
through the cut
to reach this sector of the real axis.
As the vector polarization in Eq.~(\ref{e:vecpolbartxt})
also has contributions from $J's$
at different arguments $s/4M^2_a$, $a=1,2$  due to the PV-instruction,
further cuts starting at $z=1+\Lambda^2/m^2$ and $z=1+2\Lambda^2/m^2$ have to be introduced.
However, since $\Lambda^2/m^2$ is typically $\sim 16$, both of these cuts commence too
far away to the right of the origin to essentially alter the behavior of $\Pi^{VV}_T(s)$
as determined by $J(s/4m^2)$ alone in the interval $0<s/4m^2<1$.
Hence the cusp-like structure where the values of $Re\bar\Pi^{VV}_T(s)/4m^2$
and its analytic continuation meet at $s/4m^2=1$ in Fig.~\ref{f:branch} is
a straightforward reflection of the behavior of $Re\,J(z)$ in Fig.~\ref{f:ReJ}.

The roots in $z=s/4m^2$ where
\begin{eqnarray}                                     
1+2G_2\bar\Pi^{VV}(s_\rho)=0,
\label{e:vecrts}
\end{eqnarray}
determine the poles of the $\rho$ propagator. We now verify that the vector mode
has two poles on the second sheet, one on the real axis below $4m^2$, and a second
complex pole in the lower half plane.
As pointed out in \cite{tkm91}, the real pole corresponds having the quark-antiquark pairs
making up the vector meson bound
into a virtual, (or ``antibound'') state \cite{rgn66}.

We briefly discuss this state first: according to Fig.~\ref{f:branch},
the function $\bar\Pi^{VV}_T(z)/4m^2$ is real and negative on
both of its branches for real $z$ in the interval $0<z<1$. Hence Eq.~(\ref{e:vecrts})
always has one real root lying in this interval on either the first or the second sheet depending on
where the ratio $\bar\Pi^{VV}_T(z)/4m^2$ crosses $-1/8G_2m^2$. 
The solid circles mark the intercepts of $\bar\Pi^{VV}_T(z)/4m^2$
with the value of this quantity as the interaction strength $G_2$ increases
through the indicated set of values. The value of $m$ has been kept fixed at
$264\,$MeV for this illustration. These intercepts determine
the position of the real pole of the vector mode propagator lying in the
interval $0\le s/4m^2\le 1$ as a function of the interaction strength. 
One sees from the figure that as $G_2$
increases from zero there is only a $second$ sheet real pole,
corresponding to a virtual bound state mentioned above for the $\bar qq$ pairs making up the
$\rho$. The binding energy of the virtual state decreases to zero as $G_2$
increases through the critical 
value $G^{(c)}_2=-1/2\Pi^{VV}_T(4m^2)=7.41\,$GeV$^{-2}$ in the
present case. After this, the pole reappears on the $first$ or
physical sheet where it represents a true bound state of the $\bar qq$ system
whose binding energy again increases  towards the limiting value of $2m$ with
increasing $G_2$, which it reaches\footnote{As a consequence, the ENJL
groundstate always
stays stable against the development of an RPA-like instability where
$\sqrt{s_\rho}$ becomes pure imaginary, which would be the signal for
the formation of a ``vector'' condensate.}
 as $G_2\rightarrow \infty$.
We have plotted the motion of the pole versus the interaction strength $G_2$
in Fig.~\ref{f:vpole} with $G_2$ as ordinate, in order
 to best illustrate the turn around
of the binding energy as $G_2$ passes through its critical value. For our standard
parameter set given 
in Table~\ref{table1}, this root corresponds to a virtual bound state
at the pole position $s_\rho=(0.166)\,4m^2= 0.046\,$GeV$^2$, and therefore of mass 
$\sqrt{s_\rho}=0.215$ GeV.

By contrast, the axial vector propagator has a single complex pole,
which we discuss together with the complex pole of the vector mode.
The axial vector polarization can be obtained directly from the Ward identity relation
(\ref{e:ward1}) and the regulated form of the pion form factor $\bar F_P(s)$,
that is given
in Eq.~(\ref{e:piformfac}). One has
\begin{eqnarray}
\bar\Pi^{AA}_T(s)&=&\bar\Pi^{VV}_T(s)+4f^2_p\bar F_P(s)
\label{e:axvecpolbartxt}
\end{eqnarray}                    
where
\begin{eqnarray}
\bar F_P(s)&=&\big[1-\frac{N_cm^2}{4\pi^2f^2_p}
\big\{2\sqrt{f}\coth^{-1}\sqrt{f}-2\big\}_{PV}\big].
\end{eqnarray}
Contrary to Eq.~(\ref{e:vecrts}), the condition
\begin{eqnarray}
1+2G_2\bar\Pi^{AA}_T(s_{a_1})=0,
\label{e:axvecrts}
\end{eqnarray}
that determines the poles $s_{a_1}$ of the $a_1$ propagator,
 does not have any real roots
on either sheet, since $Re\bar F(s)$ is positive on
both sheets, and overrides the negative vector polarization along the real
axis. Then $\bar \Pi^{AA}_T(s)$ is real and positive for $0<s<4m^2$ and no solutions
of Eq.~(\ref{e:axvecrts}) for real $z$ are possible.
However, both Eqs.~(\ref{e:vecrts}) and (\ref{e:axvecrts}) have complex roots
located on the second Riemann sheet which may be found by replacing both of the polarization
functions in these equations by their analytic continuations through the
cut.
In Fig.~\ref{f:cmplxpoles}, the variation of the imaginary versus the real
part of the second sheet poles of the $\rho$ and $a_1$ propagators obtained
in this way is shown with increasing
interaction strength. One observes (i) that the $\rho$ meson pole has a real part
that always lies below $4m^2$, for the indicated range of $G_2$, and 
(ii) that the imaginary parts of both complex poles are of the order larger or
equal to their real parts.

Since the poles themselves are of order unity in units of
$4m^2$, and this  is much smaller than the cutoff scale $4\Lambda^2$,
we also want to make the point that the numerical results for the $\rho$ and
$a_1$ meson poles, as well as the values of the polarization functions
in their vicinity, are actually
insensitive to the replacement of these functions
in Eqs.~(\ref{e:vecrts}) and (\ref{e:axvecrts}) by their ``unbarred'' counterparts
$\Pi^{VV}_T(s)$ and $\Pi^{AA}_T(s)$. In particular, they are insensitive
to the actual value of the cutoff employed.
This is illustrated in Table~\ref{table3}, where the two $\rho$ poles and the
single $a_1$ pole are given in units of $4m^2$
for our standard parameter set.
The differences are insignificant. One can see the reason for this more clearly
by making comparative plots of the vector and axial vector polarization functions
generated via their unsubtracted and subtracted dispersion relations.
From Fig.~\ref{f:compv}, one observes
 that these two versions of both the vector and axial vector polarizations
are essentially identical in the interval $-1 < q^2/4m^2< 1$, where the influence
of the additional branch points introduced by using PV-regulated densities have not yet made
themselves felt\footnote{It should be pointed out that
 in a low energy expansion \cite{weise90,brz94,wak89} where the
non-analytic terms
are dropped from $\bar\Pi^{VV}_T(s)$ and $\bar\Pi^{AA}_T(s)$,
 the
ENJL
model leads to approximate meson poles that are real.    One then finds
delta function spectral densities  of strengths $g^2_V$ and
$g^2_{A_1}$ that automatically obey the Weinberg 
relations $g^2_V=g^2_{A_1}$
and $g^2_V m_V^{-2}-g^2_{A_1} m_A^{-2}=f^2_\pi$ (as distinct from the  
sum rules).    To be precise,
\begin{eqnarray}
\frac{1}{4}\bar\Pi^{VV}_T(s)\approx -\frac{f^2_\pi}{6g_Am^2}s\,,
\quad \frac{1}{4}\bar\Pi^{AA}_T(s)\approx \frac{f^2_\pi}{6g_Am^2}(6m^2-s)
\nonumber
\end{eqnarray}
from Eqs.~(\ref{e:vecpolbartxt}) and (\ref{e:axvecpolbartxt}).
This approximation replaces the exact curves for $\bar\Pi^{VV}_T(s)$   
and $\bar\Pi^{AA}_T(s)$ shown in
Fig.~\ref{f:compv} by tangent lines at $s=0$. This completely suppresses all
meson dynamics associated with the branch cut.
In particular, all information regarding the properties of the second sheet
poles of the $\rho$ and $a_1$ meson
modes is lost. Clearly such an
approximation is qualitative at best, since it simply ignores the unphysical
decay channels
 that reflect the
non-confining nature of the ENJL Lagrangian. Its use leads to
meson-quark coupling constants
$g^{-2}_{\rho qq}=g^{-2}_{a_1 qq}= f^2_\pi/6g_A m^2$ as given
by the common derivative
$(-1/4)\partial\Pi/\partial s$ for $V$ and $A$, and meson masses
$m^2_V=6g_Am^2/(1-g_A)$ and $m^2_A=6m^2/(1-g_A)$, where $m^2_A/m^2_V=
g^{-1}_A$. The factor $1/4$ arises in calcuating the coupling constants
if we take the standard form \cite{sak69} of the interaction Lagrangian as
${\cal L}_{int}=g_{\rho qq}(\bar\psi\gamma_\mu \frac{1}{2}\tau^b\psi)
\rho^\mu_b$
or $g_{a_1 qq}(\bar\psi\gamma_\mu\gamma_5\frac{1}{2} \tau^b\psi)a^\mu_{1;b}$
for coupling the quark currents to the meson fields.
Because of the approximations, these meson modes have delta function
distributions as their spectral densities, of equal strengths
$g^2_V=g^2_{A_1}
=(m^2_V/g_{\rho qq})^2=f^2_\pi m^4_V/(6g_Am^2)$. Inserting this
information into
Weinberg's sum rules, one finds that his two relations quoted above
 are satisfied
(this is not yet a check of the sum rules themselves for the ENJL
model, as the
 approximations
involved are inappropriate). The
universality relation $g_{\rho qq}=f_\rho$ between the $\rho$ meson
coupling constant
and its decay constant $f_\rho$, reflecting vector meson dominance
\cite{sak69} of the
electromagnetic current, is also recovered from the vacuum to one meson
state
matrix element \cite{nar89} ($n_\mu$ is the vector meson polarization)
\begin{eqnarray}
n^{\mu}\langle 0|V^0_\mu|\rho^0\rangle
=-\frac{1}{4}g_{\rho qq}\bar\Pi^{VV}_T(m^2_V)=\frac{m^2_V}{f_\rho}
\nonumber
\end{eqnarray}
that governs the $\rho^0\rightarrow e^+e^-$ decay. After inserting the
values of
$g_{\rho qq}$ and $\bar\Pi^{VV}_T(m^2_V)$,  one finds
$g_{\rho qq}=f_\rho=f^{-1}_\pi\sqrt{6g_A m^2}$
that also can be written in the form $f_\rho f_\pi = m_V(1-g_A)^{1/2}$.
For $g_A=1/2$, this is exactly the KSRF relation \cite{KSFR66}. Then one
 also recovers
Weinberg's original
estimate $m_A/m_V=\sqrt{2}$ for the mass ratio, as well as the $\rho$
 meson mass
$m_V=\sqrt{6}\,m$ of the bosonized NJL model \cite{er86}.
For the standard parameter set of Table~\ref{table1} this low energy
expansion
 limit
gives $m_V = 1.12\,$GeV$\,(0.768)$,
$m_A=1.29\,$GeV$\,(1.230)$, and $f_\rho = 6.02\,(5.10)$ \cite{nar89}. The
experimental values are
in brackets.}.

One can state this result in yet another way: removing
the PV-instruction from the second term of the vector
polarization function $\bar \Pi^{VV}_T(s)$ in Eq.~(\ref{e:vecpolbartxt}) is
tantamount to removing the explicit cutoff dependence in the vector polarization
function, while leaving it buried implicitly in the quark mass $m$
in the first term. The values of the roots themselves are then
solely determined by the physical input parameters $f_\pi$ and $g_A$ once $m$ (which of course
contains the cutoff implicitly) has been calculated from the gap equation.
The differences between the numbers appearing in the first and second rows
of Table~\ref{table3} thus reflect the extent of the explicit cutoff dependence of the roots.
However, we learn another, equally important feature, from these
plots: both the vector and axial vector polarization functions that arise from
subtracted dispersion relations become
 negative for large space-like $q^2$. For the
standard parameter set, this has occurred at $-q^2/4m^2\sim 10$, or $\sqrt{-s}=1.7\,
$GeV, but it is clearly
a general property. This can be traced to the different asymptotic
           behavior displayed by the barred and unbarred polarization functions,
            $\bar\Pi(s)\sim -(1/s)\ln(-s)$ versus $\Pi(s)\sim s\ln(-s)$,
           see Table~\ref{table4}.
The  former expression stays positive for space-like $s$,  vanishing at
infinity, while the latter function becomes large and negative.
This feature in both polarization functions generated via once-subtracted
dispersion relations means that the unbarred versions
$1+2G_2\Pi^{VV}_T(s)$ and $1+2G_2\Pi^{AA}_T(s)$ of
Eqs.~(\ref{e:vecrts}) and (\ref{e:axvecrts}) each
develop an $additional$ real root on the physical sheet
at space-like $q^2$, corresponding to a Landau ghost \cite{iz80} and
 as shown in Fig.~\ref{f:ghosts},
i.e.
\begin{eqnarray}
1+2G_2\Pi^{VV}_T(-M^2_{\rho g})=0\,,\quad {\rm and}
\quad 1+2G_2\Pi^{AA}_T(-M^2_{a_1 g})=0.
\label{e:ghost roots}
\end{eqnarray}
We remark in passing that the same is true of the unbarred combination
 $1-2G_1\hat\Pi^{PP}(s)$ that determines
the pion pole. 
Thus the associated $\rho$ 
and $a_1$ propagators, as well as the pion propagator, contain
Landau ghosts if the polarization functions are left unbarred. For the standard parameter set,
the vector ghost poles occur at the nearly
degenerate positions
$s_{\rho g}=-18.30(4m^2)=-5.11\,$GeV$^2$ and $s_{a_1g}=-18.04(4m^2)
=-5.05\,$GeV$^2$.
The pion ghost occurs at $s_{\pi g}=-13.48(4m^2)=-3.77$GeV$^2$. The magnitudes of
the ghost poles are always of ${\cal O}(4\Lambda^2)$.

\subsubsection{Pole approximations for the vector and axial form factors}
In order to understand the connection
 between the poles of the form factors and the
peaks in the spectral densities, we now approximate the former
by single pole approximations and compare the results with the 
exact form factors.
Their single pole approximations are given by
\begin{eqnarray}
F_V(s)\approx \frac{R_\rho}{(s-s_\rho)}\,,\quad
G_A(s)\approx \frac{R_{a_1}}{(s-s_{a_1})},
\nonumber
\end{eqnarray}
with residues 
$R_\rho=[2G_2\Pi^{(\prime)VV}(s_\rho)]^{-1}$ and
$R_{a_1}=[2G_2\Pi^{(\prime)AA}(s_{a_1})]^{-1}$. The prime indicates the 
derivative
with respect to $s$ of the analytically continued polarization function. 
Here $s_\rho$ can be either the virtual bound state pole, or the 
complex pole of the
$\rho$ propagator. In principle, both poles influence the shape of the spectral function, and
$F_V(s)$ should be displayed as the sum of the contributions
coming from these two $\rho$ poles: the mass of the $\rho$ is not uniquely determined
by either one. However, the virtual pole, which for our parameter set lies at
at $s_\rho=0.166(4m^2$), is not close enough to the branch point at
$4m^2$ to exert a strong influence on $F_V(s)$, so that one expects
the complex $\rho$ pole to dominate in the present situation. One should emphasize, however,
that this circumstance is parameter-dependent: a virtual bound state just below
$4m^2$ could have a dominant influence on the vector strength function.
The residues themselves are complex numbers with moduli
$|R_\rho|=0.75\,$GeV$^2$ and $|R_{a_1}|=0.64\,$GeV$^2$
for the standard parameter set.

A comparison between the modulus squared of the
exact form factors and their approximations using only the complex $\rho$ and $a_1$ poles
is presented
in Figs.~\ref{f:veccomp} and \ref{f:aveccomp}. The form factors  $|F_V(s)|^2$ and $|G_A(s)|^2$
are faithfully reproduced by their single complex pole approximations for our parameter set.
The resultant
discrepancies  between the peak positions in the spectral densities of Fig.~\ref{f:strengths}
and the pole positions in $F_V(s)$ and $G_A(s)$ are shown in
Table~\ref{table5}. 
These come, not surprizingly, from the distortion introduced by
the kinematic threshold behavior of the one-loop
densities $\rho^{(0)V}_1$ and $\rho^{(0)A}_1$. These factors act
multiplicatively to force the spectral densities to vanish at the $\bar qq$ threshold, and
thus move what would have been a pole peak to somewhat
higher energies. This is particularly the case for the $\rho$ resonance pole
since, as seen in Fig.~\ref{f:veccomp}, its real part is ``invisible'' below
the branch point. In fact, if one ignores the threshold behavior of the
one-loop densities as a first approximation, they become proportional to $s$
and would produce peaks
in the spectral densities at 
the moduli of the pole positions $|s_\rho|$ and $|s_{a_1}|$.
Using the energies given in
Table~\ref{table5}, one obtains the estimates $\sqrt{|s_\rho|}=$ 0.63GeV and
$\sqrt{|s_{a_1}|}=$ 0.93GeV, to be compared with the spectral density
peak positions at
0.71GeV and 1.03GeV.

\subsection{Weinberg sum rules}
We now investigate the validity of Weinberg's two sum rules for the
ENJL generated spectral densities.
Taking the difference $\rho^V_1(s)-\rho^A_1(s)$ directly from
Eq.~(\ref{e:rhov}) minus Eq.~(\ref{e:rhoa}), one finds that
\begin{eqnarray}
\rho^V_1(s)-\rho^A_1(s)&=&-{1\over 4\pi}\,
Im\Big[\frac{\bar \Pi^{VV}_T(s)-\bar \Pi^{AA}_T(s)}
{\big(1+2G_2\bar \Pi^{VV}_T(s)\big)\big(1+2G_2\bar \Pi^{VV}_T(s)\big)}\Big]
\end{eqnarray}
or
\begin{eqnarray}
\rho^V_1(s)-\rho^A_1(s)={1\over \pi}Im\big[f^2_\pi\bar F_P(s)\bar F_V(s)\bar
F_A(s)\big].
\label{e:densdiff}
\end{eqnarray}
in view of the Ward identity Eq.~(\ref{e:ward1}) plus the expressions in
Eqs.~(\ref{e:vectorformfac}) and (\ref{e:axialformfac}) for the vector and axial vector
quark form factors $F_V$ and $F_A$.
Notice that the {\it renormalized}
pion decay constant $f^2_\pi=g_Af^2_p$ of Eq.~(\ref{e:fpren})
appears quite naturally in this result. The inverse moment integrals of this
difference that are required for investigating Weinberg's sum 
rules can be evaluated
in closed form using the methods of contour integration. Their
 evaluation depends
on the fact that the barred amplitudes in
\begin{eqnarray}
f(s)=f^2_\pi\bar F_P(s) \bar F_V(s) \bar F_A(s)
\label{e:formfacprod}
\end{eqnarray}
that have been constructed using barred polarization functions, are all
analytic functions of $s$ in the cut plane, with the cut extending along the
real axis from $4m^2$ to $\infty$. This is so because the poles 
of the vector and axial form factors only occur on the second sheet, and $F_P(s)$
has no poles. Hence the integrals of $f(s)/s$ and $f(s)$ taken
around any closed contour $\cal C$ on the first or physical sheet that excludes
the cut (and the pole at $s=0$ in the first case) are zero. We break up the
closed contour into
${\cal C}=C_0+C_\infty$ as shown in Fig.~\ref{f:contour},
where $C_0$ is indented to exclude the cut plus any poles as necessary,
and is closed by the contour $C_\infty$ at infinity. Then the vanishing of the
contour integrals of $f(s)/s$ and $f(s)$ leads to the relations
\begin{eqnarray}
\frac{1}{\pi}\int_{4m^2}^\infty \frac{ds}{s}Im[f(s+i\epsilon)]= f(0)
-\frac{1}{2\pi i}\int_{C_\infty}\frac{ds}{s}f(s)
\label{e:contour1}
\end{eqnarray}
and
\begin{eqnarray}
\frac{1}{\pi}\int_{4m^2}^\infty dsIm[f(s+i\epsilon)]=
-\frac{1}{2\pi i}\int_{C_\infty} ds f(s)
\end{eqnarray}
for the integrals involving $Im[f(s+i\epsilon)]$ 
taken along the upper lip of the cut.
From Eq.~(\ref{e:densdiff}), it is clear that the 
left hand sides of these two relations
are exactly the integrals that we need for the 
two Weinberg sum rules.  In the first
case, $f(0)=f^2_\pi \bar F_P(0)\bar F_V(0)\bar 
F_A(0)=f^2_\pi$ from the definition of $f(s)$ and the
normalization of the three form factors.  Moreover,
\begin{eqnarray}
f(s)\rightarrow \frac{m\langle\bar\psi\psi\rangle}{s}
\,,\quad |s|\rightarrow \infty,
\label{e:fasymp}
\end{eqnarray}
according to Eq.~(\ref{e:formfacasymp}), coupled with the information from
Table~\ref{table4} that $\bar F_V(s)\,,\bar G_A(s)\rightarrow 1$ asymptotically. Hence the
contour integral along the great circle will not contribute in the first case, but does so
for the second, since the asymptotic form of $f(s)$ shows that there is a
pole at infinity on $C_\infty$ of residue $m\langle\bar\psi\psi\rangle$.
Consequently
\begin{eqnarray}
\int_0^\infty\frac{ds}{s}\big(\rho^{V}_1(s)-\rho^{A}_1(s)\big)=f^2_\pi\,,
\quad (\rm ENJL)
\label{e:enjlw1}
\end{eqnarray}
\begin{eqnarray}
\int_0^\infty ds\big(\rho^{V}_1(s)-\rho^{A}_1(s)\big)=-m\langle\bar\psi
\psi\rangle\ne 0\,, 
\quad (\rm ENJL).
\label{e:enjlw2}
\end{eqnarray}
Thus the first sum rule is satisfied exactly by the spectral densities
 that have been
obtained by PV-regulating the one-loop densities with the concomitant 
introduction
of negative spectral density at high $s$; the second sum rule again assumes
a modified form.

As has already been remarked, the ENJL spectral densities
become well-behaved asymptotically if one replaces the one-loop densities
in Eqs.~(\ref{e:rhov}) and (\ref{e:rhoa}) by
their unregulated versions. This is equivalent to removing the bars on 
the form factors
in Eq.~(\ref{e:formfacprod}), which changes the asymptotic behavior of
$f(s)$ from what it was in Eq.~(\ref{e:fasymp}) to
\begin{eqnarray}
f(s)\rightarrow -\frac{1}{s^2}\frac{1}{\ln(-s)}\,,\quad |s|\rightarrow
\infty,
\end{eqnarray}                                    
according to Table~\ref{table4}. This is sufficiently convergent to 
prevent the
contour
integrals along $C_\infty$ from giving any contribution to either sum rule.  
However this comes at a price:
now both the vector and axial vector form factors have developed ghost 
poles at space-like $s$
that bring in contributions of their own, and both of the above sum rules will
be modified\footnote{This immediately begs the
question of by how much. The answer to this question depends on what one
 chooses to understand
by ``spectral density''. If we simply mean the unregulated versions of 
Eqs.~(\ref{e:rhov}) and
(\ref{e:rhoa}), then
\begin{eqnarray}
\int_{4m^2}^\infty \frac{ds}{s}\big(\rho^V_1(s)-\rho^A_1(s)\big)_{unreg}
=f^2_\pi+R_{\rho g}
+R_{a_1 g}=0.96 f^2_\pi,
\nonumber
\end{eqnarray}
either via contour integration,
or simply by direct numerical integration,
where $R_{\rho g}$ and $R_{a_1 g}$
are the residues of the function $f(s)/s$ of Eq.(\ref{e:formfacprod}) 
at the two
ghost poles at $s=-M^2_{\rho g}$
and $-M^2_{a_1 g}$.
Thus the unregulated
densities, which as we have remarked always remain positive definite,
saturate $96\%$ of Weinberg's first sum rule for our parameter set.
This equation is, however, open to an alternative
interpretation. Writing the
 unbarred version of $f(s)$ in Eq.~(\ref{e:formfacprod}) 
in the equivalent form
\begin{eqnarray}
f(s)=\frac{1}{8G_2}\big[\frac{1}{1+2G_2\Pi^{VV}_T(s)}-\frac{1}{1+2G_2\Pi^{AA}_T(s)}\big]
\approx \frac{C_{\rho g}}{s+M^2_{\rho g}}\quad{\rm or}\quad\frac{-C_{a_1 g}}{s+M^2_{a_1 g}},
\nonumber
\end{eqnarray}
where
$C_{\rho g}=\{16G_2^2\,[\partial \Pi^{VV}_T/\partial s]_{-M^2_{\rho g}}\}^{-1}
=0.0648\,$GeV$^4$ and $C_{a_1 g}=\{16G_2^2\,[\partial \Pi^{AA}_T/\partial s]_{-M^2_{a_1 g}} \}^{-1}
=0.0623\,$GeV$^4$ are the spectral strengths
of the ghost modes,
the residues $R_{\rho g}$ and $R_{a_1 g}$ of $f(s)/s$ are
$-C_{\rho g}M^{-2}_{\rho g}$ and $C_{a_1 g}M^{-2}_{a_1 g}$.
  The spectral densities themselves are $\delta$-function distributions
at space-like $s$ of negative strength,
\begin{eqnarray}
\rho^V_1(s)_{ghost}=-C_{\rho g}\,\delta(s+M^2_{\rho g})\quad{\rm and}
\quad 
\rho^A_1(s)_{ghost}=-C_{a_1 g}\,\delta(s+M^2_{a_1 g}) 
\nonumber       
\end{eqnarray}
if we assume the standard $(-M^2_g)-i\epsilon$ prescription
for ghost poles too.
 Then, by transferring the residue contributions from the right hand to the
left hand side of the sum rule and including them under the integral,
the sum rule may be rewritten in the equivalent form
\begin{eqnarray}
\int_{-\infty}^\infty \frac{ds}{s}\big(\rho^V_1(s)-\rho^A_1(s)\big)_
{ghosts + unreg} =f^2_\pi
\nonumber
\end{eqnarray}
so that the sum rule violation is uniquely attributable to the presence of
the ghosts. Exactly the same arguments yield
\begin{eqnarray}
\int_{4m^2}^\infty ds\big(\rho^V_1(s)- 
\rho^A_1(s)\big)_{unreg}=C_{\rho g}-C_{a_1 g}=(224\rm{MeV})^4
\nonumber
\end{eqnarray}
or
\begin{eqnarray}
\int_{-\infty}^\infty ds\big(\rho^V_1(s)-\rho^A_1(s)\big)_{ghosts + unreg}=0
\nonumber
\end{eqnarray}
for the second sum rule, since $C_{\rho g}-C_{a_1 g}$ is just the negative of
the $sum$ of the residues at the two ghost poles,
$f(M^2_{\rho g})=C_{\rho g}$ and $f(-M^2_{a_1 g})=-C_{a_1 g}$, where $f(s)$ is the
product in Eq.~(\ref{e:formfacprod}), and there is 
still no contribution from the contour $C_\infty$ at infinity.
Thus if one widens the definition of spectral density to admit contributions from
the ghost modes, both forms of Weinberg's
sum rules are recovered without modification. These results 
suggest that the unphysical ghosts represent a legitimate part of the meson mode spectrum of the
$model$ propagators when subtracted dispersion relations are used to
construct them.} by these ghost modes. Thus the unregulated ENJL densities that
are otherwise physically well-behaved
have introduced different unphysical elements into the sum rules again
by a different route.

\subsection{Gasser-Leutwyler sum rules}
The remaining sum rules can be evaluated by the same method.
We start with the
Gasser-Leutwyler sum rule in Eq.~(\ref{e:gl-2}). This has a double pole at the origin,
so that the derivative $f^\prime(0)$ of $f(s)$ at the origin appears
on the right hand side
of Eq.~(\ref{e:contour1}) instead of $f(0)$. There is also no contribution from
$C_\infty$. Thus
\begin{eqnarray}
\int_{4m^2}^\infty\frac{ds}{s^2}\big(\rho^{V}_1(s)-\rho^{A}_1(s)\big)
=f^2_\pi\big[\bar F_P^\prime(0)+\bar F_V^\prime(0)+\bar F_A^\prime(0)\big],
\label{e:njl-2}
\end{eqnarray}
where the primes indicate the
derivatives of the form factors.  Straightforward calculations yield
\begin{eqnarray}
\bar F_P^\prime(0)&=& \frac{g_A}{8\pi^2f^2_\pi}[1+O(m^2/\Lambda^2)]
\nonumber\\
\bar F_V^\prime(0)&=&\frac{1}{6m^2}(\frac{1-g_A}{g_A})
\nonumber\\
\bar F_A^\prime(0)&=&\frac{1}
{6m^2}(1-g_A)-\frac{g_A}{8\pi^2f^2_\pi}(1-g_A)[1+O(m^2/\Lambda^2)],
\end{eqnarray}
leading to\footnote{Since the ghost poles are so massive, using 
unregulated densities
in this case would add
a  completely negligible contribution
of $C_{\rho g}M^{-4}_{\rho g}-C_{a1 g}M^{-4}_{a_1 g}\sim 10^{-5}$ to the right hand
side. }
\begin{eqnarray}
\int_{4m^2}^\infty\frac{ds}{s^2}\big(\rho^{V}_1(s)-\rho^{A}_1(s)\big)=
\frac{g_A^2}{8\pi^2}+\frac{f^2_\pi}{6m^2}(\frac{1-g_A^2}{g_A}),
\label{e:enjl-2}
\end{eqnarray}
after setting $8G_2f^2_\pi=1-g_A$ again, and dropping the inconsequential
${\cal O}(m^2/\Lambda^2)$
correction to $F_P^\prime(0)$ that appears through the explicit cutoff
dependence of the pion radius parameter
in Eq.~(\ref{e:pimeanrad}).
On comparing with Eq.~(\ref{e:gl-2}), one identifies the ENJL value for $\bar l_5$
from the expression
\begin{eqnarray}
\bar l^{ENJL}_5 + \ln (m_\pi^2/\mu^2)
=1+6g_A^2+8\pi^2\frac{f^2_\pi}{m^2}(\frac{1-g_A^2}{g_A}),
\label{e:l5}
\end{eqnarray}
after adding in the chiral logarithm as in Eq.(\ref{e:lbar}) again.
  Then one
finds $\bar l_5^{ENJL} = 12.8$
for the standard parameter set.
The empirical value is $\bar l_5=13.9\pm 1.3$ \cite{gl84}.

Similar arguments suffice for checking the next sum rule,
Eq.~(\ref{e:gla0}), except that here we have to include non-chiral effects to order
${\cal O}(\hat{m})$ in the calculation of
$\rho^A_0=(1/4\pi)\,Im\,\tilde\Pi^{AA}_{L;nc}$ from
$\tilde\Pi^{AA}_{L;nc}$ of Eq.~(\ref{e:star}). Note that
$\tilde\Pi^{AA}_{L;nc}$ has acquired a pole at the physical pion mass due to
explicit chiral symmetry-breaking by the current quark mass. The behavior of
$(1/4)\tilde\Pi^{AA}_{L;nc}$ and the contribution to the spectral density
in the vicinity of the pion pole at
$s=m^2_\pi+{\cal O}(\hat{m}^2)$ are 
\begin{eqnarray}
\frac{1}{4}\tilde\Pi^{AA}_{L;nc}\approx \frac{-C_{a_L}}{s-m^2_\pi+i\epsilon}\,;
\quad \rho^A_0(s)_{\pi\,pole}=C_{a_L}\,\delta(s-m^2_\pi)\,,\quad 
C_{a_L}= f^2_\pi m^2_\pi, 
\end{eqnarray}
while the behavior on $C_\infty$ is
\begin{eqnarray}
\frac{1}{4}\tilde\Pi^{AA}_{L;nc}\approx
\frac{\hat{m}}{m}f^2_p\bar F_P(s)\approx \frac{\hat{m}\langle\bar\psi\psi\rangle}{s}
=-\frac{m^2_\pi f^2_\pi}{s}\,,\quad |s|\rightarrow \infty.
\label{e:AAasymp}
\end{eqnarray}
In obtaining these results, we have used the Goldberger-Treiman and
GMOR relations, Eqs.~(\ref{e:gtenjl}) and (\ref{e:gmor}), that are satisfied by the
ENJL coupling constants and pion mass, plus the information provided in
Table~\ref{table4}.  The pion ghost pole depicted in Fig.~\ref{f:ghosts}
does not occur, since we are
using barred polarization amplitudes.

We also need the residue of $(1/4s^2)\tilde\Pi^{AA}_{L;nc}$  at the double pole at
$s^2=0$. A short calculation gives this residue as
\begin{eqnarray}
\big\{\frac{d}{ds}\frac{1}{4}\tilde\Pi^{AA}_{L;nc}\big\}_{s=0}=\frac{1}{8\pi^2}+\frac{1}{m^2_\pi}\frac{f^4_\pi(0)}{f^2_\pi(m^2_\pi)}
\approx \frac{1}{8\pi^2}+\frac{f^2_\pi}{m^2_\pi}(1-\frac{m^2_\pi}{8\pi^2 f^2_\pi}+\cdots)
=\frac{f^2_\pi}{m^2_\pi},
\end{eqnarray}
after using the defining equation (\ref{e:physmass}) for the pion mass up to order
${\cal O}(\hat{m}^2)$. Deforming the contour in Fig.~\ref{f:contour} so that 
the pion pole on the real axis at $s=m^2_\pi \,<\,4m^2$  is also excluded, one finds
\begin{eqnarray}
\int_0^\infty\frac{ds}{s^2}\rho^A_0(s)
=\int_0^\infty\frac{ds}{s^2}\big(\rho^A_0(s)_{\pi\,pole}+\rho^A_0(s)_{cut}\big)
=\frac{C_{a_L}}{m^4_\pi}+ \big\{\frac{f^2_\pi}{m^2_\pi}+\frac{-C_{a_L}}{m^4_\pi}\big\}
=\frac{f^2_\pi}{m^2_\pi}
\end{eqnarray}
there being no contribution along $C_\infty$ in view of Eq.~(\ref{e:AAasymp}).
We notice that the sum rule is saturated by the
pion pole strength alone: the
contribution (in curly brackets) to the integrated spectral density along the cut vanishes.
This just reflects the fact that the PV-regulated
longitudinal axial density of the ENJL model again includes exactly the correct amount of negative density
at large $s$ for this sum rule to be obeyed.

The sum rules that involve the isoscalar-scalar minus isovector-pseudoscalar and isovector-scalar minus isoscalar-pseudoscalar
densities depend on the analytic behavior of the polarization differences
\begin{eqnarray}
\tilde\Pi^{S_0S_0}_{nc}(s)-\tilde\Pi^{PP}_{nc}(s)&=&\frac{\bar \Pi^{S_0S_0}_{nc}(s)-\bar \Pi^{PP}_{nc}(s)}
{\big(1-2G_1\bar\Pi^{S_0S_0}_{nc}(s)\big)\big(1-2G_1\bar\Pi^{PP}_{nc}(s)\big)}
\approx -4f^2_p \bar F_P(s)\,,\quad |s|\rightarrow \infty,
\nonumber\\
\nonumber\\
\label{e:S0-P}
\end{eqnarray}
obtained from constructing Eq.~(\ref{e:polsigmabar}) minus Eq.~(\ref{e:polstar})
and
\begin{eqnarray}
\bar\Pi^{SS}_{nc}(s)-\bar\Pi^{P_0P_0}_{nc}(s)
= -4f^2_p \bar F_P(s),
\label{e:S-P0}
\end{eqnarray}
on subtracting 
the Eq.~(\ref{e:HPP}) from Eq.~(\ref{e:HS0S0}).  This is done after formally
 upgrading  the proper polarizations
to include
non-chiral contributions too.
 As these effects will actually fall away to the order to which we are
working, this upgrading is inessential, $except$ for $\Pi^{PP}_{nc}$.
One notes that both differences behave asymptotically in the same way as the
axial vector minus
vector polarization of Eq.~(\ref{e:poldiff}).
Explicit forms of the
proper polarizations appearing in these expressions
have been given previously: $\Pi^{P_0P_0}$ in Eq.~(\ref{e:HPP}),
$\Pi^{S_0S_0}=\Pi^{SS}$ in Eq.~(\ref{e:HS0S0}), and $\tilde\Pi^{PP}_{nc}$
in Eq.~(\ref{e:polrenstar}).

In contrast to Eq.~(\ref{e:S-P0}), the difference in Eq.~(\ref{e:S0-P})
contains poles at
$s=m^2_\sigma=4m^2+{\cal O}(\hat{m})$ and $s=m^2_\pi+{\cal O}(\hat{m}^2)$,
given by the zeros of the two denominators, that determine the
masses of the $\sigma$ and $\pi$ meson modes. This structure comes about from
summing all proper polarization diagrams linked by the
interactions that are present in the isoscalar and isovector channels, a mechanism that is
absent in Eq.~(\ref{e:S-P0}).
The residues at these poles are given by the negatives $-C_\sigma$ and
$-C_\pi$ of their spectral strengths in the usual way:
\begin{eqnarray}
\rho^{S}(s)_{\sigma\,pole}=C_\sigma\delta(s-m^2_\sigma)\,,\quad C_\sigma=\frac{g^2_{\sigma qq}}{4G_1^2}
\\
\rho^{P}(s)_{\pi\,pole}=C_\pi\delta(s-m^2_\pi)\,,\quad C_\pi=\frac
{g^2_{\pi qq}}{4G_1^2}.
\end{eqnarray}
The sum rules that are still outstanding now follow without difficulty: for
 sum rule (\ref{e:gl3}), only the
residue of the integrand at zero contributes
 to the contour in Fig.~\ref{f:contour}
(that is suitably indented to avoid both the $\sigma$ and the $\pi$ poles), since
the residues at these poles just cancel against the contributions
from $\rho^{S}(s)_{\sigma\,pole}$ and $\rho^{P}(s)_{\pi\,pole}$, and there is no contribution
along $C_\infty$. The same remarks hold for evaluating the sum rule (\ref{e:gl4}), except
that the meson poles are absent. The contribution of the full scalar
polarization $\tilde\Pi^{S_0S_0}$ to the residue in the first case is
always of order
$\hat{m}^2\tilde\Pi^{S_0S_0}(0)\sim {\cal O}(\hat{m}^2)$, and can thus be
neglected to the order ${\cal O}(\hat{m})$ to which we are working.
By contrast, the pseudoscalar piece already contributes at the ${\cal O}(\hat{m})$
level,
$\hat{m}^2\tilde\Pi^{PP}(0)\sim {\cal O}(\hat{m})$.
This difference arises due to the special nature of the quasi-chiral pion mode (mass)$^2$ which
itself vanishes linearly with $\hat{m}$, while the sigma mode (mass)$^2$ approaches a constant. Thus
\begin{eqnarray}
\int_0^\infty \frac{ds}{s}\hat{m}^2\big(\rho^{S}(s)-\rho^{P}(s)\big)=
\hat{m}^2\bar\Pi^{S_0S_0}(0)-\frac{\hat{m}m}{2G_1}(1-\frac{\hat{m}}{m})=
-f^2_\pi m^2_\pi+O(\hat{m}^2),
\label{e:gl3p}
\end{eqnarray}
where the last step follows from GMOR. There is also a much more direct way of
evaluating this sum rule using the relation 
\begin{eqnarray}
\rho^A_0(s)=\frac{\hat{m}^2}{s}\rho^P(s),
\end{eqnarray}
that is readily established from the
definitions of $\rho^A_0$ and $\rho^P(s)$.
One then finds that
\begin{eqnarray}
\int_0^\infty\frac{ds}{s}\hat{m}^2\rho^P(s)&=&\int_0^\infty
ds \big(\rho^A_0(s)_{\pi\,pole}+\rho^A_0(s)_{cut}\big)
\nonumber\\
&=& C_{a_L}+\big\{-C_{a_L}+m^2_\pi f^2_\pi\big\}= m^2_\pi f^2_\pi,
\end{eqnarray}
after noticing 
that there $is$ a contribution from the contour $C_\infty$ in
this case, in view of Eq.~(\ref{e:AAasymp}) that just cancels against the residue of $(1/4)\tilde\Pi^{AA}_{L;nc}$
at the pion pole it contains.
Thus the contribution of the integral along
the cut (in curly brackets) is zero, the entire integrated strength being
determined by the pion pole in this way of calculating the integral.
Taking the negative of this result and dropping the scalar density
contribution for the same reason as before, we regain the sum rule (\ref{e:gl3p}).

Exactly the same arguments hold for obtaining the sum rule in (\ref{e:gl4}). Since there
are no meson poles in this case, the value of the integral is given by 
\begin{eqnarray}
\int_0^\infty \frac{ds}{s}\hat{m}^2\big(\tilde\rho^{S}(s)-\tilde\rho^{P}(s)\big)=
-4\hat{m}^2f^2_p\bar F_P(0)
 =-4\hat{m}^2f^2_p.
\end{eqnarray}
Equating this with $-2m^4_\pi l_7$ in conformity with
Eq.~(\ref{e:gl4}) in order to define $l_7$, we have 
\begin{eqnarray}
l_7=\frac{2\hat{m}^2f^2_p}{m^4_\pi}\sim (0.5\: {\rm to}\:
1.4)\times 10^{-3},
\label{e:l7}
\end{eqnarray}
depending on whether the ENJL or NJL parameters are used from
Table~\ref{table1}. We have used the GMOR relation to fix $\hat m$ at $3.0$MeV
or $4.5$MeV respectively, taking $m_\pi=135$MeV.
The empirical value is $\sim 5\times 10^{-3}$ \cite{gl84}.
Finally,
\begin{eqnarray}
\int_0^\infty ds \big(\rho^{S}(s)-\rho^{P}(s)\big)
= (C_\sigma-C_\pi)+\big\{-C_\sigma+C_\pi+4m\langle\bar\psi\psi\rangle\big\}
=4m\langle\bar\psi\psi\rangle
\label{e:gl5p}
\end{eqnarray}
and
\begin{eqnarray}
\int_0^\infty ds \big(\tilde\rho^{S}(s)-\tilde\rho^{P}(s)\big)
=4m\langle\bar\psi\psi\rangle,
\label{e:gl6p}
\end{eqnarray}
using by now familiar arguments. Note that there is a
contribution to both integrals coming from the contour $C_\infty$ that in fact
determines their common value.

In summary, we have extended the previous study \cite{dkl96} by showing
that all eight sum rules of interest to chiral perturbation
theory are obeyed by the PV regulated ENJL spectral densities, except for
modifications to the three zero moment sum rules that assume a value proportional to
$m\langle\bar\psi\psi\rangle$ instead of zero. Since the values of these
sum rules only involve combinations of $f_\pi$ and $m_\pi$, they continue to be
obeyed (with $f_\pi \rightarrow f_p$ and $m_\pi\rightarrow m_p$ )
by the PV-regulated NJL spectral
densities, which do not contain resonances of any kind from the
$\rho\,-\,a_1$ sector. Thus, while the vector and
axial vector modes produce resonant spectral density functions, such resonances are
not a prerequisite for the sum rules to hold.  We also comment on the
modifications of the zero moment sum rules that occurred in Eqs.~(\ref{e:weinmin2}),
(\ref{e:enjlw2}), (\ref{e:gl5p}) and (\ref{e:gl6p}). Possible modifications of
the original form of
Weinberg's sum rules, Eqs.~(\ref{e:w1}) and (\ref{e:w2}), have been investigated
by Nieh and Jackiw \cite{niehj68}, who developed criteria
for obtaining these sum rules from a specific Hamiltonian.
This in turn leads to the speculation (that remains to be explored along the lines
of the Nieh-Jackiw approach 
for example) that the modifications found here might perhaps be examples of
a more general property associated with the $dynamical$ breaking of chiral symmetry
with mass generation in the groundstate, and which
 would hold for any Hamiltonian that energetically favors the formation of a finite
condensate.

\section{Sum rules and the electromagnetic properties of the pion}
\subsection{Pion electromagnetic form factor, radius, and polarizability} 
\subsubsection{Form factor and pion radius}
The electromagnetic form factor, $F_\pi(q^2)$, of the pion is given
diagrammatically by Fig.~\ref{f:gammavert}. The diagram includes
 the $\rho^0$ exchange
renormalization of the isovector component of the quark electromagnetic current
as in Fig.~\ref{f:isospin current}. This modifies the $\gamma qq$ vertex, shown as an open
circle, as follows,
\begin{eqnarray}
-ieJ^{(em)}_\mu=-ie(\frac{1}{3}J^0_\mu+J^3_\mu)\rightarrow
-ie\big[e_q\gamma_\mu +\frac{1}{i}\Pi^{VV}_T(q^2)\frac{1}{i}D^{VV}_T(q^2)T_{\mu\nu}
\frac{\tau_3}{2}\gamma^\nu\big].
\label{e:emcurrent}
\end{eqnarray}
The charge carried by the quarks has been denoted by
$e_q=\frac{1}{2}(\frac{1}{3}+\tau_3)$, in units of $e$. 
The corresponding analytical expression for $F_\pi(q^2)$ has been derived as
Eq.~(\ref{e:ffpi}) in Appendix A, section 5:
\begin{eqnarray}
F_\pi(q^2)=\Big\{g_AF_P(q^2)+(1-g_A)+
(1-g_A)^2\frac{1}{8f^2_\pi}\Pi^{VV}_T(q^2)\Big\}F_V(q^2).
\label{e:myhope}
\end{eqnarray}
If the composite structure of the Goldstone $\pi$ meson is suppressed
by setting the contents of the curly brackets to unity, one obtains the vector dominance
model form: $F_\pi(q^2)\approx F_V(q^2)$. On the other hand, setting $g_A=1$ 
returns one to
the NJL expression $F_P(q^2)$ for the form factor.
The pion charge radius is identified from the low $q^2$ 
behavior of $F_\pi(q^2)$:
\begin{eqnarray}
F_\pi(q^2)= 1+q^2\big[\frac{g^2_A}{8\pi^2 f^2_\pi}-\frac{(1-g_A)^2}{12g_Am^2}
+\frac{1}{m^2_V}\big]+\cdots =1+\frac{1}{6}q^2\langle r_\pi^2\rangle+\cdots
\end{eqnarray}
The contribution from the derivative of the vector form factor to this
 expression has
been displayed separately as
\begin{eqnarray}
F^\prime_V(0)=(1-g_A)/6g_Am^2=1/m^2_V,
\end{eqnarray}
to show that it has exactly the structure of the vector dominance model
when expressed in terms of the $\rho$ meson mass $m_V$ as given by the 
low energy expansion of
the ENJL model. We thus obtain the following expression for the charge
radius of a chiral pion,
\begin{eqnarray}
\langle r^2_\pi\rangle&=&
\frac{3g^2_A}{4\pi^2f^2_\pi}+\frac{(1-g_A^2)}{2g_Am^2}=
\frac{1}{16\pi^2f^2_\pi}(\bar l_6-1).
\label{e:newpirad}
\end{eqnarray}
For $g_A=1$, this reduces to Tarrach's
expression \cite{tar79} for $\langle r_\pi^2\rangle$. We have introduced 
the definition of the $\bar l_6$ CHPT
coupling parameter in the last step. As a case in point, we remark that
inclusion of the leading order non-chiral contributions to $F_\pi(q^2)$
would add a chiral logarithm of the form \cite{kh95}
\begin{eqnarray}
-\frac{1}{16\pi^2f^2_\pi}\ln\,(m^2_\pi/4m^2)
\end{eqnarray}
to the middle term  of the above expression for $\langle r^2_\pi\rangle$.
This term, which arises from the leading non-chiral contribution of the 
dumbbell
diagrams to $F_\pi(q^2)$, has exactly the correct coefficient to combine
with $\bar l_6$ to give the prediction
\begin{eqnarray}
\bar l_6+\ln(m_\pi^2/\mu^2)= 1+12g_A^2+8\pi^2\frac{f^2_\pi}{m^2}\big(
\frac{1-g_A^2}{g_A}\big)
\label{e:l6}
\end{eqnarray}
at the scale $\mu=2m$.   The right hand side of this equation gives the
ENJL value for the renormalized coupling constant $l_6^r$ up to a conversion
factor \cite{gl84}.   However, there is no need to include this factor to 
compute $l_6^r$, since $\bar l_6$ is the physically interesting quantity and
this may now be computed directly.
Numerically, the calculated radius comes to $0.36$  fm$^2$ for
 our standard parameter set, and leads
to the value $\bar l_6=16.2$. The
experimental value is $\langle r^2_\pi\rangle=0.44\,\pm0.03\,$fm$^2$
\cite{dally82,amen86}, which in turn leads to the empirical value 
$\bar l_6=16.5\pm1.1$.

 The behavior of the ENJL form factor is compared with
 the experimental data \cite{piexp} for $|F_\pi(q^2)|^2$ 
 in  Fig.~\ref{f:pionf2} at both space-like and time-like $q^2$. 
Given that no parameters
 have been re-adjusted, one notes that
 the overall agreement is fair. Thus is particularly true at
 small space-like $q^2$
 where the behavior is quantitatively the same as the CHPT
 prediction \cite{bc87} (which, however, uses the experimental 
pion radius as input).
 This agreement simply reflects the reasonable value
 of $\langle r_\pi^2\rangle=0.36$fm$^2$ for
 the ENJL prediction of the
 pion radius. For larger $q^2$, and more especially in the time-like region,
 the same comments as were made in connection with
 the continuum structure of the vector and and axial vector spectral
 densities should be  borne in mind here too: The continuum
 behavior of the calculated $|F_\pi|^2$ for $q^2\geq 4m^2$ in 
Fig.~\ref{f:pionf2}
 is determined by the $\bar qq$ decay channel, not the physically expected
 $\rho\rightarrow \pi\pi$ channel.
 Thus the ENJL form factor embraces contributions in the time-like region 
from 
 a decay channel that would not be available in a confining theory.

\subsubsection{Polarizability of charged pions.}

Using the current algebra sum rule of Das, Mathur and Okubo \cite{dmo67}
 for the
axial structure constant that enters into  
 the $\pi \rightarrow e\nu\gamma$ radiative decay matrix element, Holstein has
 derived
 the following expression 
for the electric polarizability $\alpha_E$ of
charged pions \cite{holstein90}:
\begin{eqnarray}
\alpha_E =\frac{\alpha}{3m_\pi}\langle r^2_\pi\rangle
-\frac{\alpha}{m_\pi f^2_\pi}\int_0^\infty\frac{ds}{s^2}
\big(\rho^V_1(s)-\rho^A_1(s)\big).
\label{e:alphasum}
\end{eqnarray}
The first term corresponds to the ``classical'' contribution from a
charged particle of radius squared $\langle r^2_\pi\rangle$ \cite{eh73}.
Taking the pion radius squared from the Eq.~(\ref{e:newpirad})
 and the value of the integral from Eq.~(\ref{e:enjl-2}),
one obtains (we express the result directly in terms of the
ENJL $\bar l_i's$, and take their difference from Table~\ref{table2a})
\begin{eqnarray}
\alpha_E= \frac{\alpha}{m_\pi}\frac{1}{48\pi^2f^2_\pi}(\bar l^{ENJL}_6-\bar l^{ENJL}_5)=\frac{g_A^2}{8\pi^2 f^2_\pi}\frac{\alpha}{m_\pi}
=3.4\times 10^{-4}\,{\rm fm^3},
\label{e:alpharesult}
\end{eqnarray}
if $g_A=0.75$. In computing the numerical result in
Eq.~(\ref{e:alpharesult}),
we have also
 used $m_\pi=135$MeV and $f_\pi=93$MeV in addition to
 $\alpha\approx 1/137$ as input.

The first expression coincides exactly with the form
 of the result for $\alpha_E$ that is obtained by treating
the
forward scattering of soft photons by pions in chiral
perturbation theory at the  one-loop level\cite{burgi96}.
There are three additional noteworthy
features about this expression: (i) since the radius contribution and
the contribution from the integral over the
density functions occur with exactly the
same numerical coefficient when expressed in terms of the chiral
 coupling constants, all
factors common to both $\bar l_6$ and $\bar l_5$ cancel in the difference.
This feature leads to
a complete cancellation of the contribution from the $\rho$ meson degrees of
freedom, a feature that has also been obtained by Holstein \cite{holstein90}
via another method, that of 
saturating the sum rule with delta function distributions 
in conjunction with the
KSRF relation. (ii) the quark mass $m$ has fallen away due to
 this cancellation,
and (iii) the quenching effect of the $a_1$ meson degree of freedom  
only enters
via the axial form factor $g_A$. Thus Eq.~(\ref{e:alpharesult}) 
is a $prediction$ of the ENJL model that
only contains physical quantities that are known in principle. In particular,
it is a prediction for an upper limit on $\alpha_E$,
\begin{eqnarray}
\alpha_E\leq \frac{\alpha}{8\pi^2f^2_\pi}\frac{1}{m_\pi}
\approx 6\times 10^{-4}\;\rm{fm^3}.
\end{eqnarray}
The result (\ref{e:alpharesult}), and the upper limit in particular, can be compared directly with
the prediction of chiral perturbation theory at one loop that uses the empirical values of
the coupling constants $\bar l_5$ and $\bar l_6$ to obtain \cite{holstein90,burgi96}
\begin{eqnarray}
\alpha^{chpt}_E=\frac{\alpha}{m_\pi}\frac{1}{48\pi^2f^2_\pi}(\bar l_6-\bar l_5) = 2.7\times 10^{-4}\,{\rm fm^3}
\end{eqnarray}
for the polarizability of the charged pion.    One can also compare our
result with 
the estimate  $\alpha_E=(5.6\pm0.5)\times 10^{-4}$fm$^3$ that has been
obtained from Eq.~(\ref{e:alphasum})
by using the experimental pion radius as input,
and estimating the integral involving the vector and axial vector
densities using QCD sum rule methods \cite {lns94}.

The experimental
value of $\alpha_E$ is still very poorly known.
The data analysis for the combinations $\alpha_E \pm\beta_M$, where 
$\beta_M$ is the magnetic polarizability, yields values for $\alpha_E$
from $2.2\pm1.6$ \cite{jb90,bab92} through $6.8\pm1.4$ \cite{antipov83}
to $20\pm12$ \cite{aiberg86}, in units of $10^{-4}$ fm$^3$.   In addition,
these numbers depend somewhat on whether the constraint of good chiral
symmetry \cite{dh89} $\alpha_E + \beta_M =0$ is implemented or not.
Thus one may conclude that the
chiral perturbation theory prediction and the ENJL model sum 
rule prediction  agree qualitatively
with each other, and that neither result agrees with the 
experimental data to date.
We refer to \cite{pp94} for an overview of the present 
state of experiment and
comparison with theoretical calculations.

\subsection{Pion electromagnetic mass difference}
In a historically important paper from the point of view of current algebra
techniques, Das, Guralnik, Mathur, Low and Young  \cite{low67} exploited Weinberg's two
sum rules, Eqs.~(\ref{e:w1}) and (\ref{e:w2}) for massless QCD to obtain an
expression for the pion electromagnetic mass difference in terms of vector and axial
vector spectral densities. These authors showed that
the mass squared difference between the charged and uncharged chiral pions
(therefore having common rest mass zero) may be
written in the form
\begin{eqnarray}
\Delta m^2_\pi=m^2_{\pi^{\pm}}-m^2_{\pi^0}=3ie^2f^{-2}_\pi\int
 \frac{d^4 q}{(2\pi)^4 }
\frac{1}{q^2}\int_0^\infty
\frac{ds}{s-q^2}\big(\rho^V_1(s)-\rho^A_1(s)\big),
\label{e:emmassdiff}
\end{eqnarray}
to ${\cal O}(\alpha)$ in the electromagnetic coupling 
$\alpha=e^2/4\pi\approx 1/137$.
By saturating the Weinberg sum rules with low-lying vector and axial 
vector mesons, they
obtained the estimate
$\Delta m^2_\pi \approx (3\alpha/2\pi)m^2_\rho \ln\,(m^2_{a_1}/m^2_\rho)
=(37.85$MeV$)^2$ in terms of the
$\rho$ and $a_1$ meson masses, taking $m_\rho =770$MeV and the Weinberg ratio
$m_{a_1}=\sqrt 2\,m_\rho$ for the $a_1$ mass.
The current experimental value 
is $\Delta m^2_\pi=1261$MeV$^2$=$(35.51\,\rm MeV)^2$.
This calculation represented one of the first successful applications of 
QCD sum rules.

By contrast, the ENJL expression for the spectral density difference
appearing in Eq.~(\ref{e:densdiff}) allows us to evaluate
the Das {\it et al.} expression without making any 
further assumptions as to the 
resonance content of
the spectral density functions. Since, according to Eq.~(\ref{e:contour1}),
the required integral over $s$ is given by the residue $f(s)$ at
$q^2$ instead of zero, where $f(s)=f^2_\pi\bar F_P(s)\bar F_V(s)\bar F_A(s)$,
one immediately has the result
\begin{eqnarray}
\Delta m^2_\pi =3ie^2\int \frac{d^4 q}{(2\pi)^4 }\frac{1}{q^2} 
\bar F_P(q^2)\bar F_V(q^2)\bar F_A(q^2).
\label{e:emenjl}
\end{eqnarray}
A formula essentially equivalent to this expression has also been 
derived using a QCD
effective action approach \cite{brz94}. Below, we give yet another
derivation of this expression using Feynman diagrams. In the limit
 $g_A=1$, where
$\bar F_V$ and $\bar F_A $ are both unity, it reduces to the expression 
given previously
\cite{dstl95} for
$\Delta m^2_\pi$ for the NJL model, while making the {\it ad hoc}
 assumption that all the
mesons are elementary without internal structure of any kind, $\bar F_P(s)=1$,
$\bar F_{V,A}\approx (1-q^2/m^2_{\rho,a_1})^{-1}$ leads one back to the 
Das {\it et al.}
estimate.

Futher evaluation of $\Delta m^2_\pi$  has to be done numerically.
However, since the product of PV-regulated form factors in the integrand of Eq.~(\ref{e:emenjl})
goes like $-1/q^2$ according to Table~\ref{table4},
the integral over the photon momenta
diverges logarithmically, and some
form of photon momentum cutoff is required.  This contrasts with the special case of
approximate evaluation by Das {\it et al.} mentioned above, where the asumption of delta distributions
for the vector and axial vector densities that satisfy both Weinberg sum rules
guarantees a convergent result.  A logarithmic divergence also results when their
approach is extended to cover physical (i.e. non-chiral) pions \cite{glns67}.

Performing a Wick rotation to Euclidean photon momenta, one finds
\begin{eqnarray}
\Delta m^2_\pi&=&\frac{3\alpha}{4\pi}
\int_0^{\lambda^2_{ph}}dQ^2\bar F_P(-Q^2)\bar F_V(-Q^2)\bar F_A(-Q^2).
\label{e:emeval}
\end{eqnarray}
This transformation is permissible since the poles
of the integrand all lie on the second Riemann sheet and thus do not
interfere  with the deformation
of the integration contour.  The behavior of $\Delta m^2_\pi$ versus
$\lambda^2_{ph}$
is shown in Fig.~\ref{f:deltapi2}.
A  cutoff of
$\lambda^2_{ph}=(1.27\rm GeV)^2$ leads to the observed value
$\Delta m^2_\pi=1261$ MeV$^2$ of this quantity, if we use
the standard parameter set of this paper to compute
the form factors in Eq.~(\ref{e:emenjl}). 
This value of the photon momentum
cutoff requires some comment. In the ${\cal O}(4)$ regularization scheme (which we do not
employ) the Euclidean quark momenta are limited by the cutoff $\Lambda_{O(4)}$, so
that the photon momenta are restricted by $Q^2 \le 4\Lambda^2_{O(4)}$ in 
Eq.~(\ref{e:emeval}).
Since one has the rough equivalence \cite{dstl94} $\Lambda^2_{O(4)}\approx (2\ln 2)\Lambda^2\sim (
2.5\,\rm GeV)^2$, this
restriction is met by our photon cutoff. Furthermore this cutoff also agrees
satisfactorily with the scale $\mu^2=(0.95\,\rm GeV)^2$  quoted in \cite{brz94}
at which a smooth matching \cite{bar89} between Eq.~(\ref{e:emeval})
 (for a different set of input parameters however) and the short
 distance QCD contribution to
 $\Delta m^2_\pi$ can be accomplished.

\subsection{Derivation via Feynman diagrams}
While the derivation of the expression just given for $\Delta m^2_\pi$
via spectral densities is technically very concise, the underlying physics determining this
splitting is less transparent. We remedy this by using the diagram method.
One can then verify the ENJL expression (\ref{e:emenjl}) for $\Delta m^2_\pi$
explicitly by identifying the
additional set of Feynman diagrams for the pion polarization that
include  single photons in intermediate states. This method has already been described
in \cite{dstl95} for the case of the minimal NJL model.
The diagrammatic method
offers some advantages over the sum rule approach.
The electromagnetic gauge invariance can be made explicit, as well as
the consequences of the residual $U_A(1)\times U_V(1)$ chiral symmetry  as expressed by Dashen's theorem \cite{dash69},
 that keeps the neutral chiral pion mass at zero in the
presence of its own internal electromagnetic field.  It confirms that
the Das {\it et al.} formula in conjunction with the ENJL spectral densities
is correct  to  ${\cal O}(\alpha N_c)$. Also, it is straightforward to extend this method
to cover physical pions carrying non-zero mass in the absence of the
electromagnetic interaction.
An attempt to extend the  approach of Das {\it et al.} directly to cover
the case of non-zero pion mass is reported in \cite {glns67}.  

In order to identify the relevant electromagnetic
diagrams for calculating the electromagnetic mass splitting of the pion,
we include the electromagnetic interactions via minimal coupling in
the ENJL model, and study their effect to ${\cal}O(\alpha)$.  The additional
diagrams are obtained in the usual way by dressing the skeleton
pion polarization
diagrams  with single photon self-energy and vertex corrections.
The diagrammatic calculation of $\Delta m^2_\pi$ to ${\cal}O(\alpha)$  starts
out with the observation that only those diagrams involving the exchange of a single photon
 that are different in the isovector and isoscalar channels
will contribute. This immediately excludes all quark self-energy diagrams
in the pion polarization where the photon line
dresses either a quark, or an antiquark line. Such diagrams (which are however
essential to maintain  gauge invariance)  contribute
equally to the charged and neutral pion channels and thus cancel in the
mass difference calculation. They are shown in Fig.~\ref{f:selfenergy} for later
reference.
 The diagrams we thus require are 
of two types: (i) scattering diagrams where the quark-antiquark pair making up
the polarization loop interact via photon exchange in intermediate states,
and (ii) meson-pole, or ``dumbbell'' diagrams where the $\pi+a_L,\rho$ and $a_1$ meson modes
 are exchanged together with a photon between two $\gamma\pi\pi$ vertices.
These diagrams are presented in Fig.~\ref{f:emdiagram}.
In addition, $\rho^0$  exchange causes a renormalization of
the isovector component of the quark electromagnetic current
as in Eq.~(\ref{e:emcurrent}). The modified vertex
itself is again indicated as an open circle to which the photon is attached
as in Fig.~\ref{f:gammavert}.

Both sets of diagrams  in Fig.~\ref{f:emdiagram} are vertex corrections
to 
the basic one-loop irreducible diagram that determines the pion polarization
to ${\cal O}(\alpha N_c)$. At first sight,
 the appearance of such two-loop diagrams containing mesons in intermediate
states is surprizing, since they formally contain higher powers in the meson 
quark coupling constants
than the single loop diagram. However they are of the same order
${\cal O}(\alpha N_c)$  as the single loop contribution to the electromagnetic self-energy
of the pion, and therefore of ${\cal O}(\alpha)$ relative to the pion self-energy
in the absence of electromagnetic interactions. This comes about since  the meson propagators scale like $1/N_c$.
This is a special feature arising from the linked nature of the solutions for the quark and meson self-energies  in the
NJL model. 

Let  $\Pi^{T_3}_{EM}(q^2)$ denote the contribution from the
sum of diagrams in Fig.~\ref{f:emdiagram} in the channel of isospin projection  $T_3$.
Here   $T_3$  equals
$\pm 1 $ for $\pi^{\pm }$ and $0$ for $\pi^0$. Then  $\Delta m^2_\pi$ is given
by \cite{dstl95}
\begin{eqnarray}
\Delta m^2_\pi = -g^2_{\pi qq} \big\{\Pi^{\pm}_{EM}(0)-\Pi^{0}_{EM}(0)\big\}
\label{e:dmfromdiag}
\end{eqnarray}
for chiral pions.
Using an extension of the arguments presented in detail in \cite{dstl95}, one can
again show that this expression is fully gauge invariant. We can thus describe the
photon propagation in any convenient gauge. We use the Feynman gauge,
\begin{eqnarray}
-iD^{\mu\nu}_{ph}(q^2)=-i\frac{g^{\mu\nu}}{q^2}.
\label{e:phprop}
\end{eqnarray}
The translation of all the Feynman diagrams in Fig.~\ref{f:emdiagram}
is deferred to Appendix A, section 6. Here
we simply summarize the results for the two classes of diagrams:

\vspace{0.5cm}
\noindent (i) {\it Scattering diagrams.}
The scattering diagrams give rise to a contribution to the
difference (\ref{e:dmfromdiag}) that is of the form 
[Eq.~(\ref{e:scat})],
\begin{eqnarray}
\big\{\Pi^{\pm}_{EM}(0)-\Pi^{0}_{EM}(0)\big\}_{scatt}
=-ie^2\frac{f^2_p}{m^2}\int \frac{d^4 q}{(2\pi)^4q^2}F_P(q^2)
\nonumber\\
-3ie^2\frac{f^2_p}{m^2}\int \frac{d^4 q}{(2\pi)^4q^2}F_P(q^2)F^2_V(q^2)\,.
\label{e:scattxt}
\end{eqnarray}

\vspace{0.4cm}
\noindent (ii) {\it Meson pole, or dumbbell diagrams.}  By contrast, the  dumbbell diagrams only contribute to the charged channels.
There
is no contribution for $T_3=0$, due to the electromagnetic vertices
they contain.  One finds  for the sum of the $\pi+a_L$ contributions that
[Eq.~(\ref{e:pial})]
\begin{eqnarray}
\Pi^{(\pi+a_L)}_{EM}(0)=
ie^2\frac{f^2_p}{m^2}\int \frac{d^4 q}{(2\pi)^4q^2}F_P(q^2)
\label{e:pialtxt}
\end{eqnarray}
while
\begin{eqnarray}
\Pi^{(a_1)}_{EM}(0)=
3ie^2\frac{f^2_p}{m^2}\int \frac{d^4 q}{(2\pi)^4q^2}\big\{F_P(q^2)F^2_V(q^2) \frac{8G_2f^2_pF_P(q^2)}{1+2G_2\Pi^{AA}_T(q^2)}\big\}
\label{e:a1txt}
\end{eqnarray}
gives the contribution from the $a_1$ dumbbell diagram [Eq.~(\ref{e:a1})].
In obtaining the latter result, we have inserted the specific form of the
$a_1$ propagator into  Eq.~(\ref{e:a1}). Finally we note that there is no contribution from the analogous dumbbell diagram
with a $\rho$ meson exchange replacing the $a_1$ vector meson
exchange in the chiral limit. The reason for this is given in more detail
 in Appendix A.

We now add Eqs.~(\ref{e:scattxt}) through (\ref{e:a1txt}). The $\pi+a_L$ dumbbell contribution cancels
against a similar term in the scattering contribution, and we are left with
a result for the self-energy difference required in Eq.~(\ref{e:dmfromdiag})
that may be written as
\begin{eqnarray}
-g^2_{\pi qq}\big\{\Pi^{\pm}_{EM}(0)-\Pi^0_{EM}(0)\big\}&=&3ig^2_{\pi qq}e^2\frac{f^2}{m^2}
 \int \frac{d^4 q}{(2\pi)^4q^2}F_P(q^2)F^2_V(q^2)
 \Big\{1- \frac{8G_2f^2_pF_P(q^2)} {1+2G_2\Pi^{AA}_T(q^2)}\Big\}.\nonumber \\
\end{eqnarray}
At first sight,
 this expression gives the impression that the contributions from the
$\rho^0$ degrees of freedom enter quadratically through the vector form
factor $F_V(q^2)$. This impression is false.  Using the Ward 
identity of Eq.~(\ref{e:ward1}) in reverse,
one sees that
\begin{eqnarray}
\Big\{1- \frac{8G_2f^2_pF_P(q^2)} {1+2G_2\Pi^{AA}_T(q^2)}\Big\}=
\frac{1+2G_2\Pi^{VV}_T(q^2)}{1+2G_2\Pi^{VV}_T(q^2)} =G_A(q^2)/F_V(q^2).
\end{eqnarray}
Thus one power of $F_V(q^2)$ is removed. Furthermore, since $g^2_{\pi qq}=
g^2_p/g_A$, the strong interaction physics part of the factor in front of the integral becomes $f^2_pg^2_p/(m^2g_A)
=1/g_A$ that supplies precisely the correct factor to convert $G_A(q^2)$
into the axial form factor $F_A(q^2)=G_A(q^2)/g_A$. Hence
\begin{eqnarray}
\Delta m^2_\pi =3ie^2\int
 \frac{d^4 q}{(2\pi)^4 }\frac{1}{q^2} F_P(q^2) F_V(q^2) F_A(q^2).
\nonumber
\end{eqnarray}
We thus recover the same
expression for $\Delta m^2_{\pi}$ as given in Eq.~(\ref{e:emenjl})
by a completely different route. All
 that remains is to specify the regularization
procedure.

Having obtained expressions for  the  subdiagrams  making up the
pion polarization,  it becomes a simple matter to identify the underlying
physics of the pion mass splitting in more detail. To this end,
 let us introduce
the electromagnetic
 self-energies $\Sigma^{T_3}$ for the charged ($T_3=\pm1)$ and
neutral ($T_3=0$) pions. This self-energy, which has the dimensions of a
(mass)$^2$,
is obtained from the total pion polarization by including the $\pi qq$ coupling
constant $ig_{\pi qq}$ at each external pion vertex:
\begin{eqnarray}
\Sigma^{T_3}= (ig_{\pi qq})^2\Pi^{T_3}_{EM}(0).
\label{e:pionself}
\end{eqnarray}
 Consequently, a negative pion polarization corresponds to
a repulsive self-energy, and {\it vice versa}. Now this $\Pi^{T_3}_{EM}(0)$,
and therefore
$\Sigma^{T_3}$, consists of the sum of {\it all} the polarization diagrams
appearing in Fig.~\ref{f:selfenergy} plus Fig.~\ref{f:emdiagram}.
Thus we have reinterpreted this $\Pi^{T_3}_{EM}(0)$ slightly in order to
avoid introducing yet another symbol.
 Let us write out this sum
for the two charged states of the pion separately. In doing so, we notice that
the sum of diagrams from Fig.~\ref{f:selfenergy}, that are all in the nature of self-energy insertions on the quark and
antiquark lines, are common to both charge channels.
 This is so because the quarks and
antiquarks enter symmetrically.
This is not true for the diagrams in Fig.~\ref{f:emdiagram}. Here diagram
(a) of this figure represents the interaction 
via photon exchange of the $\bar q q$
pair in the polarization loop. For the like-flavored quarks making up the
$\pi^0$, this is an attractive interaction; otherwise it is repulsive. Finally the
dumbbell diagrams in Fig.~\ref{f:emdiagram}(b) only contribute in the
charged channel due to the electromagnetic
vertices that they contain. We continue to work in the chiral limit.
Then  the electromagnetic (mass)$^2$ of the $\pi^0$ is
\begin{eqnarray}
 m^2_{\pi^0}&=&(ig_{\pi qq})^2\Big[\underbrace{\Pi^{(a+b)}_{EM}(0)+
 \Pi^{(c+d)}_{EM}(0)}_{Fig.~\ref{f:selfenergy}}\Big]
+(ig_{\pi qq})^2\{\underbrace{\Pi^0_{EM}(0)}_{Fig.~\ref{f:emdiagram}(a)}\}_{scatt}
\nonumber\\
&=&\Sigma_{self}+\Sigma^0_{scatt}.
\label{e:pizeromass}
\end{eqnarray}
Further progress is made by calculating the actual values of all
these pion self-energies. This is completed in Appendix A7.
The scattering contribution represented by the last term  is known
from Eq.~(\ref{e:scatappendix}).
It is negative as anticipated (in the Feynman gauge),
and  {\it exactly cancels}
the quark self-energy contribution given by the sum in the first bracket.
One finds by direct calculation the result given in Eq.~(\ref{e:cancel}) that
\begin{eqnarray}
\Sigma_{self}=-\Sigma^0_{scatt}=ie^2g^{-1}_A \int\frac{d^4q}{(2\pi)^4q^2}
\Big\{4tr(e^2_q)+\frac{3}{2}(F^2_V(q^2)-1)\Big\}F_P(q^2).
\end{eqnarray}
Thus
$m^2_{\pi^0}=0$: the neutral pion does not
 acquire any electromagnetic mass at all from its own internal electromagnetic interactions in the chiral limit. The
 increase in its mass due to the electromagnetically heavier quarks it contains
 is  compensated  for
 exactly by the attractive interaction between these quarks.
  Notice  that this is a gauge invariant statement. While the individual terms
 appearing in Eq.~(\ref{e:pizeromass}), are not all gauge-invariant, their sum is.
 More formally this result, which is just Dashen's theorem in operation,
 has to follow, once one realises that the inclusion
of the photon gauge field via minimal coupling
breaks the
 $U(1)_V\times SU_L(2)\times SU_R(2)$ chiral symmetry of the
original ENJL
Lagrangian in Eq.~(\ref{e:lagra}) down to $U_A(1)\times U_V(1)$.
The massless $\pi^0$ is simply the Goldstone mode realization of this residual symmetry. 
As we have shown, this result is satisfied in an entirely non-trivial
way by the ENJL model. It is also a proof that our set of polarization
diagrams to ${\cal O}(\alpha N_c)$ is correct and complete.

By contrast, the charged chiral pion picks up a non-zero electromagnetic mass:
\begin{eqnarray}
 m^2_{\pi^\pm}=(ig_{\pi qq})^2\Big[\underbrace{\Pi^{(a+b)}_{EM}(0)
 +\Pi^{(c+d)}_{EM}(0)}_{Fig.~\ref{f:selfenergy}}\Big]
 +(ig_{\pi qq})^2\{\underbrace{\Pi^\pm_{EM}(0)}_{Fig.~\ref
{f:emdiagram}(a)}\}_{scatt}
 \nonumber\\
 +
  (ig_{\pi qq})^2\Big[\underbrace
{\Pi^{(\pi+a_L)}_{EM}(0)+\Pi^{(a_1)}_{EM}
(0)}_{Fig.~\ref{f:emdiagram}(b)}\Big],
 \label{e:pizchargedmass}
 \end{eqnarray}
or
\begin{eqnarray}
m^2_{\pi^\pm} =\Sigma_{self}+\Sigma^\pm_{scatt}+\Sigma^\pm_{dumbbells}
\end{eqnarray}
using a self-evident notation for the various contributions to the self-energy of the
charged pion. In this expression, both the quark self-energy contribution as well as
the scattering contribution, are positive, 
the latter being so because the EM interaction between $\bar qq$ pairs of
different flavors ($\bar ud$ or $\bar du$) is repulsive.
Thus the last term arising from the sum dumbbell diagrams must
be negative, and, as we will see, strongly so, in order to cancel out most of the repulsive
self-energy associated with the first two terms. We naturally regain
Eq.~(\ref{e:dmfromdiag}) again by forming the difference
$m^2_{\pi^\pm}- m^2_{\pi^0}$. 
  As we have seen, only the charged pion contributes to this difference.
In Fig.~\ref{f:dashenoper}, we display the various contributions
to the electromagnetic (masses)$^2$ of the $\pi^0$ and $\pi^\pm$,
using the same parameter set as for evaluating Eq.~(\ref{e:emenjl}). In examining this figure, it  should again be borne in mind
that the individual
shifts are gauge-dependent (we have used the Feynman gauge).
Only the  total electromagnetic (masses)$^2$, and therefore their 
splitting, are  gauge-invariant quantities.

These results  only hold in the chiral limit of vanishing pion mass.
Their  extension to the non-chiral case of physical pions requires re-evaluation  of the same set of diagrams
with the external pion placed on its mass shell at $k^2=m^2_\pi$ already
in the absence of its own internal  electromagnetic field.
This is a tedious task,
even for the minimal NJL model \cite{dstl95}, and no less so in chiral
perturbation theory \cite{urech95}. However, the main conclusions reached
by including non-chiral effects \cite{dstl95} carry over without change into the
present calculation, and can be restated quite simply. 
The expression (\ref{e:emmassdiff}) accounts for the major  part of the
electromagnetic mass shift. In addition, the assumption of a finite
current quark mass in Eq.~(\ref{e:lagra}) breaks the chiral symmetry explicitly.
This gives the pions of the
model a finite  mass  $m^2_\pi\sim \hat m$ as per Eq.~(\ref{e:enjlms}).
The subleading (in $m^2_\pi$) contribution to the electromagnetic
splitting from this source
comes from expanding the scattering and dumbbell diagrams
in Fig.~\ref{f:emdiagram} in terms of $m^2_\pi$.
This gives rise to ${\cal O}(\alpha m^2_\pi)$ correction terms
in the neutral channel, but  an additional chiral logarithm 
 $\sim\alpha m^2_\pi\ln m^2_\pi$ appears in the charged channels.
 The latter term, which is identical to the correction found in CHPT
 \cite{urech95},
 arises
specifically from the pion dumbbell diagram in Fig.~\ref{f:emdiagram}.
It dominates the non-chiral correction. Since the chiral limit
of this diagram, as given by Eq.~(\ref{e:pialtxt}), is the same for both
versions of the  NJL model, we can  take the precise
form of the chiral logarithm correction from the former
calculation \cite{dstl95} and write
\begin{eqnarray}
\big\{\Delta m^2_{\pi}\big\}_{non-chiral}= \Delta m^2_{\pi} -\frac{3\alpha}{4\pi}
\ln(m^2_\pi/m^2_\sigma)+O(\alpha m^2_\pi)+\cdots
\end{eqnarray}
 for the ENJL model too.
 The chiral logarithm
  correction in this equation  coincides exactly with the
result obtained in chiral perturbation theory \cite{urech95} (at scale $\mu^2=
m^2_\sigma$).   Numerically, the chiral logarithm
and ${\cal O}(\alpha m^2_\pi)$  terms
constitute typically $10\,\%$ and $1\,\%$ of the leading chiral mass
splitting term.   Unequal current quark masses, or $\pi^0 \eta$ or
$\eta^\prime$ mixing in the $U_L(3)\times U_R(3)$ flavor extension of the
ENJL model (including the 't Hooft term \cite{th76} to remove the unwanted $U_A(1)$ symmetry),
are additional sources  of chiral symmetry breaking that contribute
to $\Delta m_\pi^2$. Both effects contribute at the few percent
level. One knows this from a direct estimate \cite{dstl95} of
the up-down current quark mass difference contribution
$\sim (\hat m_u-\hat m_d)^2$  to $\Delta m^2_\pi$, and from a rather general discussion
given in \cite{gtw79} for the mixing effects.
Bearing these results in mind, one is led to regard
the mass of the charged pion, and especially
the $\pi^\pm-\pi^0$ mass difference, as an essentially  electromagnetic
phenomenon.

\section{Summary and Discussion}
In this paper, we have  established that the two sum rules of Weinberg
for massless QCD, plus the
six sum rules introduced by Gasser and Leutwyler
in connection with chiral perturbation theory, are all obeyed by 
 spectral densities  that are
 generated by  the Nambu-Jona-Lasinio model Lagrangian
 in its minimal, as well as in its extended form, and which
 includes vector and axial vector degrees of freedom. The only exceptions to
 this statement occur for the zero moment sum rules [Eqs.~(\ref{e:enjlw2}),
 (\ref{e:gl5p}) and (\ref{e:gl6p})] that acquire a finite value proportional to
 the quark condensate density that represents the order parameter
 of the model. We suggest that this finding might perhaps be an example of a more
 general feature of
 any chiral Hamiltonian that energetically favors a chirally broken
 groundstate with a finite quark condensate.

 The NJL model in either form is both non-renormalizable
 and non-confining. Thus the statements regarding sum rules
 are subject to stating what sort of regularization precedure has
 been followed.   We show in particular
 that the sum rules hold for  spectral densities   computed from  polarization
 amplitudes which have been regulated according to the Pauli-Villars
 prescription. As might be anticipated, the PV-scheme introduces
 an unphysical behavior into the spectral densities at high energies,
 where these densities become negative  instead of remaining positive
 definite.  Nevertheless
 this happens in such a way that the sum rules continue to be saturated.
 We interpret this rather surprizing conclusion to be a consequence of the 
underlying
 chiral symmetry properties of the Lagrangian and its faithful representation
 under the PV-scheme.   As a corrollary to this statement, we find that
 the violation, due to PV-regularization, of the positive definite 
nature of the model densities
 becomes less  and less important for the sum rules involving higher and 
higher
 inverse moments, such as the sum rule determining the pion polarizability,
 for example.

 While the sum rules alluded to above also hold for the spectral densities
 belonging to the minimal NJL model that does not contain any
 explicit vector meson modes, this is perhaps less interesting from the point
 of view of physical applications. The sum rules involve {\it inter alia}
 explicit reference to both vector and axial vector densities, and so it is of
 direct physical
 interest to consider such degrees of freedom explicitly. With this in mind,
 we have
 revisited the problem of incoporating these modes into the NJL model in 
order to determine
 the behavior of the associated spectral densities and their masses.
 The basic effects of these modes are
(a) to renormalize the pion decay constant and $\pi q q$ coupling constant
in a complementary fashion such that the Goldberger-Treiman relation is 
retained, and (b)
 to concentrate the vector and axial vector
 strengths of the one-loop spectral densities of the minimal NJL model into
 two
 well-defined peaks at low energies that lie just above the $\bar qq$ 
continuum
 threshold. This redistribution of strength takes place in such a way, however,
 that all the
 sum rules continue to be obeyed. If the positions of these peaks are taken to define the $\rho$ and $a_1$ meson masses of the
 ENJL model, their numerical positions are found to be in qualitative accord with the experimental
 masses of the physical $\rho$ and $a_1$ mesons, suggesting that this
  identification has physical meaning.

  We have combined the sum rules  techniques developed in this paper,
  together with
  other calculated information to identify  ENJL expressions for the empirical coupling
  constants introduced by Gasser and Leutwyler into
  chiral perturbation theory, to evaluate the current algebra expression of
  Das {\it et al.} for the charged to neutral electromagnetic
  mass splitting of the pion, and 
  to compute the charged pion's polarizability
  from Holstein's sum rule.
  In extracting the coupling constants of chiral perturbation theory,
  it has been borne in mind that the physically relevant scale-invariant
 constants,
  $\bar l_i$,
  are not well-defined in the chiral limit because of the chiral logarithms
  they contain. However, the combination $\bar l_i+\ln(m^2_\pi/\mu^2)$ is
  well-defined, being proportional to the renormalized $l^r_i$ of Gasser
 and Leutwyler,
  and it is this combination that one can extract from the ENJL model.
  In fact, it is possible to go further and extract the $\bar l_i$'s
  themselves
  due to the following circumstance. One can show that the leading
  order non-chiral corrections to the ENJL amplitudes are precisely in the
 form of chiral
  logarithms that occur with the correct coefficients to reproduce the 
combination given above
  (at the scale of the sigma meson mass of the model).
  The appearance of the correct chiral logarithm has been demonstrated
 explicitly
  for the case of $\bar l_6$, for example, which must be identified  from a 
calculation of the
  pion radius.  One is thus fully justified in solving for the $\bar l_i$ 
themselves
  from the ENJL predictions of the $l^r_i$. This we have done and 
have obtained
  reasonable agreement with the empirical values of the $\bar l_i$'s 
  as set out in Table~\ref{table2b}.

  The sum rule calculation of the pion polarizability $\alpha_E$ made direct
 use of the
  ENJL model generated spectral densities, as well as the ENJL expression
 for the
  pion radius.  The resulting expression for $\alpha_E$, which coincides
 exactly in form
  with the CHPT result for this quantity, only depends on the difference
  $ \bar l^{ENJL}_6-\bar l^{ENJL}_5$.
  This allows  common terms in the two ENJL values for the
  coupling constants to cancel, and leads 
  to  an upper limit of $\alpha/(8\pi^2m_\pi f^2_\pi)\approx
  6\times 10^{-4}$fm$^3$ on $\alpha_E$ in terms of the
  physical mass and decay constant of the pion.
   This is an important result, since it represents a $prediction$ of the
  ENJL model approach that may be
 compared directly with the chiral perturbation
  theory prediction ($\alpha^{chpt}_E=2.7\times
 10^{-4}$fm$^3$), as well as the
  result of Ref.~\cite{lns94}
 ($\alpha^{QCD}_E=5.6\pm 0.5\times 10^{-4}$fm$^3$) that is
  based on
  an  evaluation of Holstein's sum rule using  approximate QCD 
  spectral densities and methods.
  We also showed that the limiting value of $\alpha_E$ quoted above is
  reduced by $g^2_A$, the quark axial form factor squared,
  in the presence of  
   axial vector mesons accompanying the pion. Such  meson clouds lead to
   a less polarizable system: for $g_A=0.75$,
 one finds $\alpha_E=3.4\times 10^{-4}$
   fm$^3$ that is nicely bracketed by the
 chiral perturbation theory and QCD sum rule
   predictions. By contrast, the $\rho$ mesons do not change the pion
   polarization at all. Their contributions 
 cancel out completely, in complete agreement
   with the findings of Holstein \cite{holstein90},
 who employed delta function
   spectral densities together with Weinberg's vector meson mass relation
   and the KSRF relation.

   Finally we must caution that the 
   the limiting value on $\alpha_E$  quoted above is subject to
   a multiplicative correction of the form $[1+{\cal O}(m^2_\pi) +{\cal O}(1/N_c)]$ 
   where the additional modifications arise from neglected non-chiral effects
   of ${\cal O}(m^2_\pi)$ in the
   sum rule itself, and neglected chiral corrections of ${\cal O}(1/N_c)$
   in the evaluation of the spectral densities that enter the sum rule.
    The calculation of such  higher order correction terms within the framework
    of the ENJL   presents an important but formidable challenge, especially since the
   chiral perturbation  theory
   results for $\alpha_E$ to two loops, consisting of the calculation and summation
   of over 100 diagrams(!),  are now available \cite{burgi96} for direct comparison.

\acknowledgments
One of us (R.H.L.) would like to thank Prof. H.A. Weidenm\"uller for the
hospitality at the Max-Planck-Institut f\"ur Kernphysik.
This work has been supported in part by the Deutsche Forschungsgemeinschaft
DFG under the contract number Hu 233/4-4, and by the German Ministry for
Education and Research (BMBF) under contract number 06 HD 742.

\appendix
\section{Regularization of Feynman Diagrams}
\label{a:A}
In this appendix,
 we discuss the
 regularization of the various Feynman loop diagrams that have
appeared in the main text.
\subsection{Gap equation for the light quark masses}
The gap equation in the chiral limit $\hat{m}=0$ that generates the quark mass self-consistently in both the NJL and ENJL
versions of the model Lagrangian in the mean field, or Hartree approximation has been written down in
Eq.~(\ref{e:gap}), and its extension to include a non-vanishing current quark
mass in Eq.~(\ref{e:ncgap}). Either equation contains a single quark loop that diverges
quadratically. Regulating this loop according to the
Pauli-Villars prescription for the reasons explained in the main text, we get
\begin{eqnarray}
m-\hat{m}&=&(16G_1N_c m)\Big\{\sum_a C_a \int\frac{d^4p}{(2\pi)^4}
\frac{i}{p^2-M^2_a}\Big\}
\nonumber\\
&=&(16G_1N_c m)
\Big\{\frac{1}{(4\pi)^2}\sum_aC_aM^2_a\ln\frac{M^2}{m^2}\Big\},
\label{e:pvgap}
\end{eqnarray}
provided that 
\begin{eqnarray}
\sum_{a=0}^2 C_a=0\quad {\rm and}\quad \sum_{a=0}^2 C_aM^2_a=0\,,
\label{e:pvc}
\end{eqnarray}
that may be satisfied by
the standard choice \cite{iz80}
\begin{eqnarray}
C_a=(1,\,1,\,-2)\quad{\rm and}\quad M^2_a=m^2+\alpha_a\Lambda^2\,,
\quad \alpha_a=(0,\,2,\,1),
\end{eqnarray}
where $\Lambda$ is the regulating parameter of the PV scheme. 
We designated the
solution of Eq.~(\ref{e:pvgap}) in the presence of $\hat{m}$ as $m=m^*$ in the main text.
One has the useful relation 
\begin{eqnarray}
\sum_a C_aM^2_a\ln
\frac{M^2_a}{m^2}=\frac{\pi^2}{G_1N_cm}(m-\hat{m})=-\frac{2\pi^2}
{N_c}\frac{\langle\bar \psi\psi\rangle}{m},
\label{e:densconnect}
\end{eqnarray}
after using the mean
 field relation $m-\hat{m}=-2G_1\langle\bar\psi\psi\rangle$ between $m$
and the condensate density $\langle\bar\psi\psi\rangle$ of the quarks.

\subsection{The one-loop integral $I(q^2)$ }
\subsubsection{Evaluation}
The integral $I(q^2)$ given in Eq.~(\ref{e:eye}) appears in the pion weak decay constant
$f_p^2=-4N_cm^2iI(0)$, in the pion form factor $F_P(q^2)=iI(q^2)/iI(0)$, and
thus
the pseudoscalar polarization loop as given by Eq.~(\ref{e:HPP}). To evaluate
it we introduce the standard Feynmann parametrization \cite{iz80} to find
\begin{eqnarray}
I(q^2)&=&\int\frac{d^4p}{(2\pi)^4}\frac{1}{[(p+q)^2-m^2][p^2-m^2]}
\nonumber
\\
\nonumber
\\
&=&\frac{i}{(4\pi)^2}\int_0^{\infty} d\alpha\,
I_1([m^2-q^2\alpha(1-\alpha)],
\label{e:eyea}
\end{eqnarray}
where
\begin{eqnarray}
I_1(q^2)=\int_0^\infty\frac
{d\rho}{\rho}e^{-i\rho[m^2-q^2\alpha(1-\alpha)]}.
\label{e:eye1}
\end{eqnarray}
The last integral is logarithmically divergent. We regulate it using the
Pauli-Villars [PV] prescription again. Denote the
 corresponding PV-regulated quantity with a bar. Then
\begin{eqnarray}
\bar I_1(q^2)=\lim_{\eta\to 0}\sum_{a=0}C_a
\int_\eta^\infty {d\rho\over \rho}e^{-i\rho[M^2_a-q^2\alpha(1-\alpha)]}
=-\sum_{a=0}C_{a}\ln [M^2_a-q^2\alpha (1-\alpha)].
\label{e:eye1pv}
\end{eqnarray}
Hence
\begin{eqnarray}
i\bar I(s)=i\bar I(0)
+\frac{1}{(4\pi)^2}\sum_{a=0} C_a\int_0^1 d\alpha 
\ln[1-{s\over M^2_a}\alpha(1-\alpha)],
\label{e:eyeinter}
\end{eqnarray}
where
\begin{eqnarray}
i\bar I(0)=\frac{1}{(4\pi)^2} \sum_{a=0} C_a \ln{M^2_a\over m^2},
\label{e:eyebar0}
\end{eqnarray}
after calling $q^2=s$. The remaining integral is elementary. For space-like
momentum transfers squared $-\infty\,<\,s\,<0$, it is given by
\begin{eqnarray}
\int_0^1 d\alpha \ln[1-\frac{s}{m^2}\alpha(1-\alpha)]=2J(z)-2\,,
\label {e:intJ}
\end{eqnarray}
where
\begin{eqnarray}
J(z)&=&\sqrt{f}\coth^{-1}\sqrt{f}=\sqrt{1-{1\over z}}\ln[\sqrt{-z}+\sqrt{1-z}]
\label{e:JC}
\\
&\sim& \ln\sqrt{-z}\,(1-\frac{1}{2z}-\frac{1}{8z^2}
+\cdots)+\ln\,2-({1\over 4}+\frac{1}{2}\ln\,2)\frac{1}{z}
\nonumber\\
&+&(\frac{1}{32}-\frac{1}{8}\ln\,2)\frac{1}{z^2}+\cdots\,,\quad
 |z|\rightarrow \infty,
\label{e:jasymp}
\end{eqnarray}
after setting $z=s/4m^2$ and
\begin{eqnarray}
f=1-{1\over z}.
\end{eqnarray}
The PV-regulated version of $iI(s)$ for space-like $q^2$ is thus
\begin{eqnarray}
 i\bar I(s)=\frac{1}{(4\pi)^2}\Big[ \sum_{a=0}^2\ln\,\frac{M^2_a}{m^2}+
\big\{(2\sqrt{f}\,\coth^{-1}\sqrt{f}-2)\big\}_{PV}\Big],
\label{e:eyespace}
\end{eqnarray}
where the PV-subscript signifies the replacement \cite{iz80}
\begin{eqnarray}
\big\{(2\sqrt{f}\,\coth^{-1}\sqrt{f}-2)\big\}_{PV}\rightarrow
\sum_{a=0}^2C_a(2\sqrt{f_a}\,\coth^{-1}\sqrt{f_a}-2)
\label{epvnotation}
\end{eqnarray}
in terms of the variable
\begin{eqnarray}
f_a=1-\frac{4M^2_a}{s}.
\label{e:fa}
\end{eqnarray}
In accordance with Feynman's causal prescription \cite{iz80},
$i\bar I(s)$ has to be continued to time-like $q^2=s\,>0$
by passing along the upper lip of
the cut starting at the branch point at $s=4m^2$ and running to $\infty$. This in turn means that
the square roots in logarithmic function determining $J(z)$ have to be chosen
such that
\begin{eqnarray}
\sqrt{-z}&=&|z|^{1/2}e^{i(\phi-\pi)/2},\qquad 0\:<\:\phi<\:2\pi\\
\sqrt{1-z}&=&|1-z|^{1/2}e^{i(\theta-\pi)/2},\qquad  0\:<\theta\:<\:2\pi
\label{e:physheet}
\end{eqnarray}
on the ``first'', or physical sheet of $J(z)$, in terms of the complex
variables $z=|z|\exp\,i\phi$ and $z-1=|z-1|\exp\,i\theta$. The further branch points at $z=s/4M^2_a$, $a=1,2$, that are introduced into $J(s/4m^2)_{PV}$ and
thus $i\bar I(s)$ by the regulating masses of the Pauli-Villars
scheme are handled in the same way.
The schedule of forms for the multivalued complex function $J(z)$ can now be written down from
\begin{eqnarray}
J(z)=|\frac{z-1}{z}|^{1/2}e^{i(\theta-\phi)/2}\ln\big[
|z|^{1/2}e^{i(\phi-\pi)/2}
+|1-z|^{1/2}e^{i(\theta-\pi)/2}\big].
\end{eqnarray}
Along the real axis $z=x$ of the first sheet $J(x)$ thus behaves like
\begin{eqnarray}
J(x)&=&\left\{\begin{array}{lc}
(1-{1\over x})^{1/2}\ln(\sqrt{-x}+\sqrt{1-x})
=\sqrt f\coth^{-1}\sqrt f,&-\infty\:<\:x\:<0 \\
 & \\
({1\over x}-1)^{1/2} \sin^{-1}\sqrt{x}=
\sqrt{-f}\cot^{-1}\sqrt{-f},&0\:<x\:<\:1\\
 & \\
(1-{1\over x})^{1/2}\,\left(\ln[\sqrt{x}+\sqrt{x-1}]-i{\pi\over 2}\right)=
\sqrt f\,(\tanh^{-1}\sqrt f-i{\pi\over 2}),&1\:<\: x\:<\:\infty,
\end{array}\right.
\nonumber\\
&&
\label{e:J1c}
\end{eqnarray}
where $f=1-1/x$. The appearance of an imaginary part for $x>1$ is associated
with the (unphysical) $\bar qq$ decay threshold that is one of the drawbacks of the
NJL model.

In order to discuss the possible poles of the vector and axial mode
 propagators  in ENJL, we also require
the analytic continuation, $\tilde J(z)$ of $J(z)$
through the cut onto the second sheet. This sheet is defined by letting the 
angles $\phi$ and
$\theta$ assume values in the angular interval $-2\pi < \theta,\phi<0$. 
Except for the
sector $1\:<x\:<\infty$ along the real axis of the second sheet where the
two functions coincide by construction, the analytic continuation $\tilde
 J(x)$ is a different
function on the second sheet. In particular, its behavior along the real
axis of that sheet is as follows:
\begin{eqnarray}
\tilde J(x)=\left\{\begin{array}{lc}
J(x)-i\pi\sqrt f,& -\infty\:<\:x\:<0\\
 & \\
J(x)-\pi\sqrt{-f},& 0\:<\:x\:<\:1\\
 & \\
J(x), & 1\:<\: x\:<\:\infty.
\end{array}\right.
\nonumber\\
\label{e:Jtildc}
\end{eqnarray}
We note in particular that $\tilde J(x)$ like $J(x)$, is also real for
$0\,<x\,<1$ but picks up an imaginary part in the
sector $-\infty\:<x\:<0$. Their behavior at the origin as $x\rightarrow 0^+$
is also different: $J(x)\rightarrow x^{-1/2}\sin x^{1/2}\rightarrow 1$,
but $\tilde J(x)\rightarrow -\pi x^{-1/2} \rightarrow -\infty$.
Thus $ReJ(x)$ and $Re\tilde J(x)$ coincide for $-\infty\,<\,x\,<0$ and $1\,>x\,>\infty$,
but
differ for $0\,<x\,<1$, while $ImJ(x)$ and $Im\tilde J(x)$ differ in the
first interval but coincide in the latter two.
All of these properties are essential for appreciating the analytic
structure of the various polarization functions to follow.
The real parts of $J(x)$ and $\bar J(x)$ are plotted for real arguments
in Fig.~\ref{f:ReJ}.

Returning to the problem at hand, we can now write down a general form for
the PV-regulated version of $iI(q^2)$ for all $q^2=s$ as
\begin{eqnarray}
i\bar I(s)= \frac{1}{(4\pi)^2}\Big[\sum_{a=0}^2C_a \ln{M^2_a\over m^2}
+\big\{(2J(s/4m^2)-2)\big\}_{PV}\Big],
\label{e:eyepv}
\end{eqnarray}
where 
\begin{eqnarray}
\big\{(2J(s/4m^2)-2)\big\}_{PV}\rightarrow
\sum_{a=0}^2C_a\big\{2J(s/4M^2_a)-2\big\},
\label{e:pvnotation}
\end{eqnarray}
as before. The form of $J$ appropriate to the sheet, and interval that its
argument falls in on that sheet,
has to be selected from Eqs.~(\ref{e:J1c}) or (\ref{e:Jtildc}).
 In particular, for
time-like $q^2=s\ge 4m^2$, the explicit form of $Im[i\bar I(s+i\epsilon)]$ 
along the upper lip of the cut plane is
\begin{eqnarray}
Im[i\bar I(s)]
&=&-\frac{1}{16\pi}\big\{\sqrt{1-{4m^2\over s}}\,\theta(s-4m^2)\big\}_{PV}
\nonumber
\\
&=&-\frac{1}{16\pi}\sum_{a=0}^2C_a\sqrt{1-{4M^2_a\over
s}}\,\theta(s-4M^2_a).
\label{e:imeye}
\end{eqnarray}
Thus, apart from the contribution starting at $s=4m^2$,
the imaginary part of $i\bar I(s)$ also contains additional contributions of
different weights coming from the artificial thresholds that are
introduced by the regulating masses of the Pauli-Villars scheme at
$s=4m^2(1+\Lambda^2/m^2)$ and $s=4m^2(1+2\Lambda^2/m^2)$. This in turn
means that $Im [i\bar I(s)]$ need not remain negative definite for $s$ large
enough, as could perhaps have been anticipated from the 
introduction of indefinite metric sectors of Hilbert space 
by the PV-regulation
procedure \cite{iz80}.

\subsubsection{Asymptotic behavior and dispersion relations}
The asymptotic behavior of the function $J(z)$ is given in
Eq.~(\ref{e:jasymp}) which shows that it diverges logarithmically.
Thus if the PV instruction were removed from the second factor 
in Eq.~(\ref{e:eyepv}),
then the resulting expression for $iI(s)$ would also diverge
logarithmically, $iI(s)\sim \log(-s)$,
with $s\rightarrow \infty$ in the complex plane.
The PV regularization instruction alters this behavior drastically to a
$1/s$ convergent one as we show below. The change-over is, 
not surprizingly, related
to keeping the imaginary part PV regulated as in Eq.~(\ref{e:imeye})
(which is finite and does not need regularization anyway), which then
determines
what sort of dispersion relation the expression for $iI(s)$ can satisfy
in a consistent fashion.
The change in asymptotic behavior in the present case comes about because upon PV regulating
$J(s/4m^2)$, the conditions that $\sum_a C_a$ and $\sum_a C_aM^2_a$ 
must vanish,
suppress both the $\ln(-s)$ and the $(1/s)\ln(-s)$ terms in Eq.~(\ref{e:jasymp}), leaving only a $1/s$ and the sub-leading
logarithm $(1/s^2)\ln(-s)$ to survive:
\begin{eqnarray}
\bar J(s/4m^2)&=&\sum_a C_a J(s/4M^2_a)\sim
-{1\over 2}\sum_a C_a\ln{M^2_a\over m^2}
+{1\over s}(\sum_a C_a M^2_a\ln{M^2_a\over m^2})
\nonumber
\\
&-&{1\over s^2}\ln(-s/ 4m^2)\, (\sum_aC_aM^4_a)+O(1/s^2).
\label{e:jaybar}
\end{eqnarray}
Inserting this form of $\bar J$ into Eq.~(\ref{e:eyepv}) and using the
result that $\sum_aC_aM^4_a =2\Lambda^4$ to simplify the coefficient of the
$(1/s^2)\ln(-s)$ term, one obtains
the behavior of $i\bar I(s)$ at large $s$ as
\begin{eqnarray}
i\bar I(s)=
{1\over 8\pi^2 s}\sum_a C_a M^2_a\ln {M^2_a\over m^2}
-\frac{\Lambda^4}{4\pi^2}{1\over s^2}\ln(-s)+{\rm O}(1/s^2),
\label{e:eyebarasymp}
\end{eqnarray}
instead of $\sim \ln(-s)$. Thus the function $i\bar I(s)$
is rendered sufficiently
convergent to satisfy an unsubtracted dispersion relation of the form
\begin{eqnarray}
i\bar I(s)={1\over \pi}\int_{4m^2}^\infty dt\frac{Im [i\bar I(t)]}{t-s}.
\label{e:eyedisp}
\end{eqnarray}
It is instructive to verify this result explicitly, for the calculation
shows how the PV-regularization of the imaginary part conspires to allow the
dispersion relation to hold.  One finds, using Eq.(\ref{e:imeye}), that
\begin{eqnarray}
i\bar I(s)=
-\frac{1}{(4\pi)^2}\lim_{\mu^2 \to \infty}\sum_{a=0}^2C_a
\int_{4M^2_a}^{\mu^2}\frac{dt}{t-s}\sqrt{1-\frac{4M^2_a}{t}}\theta(t-4M^2_a).
\end{eqnarray}
The auxillary cutoff $\mu^2 $ has been introduced as a calculational device
since the individual integrals diverge before carrying out the PV summation.
Thse can be evaluated explicitly using the change of
variable $t=4M^2_a\cosh^2\phi$. Assuming $s<0$ for definiteness, one obtains
\begin{eqnarray}
i\bar I(s)=\frac{1}{(4\pi)^2}\lim_{\mu^2\to\infty}\sum_{a=0}^2C_a
\Big[2\sqrt{f_a}\coth^{-1}\sqrt{f_a(1-\frac{4M^2_a}{\mu^2})^{-1}}
-2\cosh^{-1}\sqrt{\frac{\mu^2}{4M^2_a}}\,\Big].
\label{e:ibardisp}
\end{eqnarray}
Taking the limit $\mu^2\rightarrow\infty$ is harmless in the argument of the inverse hyperbolic
cotangent function, but the second term diverges logarithmically.
The PV-instruction removes this divergence and replaces it by a finite constant in the standard way as follows,
\begin{eqnarray}
 -2 \lim_{\mu^2\to\infty}\sum_{a=0}^2C_a\cosh^{-1}\sqrt{\frac{\mu^2}{4M^2_a}}
\rightarrow &-&\lim_{\mu^2\to\infty}\sum_{a=0}^2C_a\ln\frac{\mu^2}{M^2_a}
\nonumber\\
&=&- \lim_{\mu^2\to\infty}\sum_{a=0}^2C_a\ln{\frac{\mu^2}{m^2}}
+\sum_{a=0}^2C_a\ln\,\frac{M^2_a}{m^2},
\end{eqnarray}
since, according to the  PV regulation
rules, the instructions of Eqs.~(\ref{e:pvc})
take precedence over taking the limit $\mu^2\to\infty$.
 Inserting this result back
into Eq.~(\ref{e:ibardisp}) and reinstating an irrelevant $-2$ under the
PV-regulation bracket, we regain Eq.~(\ref{e:eyespace}).

On the other hand, if the PV instruction on the imaginary part given by
Eq.~(\ref{e:imeye}) is {\it dropped}, then $iI(s)$ diverges logarithmically
as we have seen. Then, not $iI(s)$, but only $iI(s)/s$, is sufficiently
convergent to give no contribution upon integration around a
closed contour $C_\infty$ at infinity in the complex $s$ plane.
 Hence the dispersion
relation for
$this$ representation of $iI(s)$ requires {\it one} subtraction,
\begin{eqnarray}
{iI(s)}={iI(0)}+{s\over \pi}\int_{4m^2}^\infty dt\frac{Im [ iI(t)]}
{t(t-s)}
\label{e:oncesubdisp}
\end{eqnarray}
where only the (formally) infinite constant $iI(0)$
 now requires regularization. 
This ``dispersive'' type of regularization prescription was followed in
Ref.~\cite{brz94} for example, in conjunction with a proper time cutoff.
If we opt for PV-regulating $iI(0)$, then Eq.~(\ref{e:oncesubdisp}) leads
back to Eq~(\ref{e:eyeinter}), or equivalently to
Eq.~(\ref{e:eyespace}) again, but with the PV-instruction on the second
term in either of these equations removed, 
if the unregulated imaginary part of $iI(s)$ as given by Eq.~(\ref{e:imeye})
is used. One then has
\begin{eqnarray}
{s\over\pi}\int_{4m^2}^\infty dt
\frac{Im [iI(t)]}{t(t-s)}=-{s\over (4\pi)^2}\int_{4m^2}^\infty\frac{dt}{t}
\frac{\sqrt{1-\frac{4m^2}{t}}}
{(t-s)}
={1\over{(4\pi)^2 }}\int_0^1 d\alpha\,
\ln\big[1-\frac{s}{m^2}\alpha(1-\alpha)\big]
\end{eqnarray}
under the successive transformation of variables $u=(1-4m^2/t)^{1/2}$,
 followed
by $\alpha=(1-u)/2$ after an integration by parts. Hence
\begin{eqnarray}
{iI(s)}={i\bar I(0)}+{1\over{(4\pi)^2 }}\int_0^1 d\alpha\,\ln\big[1-
\frac{s}{m^2}\alpha(1-\alpha)\big]
\end{eqnarray}
that reproduces Eq.~(\ref{e:eyeinter}).

The origin of the difference between Eq.~(\ref{e:eyeinter}) and the above expression for $iI(s)$
thus lies entirely in the way the imaginary part is treated. This in turn means that either
an unsubtracted, or a once-subtracted dispersion relation
is satisfied by the regulated function. Not surprisingly, these two cases
can lead to regulated functions that
behave very
differently at large momentum transfers. A summary of the asymptotic behavior
of functions that involve amplitudes regulated via these two methods
is given in Table~\ref{table4}.

\subsection{Pion weak decay constant, form factor and pseudoscalar 
polarization loop
in the NJL model}

It is now a simple matter to write down expressions for
 the regulated pion weak
decay constant, the form factor, and thus the one-loop pseudoscalar
 polarization
function $\Pi^{PP}(q^2)$ in terms of $i\bar I(q^2)$. One has
\begin{eqnarray}
f^2_p=-4N_c m^2i\bar I(0)=-\frac{N_cm^2}{4\pi^2}\sum_a C_a\ln{ M^2\over m^2}
\label{e:regpicoupling}
\end{eqnarray}
and
\begin{mathletters}
\begin{eqnarray}
& &\bar F_P(s)=\frac{i\bar I(s)}{i\bar I(0)}=[1-\frac{N_cm^2}{4\pi^2f^2_p}
\big\{(2J(s/4m^2)-2)\big\}_{PV}]
\\
&\sim& -\frac{N_cm^2}{2\pi^2 f^2_p}
\left(\sum_a C_a M^2_a\ln{M^2_a\over m^2}\right){1\over s}
+\frac{N_c m^2}{\pi^2f^2_p}\frac{\Lambda^4}{s^2}\ln(-s)
\,,\quad |s| \rightarrow \infty.
\nonumber\\
\label{e:piformfac}
\end{eqnarray}
\end{mathletters}
With the aid of the relation (\ref{e:densconnect}) between the quark 
mass and the
condensate density that follows from the
PV-regulated version of the gap equation, the asymptotic behavior of
$\bar F_P(s)$ can be recast in the transparent form
\begin{eqnarray}
f^2_p \bar F_P(s)\sim \frac{m\langle\bar\psi\psi\rangle}{s}
+\frac{N_c m^2}{\pi^2}\frac{\Lambda^4}{s^2}\ln(-s)
\,,\quad |s|\rightarrow \infty.
\label{e:formfacasymp}
\end{eqnarray}
For time-like $s\ge 4m^2$, the imaginary part of $\bar F_P(s)$ is obtained
from Eq.~(\ref{e:imeye}):
\begin{eqnarray}
Im\,\bar F_P(s)=\frac{m^2N_c}{4\pi f^2_p}
\big\{\sqrt{1-\frac{4m^2}{s}}\theta(s-4m^2)\big\}_{PV}.
\label{e:imformfac}
\end{eqnarray}
The dispersion relation Eq.~(\ref{e:eyedisp}) for $i\bar I(s)$ immediately
translates into a dispersion relation for the form factor:
\begin{eqnarray}
\bar F_P(s)={1\over \pi}\int_{4m^2}^\infty dt\frac{Im [\bar F_P(t)]}{t-s}.
\label{e:formfacdisp}
\end{eqnarray}
Note that using the PV-regulated form $Im \bar F_P(t)$ automatically
 guarantees
the condition $\bar F_P(0)=1$, since we have explicitly demonstrated that
the dispersion relation satisfied by
$i\bar I(s)$ correctly reproduces $i\bar I(0)$ according to
Eq~(\ref{e:ibardisp}).

The pion electromagnetic radius is determined by the derivative
$\bar F _P^\prime(s)$ at $s=0$. This derivative is most simply obtained
by evaluating
\begin{eqnarray}
\bar F^\prime_P(0)= {1\over \pi}\int_{4m^2}^\infty
{dt\over t^2}Im [\bar F_P(t)]={1\over 6}\frac{N_c}{4\pi^2f^2_p}
\sum_aC_a\frac{m^2}{M^2_a}={1\over 6}\langle r^2_p\rangle_{PV}\,,
\end{eqnarray}
using the same change of variable $t=4M^2_a\cosh^2\phi$ as before.
 No auxillary cutoff is
necessary in this case since each integral converges separately
 due to the extra
$t$ in the denominator of the integrand.
For $N_c=3$, the mean field approximation to the pion radius is 
thus found to
be given by
\begin{eqnarray}
\langle r^2_p\rangle_{PV}=\frac{3}{4\pi^2 f_p^2}\sum_a
C_a{m^2\over M^2_a}\approx \frac{3}{4\pi^2 f_p^2},
\label{e:pimeanrad}
\end{eqnarray}
that essentially coincides with the result of Tarrach \cite {tar79} given on the right
of this equation.

Finally we can write down the explicit form of the pseudoscalar polarization
and its asymptotic behavior using Eq.~(\ref{e:piformfac}). We take the non-chiral
case given in Eq.~(\ref{e:polrenstar}):
\begin{eqnarray}
\bar\Pi^{PP}_{nc}(s)&=&\frac{(m-\hat{m})}{2G_1m}+\frac{f^2_\pi(s)}{m^2}s
\nonumber\\
&&\sim \frac{N_c\Lambda^4}{\pi^2}\frac{1}{s}\ln(-s)\,,
\quad |s|\rightarrow \infty.
\label{e:psasymp}
\end{eqnarray}
Notice that the constant term of the combination $(sf^2_\pi(s)/m^2)\rightarrow
sf^2_p\bar F_P(s)\rightarrow \langle\psi\psi\rangle/m
=-(m-\hat{m})/2G_1m$ just cancels the constant term 
in $\Pi^{PP}$ to leading order according to Eq.~(\ref{e:formfacasymp}),
so that one has to include the next (logarithmic) order
in $\bar F_P(s)$ as determined by Eq.~(\ref{e:eyebarasymp}) to obtain the
stated result. The asymptotic behavior of the regulated pseudoscalar
polarizations in the chiral limit for either the NJL model
[Eq.~(\ref{e:HPP})] or the ENJL model [Eq~(\ref{e:polren})]
are also given by Eq.~(\ref{e:psasymp}).

\subsection{Evaluation and properties of the vector and axial vector polarization loops}
\subsubsection{The vector polarization $\Pi^{VV}_T(q^2)$}
We set $\mu=\nu$ in  Eq.~(\ref{e:vectorpol}) and then contract on $\mu$ in
order to exploit the fact that the vector polarization is expected to be purely
transverse. Then
\begin{eqnarray}
g^{\mu\nu}\Pi^{VV}_{\mu\nu;ab}=3\Pi^{VV}_T\delta_{ab}=
2iN_c\delta_{ab}\int\frac{d^4p}{(2\pi)^4}tr[\gamma^\mu S(p+q)\gamma_\mu
S(p)],
\end{eqnarray}
after doing the color and flavor traces. Using the fact that
\begin{eqnarray}
tr[\gamma^\mu(\not\!p+\not\!q+m)\gamma_\mu(\not\!p+m)]&=&16m^2-8p\cdot
(p+q),
\end{eqnarray}
one has 
\begin{eqnarray}
\Pi^{VV}_T(q^2)&=&{16\over 3}iN_c\int\frac{d^4p}{(2\pi)^4}
\frac{2m^2-p\cdot(p+q)}{[(p+q^2-m^2][p^2-m^2]}
\nonumber\\
&=&-16 q^2\frac{N_c}{(4\pi)^2}\int_0^1d\alpha\,\alpha(1-\alpha)\int_0^\infty
\frac{d\rho}{\rho}e^{-i\rho[m^2-q^2\alpha(1-\alpha)]}
\nonumber\\
&-&\frac{32i}{3}\frac{N_c}{(4\pi)^2}\int_0^1 d\alpha\,\int_0^\infty
d\rho\,\frac{d}{d\rho}
{\frac{e}{\rho}}^{-i\rho[m^2-q^2\alpha(1-\alpha)]}.
\label{e:vvb4pv}
\end{eqnarray}
The $\rho$ integrals diverge like $\ln\rho$ and $1/\rho$ respectively 
at the origin. Under
PV-regularization, the first integral is just $\bar I_1(q^2)$ as given by
Eq.~(\ref{e:eye1pv}), while the second integral makes no 
contribution at all since
\begin{eqnarray}
\lim_{\eta\to 0}\sum_{a=0}^2C_a\int_{\eta}^\infty
d\rho\,\frac{d}{d\rho}
{\frac{e}{{\rho}}^{-i\rho[M^2_a-q^2\alpha(1-\alpha)]}}=
-\lim_{\eta\to 0}\sum_{a=0}^2C_a\big[{1\over \eta}-i
[M^2_a-q^2\alpha(1-\alpha)]\big]
\rightarrow 0.
\end{eqnarray}
Thus, with $s=q^2$ again,
\begin{mathletters}
\begin{eqnarray}
\bar\Pi^{VV}_T(s)&=&-16 s\frac{N_c}{(4\pi^2)}\int_0^1d\alpha\,
\alpha(1-\alpha)\bar I_1(s)
\\
&=&\frac{N_cs}{6\pi^2}\big[\sum_aC_a\ln\frac{M^2_a}{m^2}+
\big\{{1\over 3}+(3-f)(\sqrt{f}\coth^{-1}\sqrt{f}-1)\big\}_{PV}\big]
\label{e:vecpolbar}
\\
&=& \frac{N_c s}{6\pi^2}\sum_aC_a\ln\frac{M^2}{m^2}
=-{2\over 3}{f^2_p\over m^2}s
\,, \quad s\rightarrow 0
\label{e:vecpolbar0}
\\
&\sim& -{1\over s}\ln(-s)\,,\quad |s|\rightarrow \infty.
\end{eqnarray}
\end{mathletters}
with the indicated limiting behavior
near the origin and on a great circle in the complex plane.
Eq.~(\ref{e:regpicoupling}) has been used to re-introduce the mean 
field pion decay
constant.
In obtaining Eq.~(\ref{e:vecpolbar}), we have used the intermediate result that
\begin{eqnarray}
\int_0^1d\alpha\,\alpha(1-\alpha)\ln[1-\frac{s}{m^2}\alpha(1-\alpha)]=
{1\over 6}\big[\,{1\over 3}+(3-f)(\sqrt{f}\coth^{-1}\sqrt{f}-1)\big]
\end{eqnarray}
for space-like $s$, but which may immediately be extended to cover
all $s$, real or complex, using the definition of the function $J(z)$ 
in Eq.~(\ref{e:JC}).

Thus like $i\bar I(s)$, only the PV-regulated vector polarization is
 sufficiently convergent on the great circle to satisfy an unsubtracted
dispersion relation,
\begin{eqnarray}
{1\over 4}\bar \Pi^{VV}_T(s)={1\over 4 \pi}\int_{4m^2}^\infty dt
 \frac{Im[\bar \Pi^{VV}_T(t)]}{t-s}
=-\int_{4m^2}^\infty dt\;\frac{{\rho}^{(0)V}_1(t)}{t-s}
\label{e:vectorunsub}
\end{eqnarray}
where
\begin{eqnarray}
{\rho}^{(0)V}_1(s)=-{1\over 4\pi}\,Im[\bar\Pi^{VV}_T(s)] 
\end{eqnarray}
is the vector spectral density function. Its explicit form is found with the
help of Eq.~(\ref{e:J1c}) to be
\begin{eqnarray}
\rho^{(0)V}_1(s)=\frac{N_cs}{24\pi^2}\big\{\sqrt{1-{4m^2\over s}}
(1+{1\over 2}{4m^2\over s})\theta(s-4m^2)\big\}_{PV}.
\label{e:vecdens}
\end{eqnarray}
Again we comment that this expression is no longer positive definite for all
$s$ for the same reason as before, due
to the PV-instruction. One has a linearly divergent
behavior $\rho^V_1(s)\sim s$ as $s$ increases if the PV-instruction is removed.
If it is kept, however, a short calculation shows that
\begin{eqnarray}
\rho^{(0)V}_1(s)\sim -\frac{N_c}{2\pi^2}\frac{\Lambda^4}{s}\,,\quad 
s\rightarrow \infty,
\end{eqnarray}
which is convergent at the expense of becoming negative at large $s$. However,
such a behavior seems unavoidable if at the same 
time one wants the vector polarization
as given by the NJL model to obey an unsubtracted dispersion relation. 
Indeed, if
the PV-instruction is removed in Eq.~(\ref{e:vecpolbar}) 
(and therefore also in
Eq.~(\ref{e:vecdens}), rendering $\rho^{(0)V}_{1}(s)$ positive definite)
then $\Pi^{VV}_T(s)$ diverges like $s\ln(-s)$, so that two subtractions are
required in the  dispersion relation for this $\Pi^{VV}_T(s)$
(or equivalently, a once-subtracted one for the combination $\Pi^{VV}_T(s)/s$
 - see, for example,
Ref.~\cite{brz94}),
\begin{eqnarray}
\Pi^{VV}_T(s)=s\Pi^{(\prime)VV}_T(0)+\frac{s^2}{\pi}\int_{4m^2}^\infty dt
\frac{Im[\Pi^{VV}_T(t)]}{t^2(t-s)},
\label{e:vector2sub}
\end{eqnarray}
where the prime denotes the value of the derivative at the origin.  As with the
case of $i\bar I(s)$ and $iI(s)$, one can again demonstrate explicitly by direct
integration using either the
PV-regulated, or the unregulated $\rho^{(0)V}_{1}(s)$, that the dispersion
relations (\ref{e:vectorunsub}) or (\ref{e:vector2sub}) lead back to the expression
for $\Pi^{VV}_T(s)$ in Eq.~(\ref{e:vecpolbar}) with the PV-instruction either
present or absent on the second factor.

\subsubsection{The longitudinal and transverse axial vector polarization $\Pi^{AA}_L(q^2)$ and
$\Pi^{AA}_T(q^2)$}
We proved in Eq.~(\ref{e:fullong}) that the longitudinal axial vector polarization
must vanish when all interactions are considered. At one-loop level, however,
it does not vanish and is given by Eq.~(\ref{e:ward1}), i.e.
\begin{eqnarray}
\bar\Pi^{AA}_L(q^2)=4f^2_p\bar F_P(q^2)
\label{e:HAAreg}
\end{eqnarray}
after regularization, where $\bar F_P$ is taken from Eq.~(\ref{e:piformfac}) and
$f^2_p$ from Eq.~(\ref{e:regpicoupling}). We use
Eq.~(\ref{e:ward1}) to determine the regulated transverse axial
polarization as
\begin{mathletters}
\begin{eqnarray}
\bar\Pi^{AA}_T(s)&=& \bar\Pi^{VV}_T(s)+4f^2_p\bar F_P(s)
\label{e:axialpolreg}
\\
&=& 4f^2_p\,,\quad s\rightarrow 0
\\
&\sim&-{1\over s}\ln(-s)\,,\quad s\rightarrow\infty.
\end{eqnarray}
\end{mathletters}
Again $\Pi^{AA}_T(s)$ will satisfy an unsubtracted dispersion relation like
Eq.~(\ref{e:vectorunsub}) containing the axial vector spectral density
that is related to $\rho^{(0)V}_1(s)$.     This can be seen 
 by taking the imaginary part of
Eq.~(\ref{e:axialpolreg}) and using Eq.~(\ref{e:imformfac}):
\begin{eqnarray}
\rho^{(0)V}_1(s)-\rho^{(0)A}_1(s)&=&\frac{f^2_p}{\pi}Im[\bar F_P(s)]
=\frac{N_c m^2}{4\pi^2}
\big\{\sqrt{1-{4m^2\over s}}\,\theta(s-4m^2)\big\}_{PV}
\nonumber
\\
&\sim&-\frac{N_cm^2}{\pi^2}\frac{\Lambda^4}{s^2}\,,\quad s\rightarrow
\infty.
\label{e:axvecdens}
\end{eqnarray}
Hence the density difference that enters into the two Weinberg
sum rules converges like $1/s^2$, i.e. faster than either density separately
and renders both sum rules convergent for the PV-regulated spectral densities.
By contrast, with the PV instruction removed, $\rho^{(0)V}_1(s)-\rho^{(0)A}_1(s)\sim
N_cm^2/(4\pi^2)$ asymptotically, and neither sum rule would converge.
The regulated axial density itself is given by
\begin{eqnarray}
\rho^{(0)A}_1(s)=\frac{N_cs}{24\pi^2}\sum_{a=0}^2C_a
\big\{\sqrt{1-{4M_a^2\over s}}\big((1-{4M_a^2\over s})+\frac{6}{s}
(M^2_a-m^2)\big)
\theta(s-4M^2_a)\big\}.
\label{e:axvecdensexp}
\end{eqnarray}

The first Weinberg sum rule is explicitly satisfied
by the spectral densities of the minimum NJL model.
We have
\begin{eqnarray}
\int_0^\infty {ds\over s}\,(\rho^{(0)V}_1(s)-\rho^{(0)A}_1(s))
&=&\frac{f^2_p}{\pi}\int_{4m^2}^\infty {dt\over t}Im[\bar F_P(t)]
\nonumber
\\
&=&f^2_p\bar F_p(0)=f^2_p.
\label{e:weinbergmin1}
\end{eqnarray}
after using the dispersion relation, Eq.~(\ref{e:formfacdisp}), at $s=0$ to deduce the
value of the integral. 
To check the second
sum rule we require
\begin{eqnarray}
\int_0^\infty {ds}\,(\rho^{(0)V}_1(s)&-&\rho^{(0)A}_1(s))=\frac{m^2N_c}{4\pi^2}
\lim_{\mu^2\to\infty}\sum_aC_a\int_{4M^2_a}^{4\mu^2}
{ds}\;\sqrt{1-\frac{4M^2_a}{s}}\;\theta(s-4M^2_a)
\nonumber
\\
&=&\frac{m^2N_c}{\pi^2}\lim_{\mu^2\to\infty}\sum_aC_a
\big[\mu^2}({1-{M^2_a\over \mu^2})^{1\over 2}
-M^2_a\cosh^{-1}(\frac{\mu^2}{M^2_a})^{1\over 2}\big]
\nonumber
\\
&=&\frac{m^2N_c}{2\pi^2}\sum_aC_aM^2_a\ln\frac{M^2_a}{m^2}\,,
\quad\mu^2\rightarrow\infty
\end{eqnarray}
or
\begin{eqnarray}
\int_0^\infty {ds}\,(\rho^{(0)V}_1(s)&-&\rho^{(0)A}_1(s))
=-m\langle\bar\psi\psi\rangle,
\label{e:weinbergmin2}
\end{eqnarray}
where the last step follows from Eq.~(\ref{e:densconnect}).

\subsection{Electromagnetic form factor of the pion in the ENJL model}
In the absence of vector and axial vector meson fields, the electromagnetic form
factor of the pion is given by the $F_P(q^2)$ of Eq.~(\ref{e:ffpimean}) to
${\cal O}(N_c)$ in the mean field approximation. In the presence of
 the vector 
fields, this is no
longer the case, since the axial form factor $G_A(0)=g_A<1$ differs from unity. 
This renormalizes the $\pi qq$ coupling constant, $g_p\rightarrow 
g_{\pi qq}=g_p/\sqrt{g_A}$,
that in turn renormalizes the pion charge so that $F_P(0)\ne 1$. However, as we 
now show, a compensatory
renormalization of the pion field also occurs due to the pseudovector
coupling and pseudoscalar-pseudovector interference terms introduced by the
axial vector meson field that restores the
total charge of the pion to its physical value. The inclusion  of the vector meson
fields does not alter this conclusion since $F_V(0)=1$.

The set of diagrams that determines the charged pion's
electromagnetic form factor in the ENJL model is obtained by coupling
the photon to both the quark and antiquark lines making up the $\gamma\pi\pi$
triangle diagram. Since the electromagnetic current is renormalized as per
Eq.~(\ref{e:emcurrent}), the photon vertex contains a contribution from the $\rho^0$
exchange diagram too. The resulting $\gamma\pi\pi$ vertex thus has two
contributions as depicted in  Fig.~\ref{f:gammavert}.
We work in the chiral limit of zero pion mass. Then the
contribution from the first diagram of Fig.~\ref{f:gammavert} to $\pi qq$
vertex can be written as a four vector of the form
\begin{eqnarray}
\frac{1}{i}\Gamma^{(em)}_\mu(k^\prime,k)=
(ig_{\pi qq})^2\lambda_\pi(0)\frac{1}{i}V^{(em)}_\mu(k^\prime,k).
\label{e:defv}
\end{eqnarray}
Adding in the second diagram replaces $\Gamma^{(em)}_\mu (k^\prime,k)$ by
\begin{eqnarray}
\Gamma^{(em)}_\mu (k^\prime,k)&&\rightarrow \Gamma^{(em)}_\mu (k^\prime,k)
-\Pi^{VV}_{\mu\rho}(q^2)D^{VV;\rho\sigma}(q^2)\Gamma^{(em)}_\sigma (k^\prime,k)
\nonumber\\
&&=\big[L_{\mu\nu}+F_V(q^2)T_{\mu\nu}\big]\Gamma^{(em);\nu}(k^\prime,k)
\nonumber\\
&&=F_V(q^2)\Gamma^{(em)}_\mu (k^\prime,k)\, .
\label{e:modgam}
\end{eqnarray}
 The last expression is only valid for on-shell pions.
The electromagnetic
form factor $F_\pi(q^2)$ of the pion can then be identified from
\begin{eqnarray}
e_\pi F_\pi(q^2)(k+k^\prime)_\mu=\Gamma^{(em)}_\mu (k^\prime,k)F_V(q^2)
\label{e:enjlpiform}
\end{eqnarray}
We thus only need to evaluate the contribution from the first diagram.
Its spinor structure can be read off from Eq.~(\ref{e:pimode}) as
$P^a\cos\theta\pm i(\hat{k}\cdot A^a)\sin\theta$ for outgoing (incoming) pions
of isospin $a$ using the abbreviations $P^a=i\gamma_5\tau^a$
and $A^a_\mu=\gamma_\mu\gamma_5\tau^a$.
 We first construct the fully
off-shell version of $-iV^{(em)}_\mu(k^\prime,k)$ and then
specialize to on-shell chiral pions to obtain the charge form factor.
One finds
\begin{eqnarray}
\frac{1}{i}V^{(em)}_\mu (k^\prime,k)
&=&\cos\theta^\prime\cos\theta \frac{1}{i}V^{PP}_\mu (k^\prime,k)
+\cos\theta\sin\theta^\prime\; V^{A_\nu P}_{\mu}(k^\prime,k)\hat{k}^{\prime\nu}
\nonumber\\
&-& \cos\theta^\prime\sin\theta\; V^{PA_\nu}_{\mu}(k^\prime,k)\hat{k}^{\nu}
+\sin\theta^\prime\sin\theta\; \hat{k}^{\prime\nu}\frac{1}{i}V^{A_\nu A_\rho}_{\mu}(\hat{k}^\prime,k)k^\rho 
\label{e:newff}
\end{eqnarray}
for absorbing a photon of four momentum $q=k^\prime-k$. The individual amplitudes
in this expression
arise from the various combinations of pseudoscalar and axial vector vertices
that are introduced through
the revised $\pi qq$ interaction. They  are given by  
\begin{mathletters}
\begin{eqnarray}
V^{PP}_\mu (k^\prime,k) &=&(-2ie_\pi N_c)\int\frac{d^4 p}{(2\pi)^4}
tr[\gamma_5 S(p+k^\prime)\gamma_\mu S(p+k)\gamma_5 S(p)]
\\
V^{PA_\nu}_{\mu}(k^\prime,k)&=&(-2e_\pi N_c)\int\frac{d^4 p}{(2\pi)^4}
tr[\gamma_5 S(p+k^\prime)\gamma_\mu S(p+k)\gamma_\nu\gamma_5 S(p)]
\label{e:PAnu}
\\
\nonumber\\
V^{A_\nu A_\rho}_{\mu}(k^\prime,k)&=&(2ie_\pi N_c)\int\frac{d^4 p}{(2\pi)^4}
tr[\gamma_\nu\gamma_5 S(p+k^\prime)\gamma_\mu S(p+k)\gamma_\rho\gamma_5
S(p)].
\end{eqnarray}
\end{mathletters}
The symbol $e_\pi =eT_3$ is the charge carried by the pion. 

A direct calculation gives \cite{dstl95}
\begin{eqnarray}
V^{PP}_\mu (k^\prime,k)=
-e_\pi\frac{f^2_p}{m^2}F_P(q^2)\,(k+k^\prime)_\mu\,, {\rm on shell},
\label{e:VPP}
\end{eqnarray}
for the first amplitude, 
a result that also extends to the $1/2$ off-shell amplitude in the
chiral limit.   On the other hand,
repeated use of the Ward identity
\begin{eqnarray}
\not\!k\gamma_5=2m\gamma_5+S^{-1}(p+k)\gamma_5  + \gamma_5S^{-1}(p)
=-[2m\gamma_5+\gamma_5 S^{-1}(p+k) + S^{-1}(p)\gamma_5]
\end{eqnarray}
leads to the results
\begin{eqnarray}
V^{PA_\nu}_{\mu}(k^\prime,k)k^{\nu}&=&-2imV^{PP}_\mu (k^\prime,k)
+e_\pi\Pi^{PA}_\mu(k^{\prime 2})
\nonumber\\
&=&-2imV^{PP}_\mu (k^\prime,k)-2i\frac{e_\pi}{m}f^2_p(k^\prime)
\,k^\prime_\mu,
\label{e:VPA}
\end{eqnarray}
using Eq.~(\ref{e:HAP}) for $\Pi^{PA}_\mu$, and
\begin{eqnarray}
&&k^{\prime \nu} V^{A_\nu A_\rho}_{\mu}(k^\prime,k)=
2im V^{PA_\rho}_{\mu}(k^\prime,k)+e_\pi\big[\Pi^{AA}_{\mu\rho}(k^2)
-\Pi^{VV}_{\mu\rho}(q^2)\big].
\end{eqnarray}
It follows that
\begin{eqnarray}
k^{\prime\nu}V^{A_\nu A_\rho}_{\mu}(k^\prime,k)k^\rho
&=&2imV^{PA_\nu}_{\mu}(k^\prime,k)k^{\nu}
+e_\pi\big[\Pi^{AA}_{\mu\nu}(k^2)
-\Pi^{VV}_{\mu\nu}(q^2)\big]k^\nu.
\end{eqnarray}
Only the longitudinal part of the axial polarization contributes
 to this expression, since
$\Pi^{AA}_{\mu\nu}(k^2)k^\nu $$= \Pi^{AA}_{L}(k^2)k_\mu = 4f^2_p(k^2)k_\mu$.
However, there is also a contribution from the (purely transverse)
 vector polarization,
since its argument involves the momentum transfer $q_\mu=(k^\prime-k)_\mu$
that is different from $k_\mu$.   Then $T_{\mu\nu}(q^2)k^\nu = T_{\mu\nu}
(q^2)k^{\prime\nu}$, since $T_{\mu\nu}(q^2) = g_{\mu\nu} - q_\mu q_\nu/q^2$,
and 
\begin{eqnarray}
\Pi^{VV}_{\mu\nu}(q^2)k^\nu=\Pi^{VV}_T(q^2)T_{\mu\nu}(q^2)k^\nu
=\Pi^{VV}_T(q^2)(q^2)k^{\prime\nu}
= \frac{1}{2}\Pi^{VV}_T(q^2)T_{\mu\nu}(q^2)(k+k^\prime)^\nu.
\end{eqnarray}
Hence
\begin{eqnarray}
k^{\prime\nu}V^{A_\nu A_\rho}_{\mu}(k^\prime,k)k^\rho
=4m^2V^{PP}_\mu(k^\prime,k)&+&4e_\pi[f^2_p(k^{\prime 2})k^\prime_\mu+
f^2_p(k^{2})k^\mu]
\nonumber\\
&+&\frac{1}{2}e_\pi\Pi^{VV}_T(q^2)T_{\mu\nu}(q^2)(k+k^\prime)^\nu.
\end{eqnarray}
Collecting terms, one finds that the final form of the off-shell
 electromagnetic
vertex is
\begin{eqnarray}
V^{(em)}_\mu(k^\prime,k)&=&
(\cos\theta^\prime-\frac{2m}{\sqrt{k^{\prime 2}}}\sin\theta^\prime)
V^{PP}_\mu(k^\prime,k)
(\cos\theta-\frac{2m}{\sqrt{k^{2}}}\sin\theta)
\nonumber\\
&-&2\frac{e_\pi}{m}f^2_p(k^{\prime 2})k^\prime_\mu(\cos\theta^\prime-
\frac{2m}{\sqrt{k^{\prime 2}}}\sin\theta^\prime)
\frac{\sin\theta}{\sqrt{k^2}}
\nonumber\\
&-&2\frac{e_\pi}{m}f^2_p(k^{2})k_\mu(\cos\theta-\frac{2m}
{\sqrt{k^{2}}}\sin\theta)
\frac{\sin\theta^\prime}{\sqrt{k^{\prime 2}}}
\nonumber\\
&-&\frac{1}{2}e_\pi\Pi^{VV}_T(q^2)T_{\mu\nu}(q^2)(k+k^\prime)^\nu\,
\frac{\sin\theta^\prime}
{\sqrt{k^{\prime 2}}}\frac{\sin\theta}{\sqrt{k^{2}}},
\label{e:fullemvertex}
\end{eqnarray}
after using the symmetry property $V^{A_\nu P}_{\mu}(k^\prime,k)=
-V^{P A_\nu}_{\mu}(k,k^\prime)$. The on-shell value for chiral pions,
$k^{\prime 2}= k^2=m^2_\pi=0$, is obtained by inserting the limiting
values $\cos\theta\rightarrow 1$ and
$(2m/\sqrt{k^2})\sin\theta\rightarrow 8G_2f^2_\pi=(1-g_A)$ which come from Eqs.~(\ref{e:thetasm}) and hold
for both the primed and unprimed quantities. Then
\begin{eqnarray}
V^{(em)}_\mu(k^\prime,k)=-e_\pi\frac{f^2_\pi}{m^2}\Big\{g_AF_P(q^2)+
(1-g_A)+\frac{(1-g_A)^2}{8f^2_\pi}\Pi^{VV}_T(q^2)\Big\}(k+k^\prime)_\mu
\end{eqnarray}
is manifestly gauge invariant. In conjunction with Eqs.~(\ref{e:defv}) and
(\ref{e:enjlpiform}), it leads to the identification
\begin{eqnarray}
F_\pi(q^2)=\Big\{g_AF_P(q^2)+(1-g_A)+(1-g_A)^2\frac{1}{8f^2_\pi}
\Pi^{VV}_T(q^2)\Big\}F_V(q^2),
\label{e:ffpi}
\end{eqnarray}
after invoking the Goldberger-Treiman relation $f_\pi g_{\pi qq}=m$ once more.
Thus the electromagnetic form factor of the pion has the required property,
$F_\pi(0)=1$, that avoids charge renormalization.

\subsection{Evaluation of the ENJL Feynman diagrams that enter into
$\Delta m^2_\pi$}
In this section, we obtain closed forms for the Feynman
diagrams shown in Fig.~\ref{f:emdiagram} that enter into $\Delta m^2_\pi$.
We start with the scattering
diagram in Fig.~\ref{f:emdiagram}(a). Translating the diagram,
one finds in the chiral limit for the external pion that
\begin{eqnarray}
\frac{1}{i}\Big\{\Pi^{T_3}_{EM}(0)\Big\}_{scatt}=
ie^2N_c\int \frac{d^4 q}{(2\pi)^4q^2}
\Big\{tr(T^\dagger_f e_qT_ie_q)g_{\mu\nu}
\nonumber\\
+tr(T^\dagger_f\frac{\tau_3}{2}T_i\frac{\tau_3}{2})
\Big(\Pi^{VV}_T(q^2)D^{VV}_T(q^2)\Big)^2T_{\mu\nu}
\nonumber\\
-\big[tr(T^\dagger_fe_qT_i
\frac{\tau_3}{2})+tr(T^\dagger_f\frac{\tau_3}{2}T_ie_q)\big]
\Pi^{VV}_T(q^2)D^{VV}_T(q^2)T_{\mu\nu}\Big\}I^{\mu\nu}(q^2),
\nonumber\\
\label{e:scatappendix}
\end{eqnarray}
with
\begin{eqnarray}
I^{\mu\nu}(q^2)=\int \frac{d^4 p}{(2\pi)^4}tr\big[\gamma_5 
S(p+q)\gamma_\mu S(p)\gamma_5
S(p)\gamma_\nu S(p+q)\big].
\end{eqnarray}
As before,
 $e_q=\frac{1}{2}(\frac{1}{3}+\tau_3)$ is the quark electric charge
in units of $e$, and
$T^\dagger_f$ and $T_i$ are isospin creation and destruction operators for
the initial and final pion channels. The
photon propagator is in the Feynman gauge of Eq.~(\ref{e:phprop}).
 In obtaining these results, we have used the fact that
the effective $\gamma qq$ vertex, including $\rho^0$ exchange,
is given by Eq.~(\ref{e:emcurrent}), i.e.
\begin{eqnarray}
-ieJ^{(em)}_\mu(q^2)=-ie\big[e_q\gamma_\mu+
\frac{1}{i}\Pi^{VV}_T(q^2)\frac{1}{i}D^{VV}_T(q^2)T_{\mu\nu}
(\frac{1}{2}\tau_3\gamma^\nu)\big].
\end{eqnarray}
The values of the isospin traces in $\Pi^{T_3}_{EM} $ of course depend on the
charge channel $T_3$ in question.
They are listed in Table~\ref{table of trace}.
We are, however, only interested in the difference
$\Pi^{\pm}_{EM}-\Pi^{0}_{EM}$.
With the help of the isospin table, one finds 
the surprisingly simple result for
this difference,
\begin{eqnarray}
\frac{1}{i}\Big\{\Pi^{\pm}_{EM}-\Pi^{0}_{EM}\Big\}_{scatt}=
 -ie^2N_c \int \frac{d^4 q}{(2\pi)^4q^2}\Big\{L_{\mu\nu}+T_{\mu\nu}
 F^2_V(q^2)\Big\}I^{\mu\nu}(q^2),
\end{eqnarray}
after using $g_{\mu\nu}=L_{\mu\nu}+T_{\mu\nu}$  and inserting the explicit form of the
transverse $\rho^0$ meson mode propagator from Eq.~(\ref{e:vecprot}), plus the definition of
the vector form factor from Eq.~(\ref{e:vectorformfac}).  Finally we compute
$I^{\mu\nu}(q^2)$. From its spinor structure, $I^{\mu\nu}(q^2)$ has to be a linear
combination of the longitudinal and transverse projections with coefficients that
are only functions of $q^2$. Upon evaluating the integral, one in fact
finds that  the two coefficients are equal, and
\begin{eqnarray}
I^{\mu\nu}(q^2)=  -4I(q^2)g^{\mu\nu}
\end{eqnarray}
where $I(q^2)$ is defined by Eq.~(\ref{e:eyea}).  Consequently,
\begin{eqnarray}
\frac{1}{i}\Big\{\Pi^{\pm}_{EM}-\Pi^{0}_{EM}\Big\}_{scatt}&=&
4ie^2N_c \int \frac{d^4 q}{(2\pi)^4q^2}I(q^2)
+12ie^2N_c\int \frac{d^4 q}{(2\pi)^4q^2}I(q^2)F^2_V(q^2).
\nonumber
\end{eqnarray}
We eliminate  $I(q^2)$ in favor of the pion form factor
$F_P(q^2)$ and pion decay constant $f_p$, Eqs.~(\ref{e:ffpimean})
 and (\ref{e:fpimean}),
 that refer to the mean field calculation, to obtain the
result quoted in the main text as Eq.~(\ref{e:scattxt}),
\begin{eqnarray}
\frac{1}{i}\Big\{\Pi^{\pm}_{EM}-\Pi^{0}_{EM}\Big\}_{scatt}=
 -e^2\frac{f^2_p}{m^2}\int \frac{d^4 q}{(2\pi)^4q^2}F_P(q^2)
 \nonumber\\
 -3e^2\frac{f^2_p}{m^2}\int \frac{d^4 q}{(2\pi)^4q^2}F_P(q^2)F^2_V(q^2)
\label{e:scat}
\end{eqnarray}
The next task is to evaluate the dumbbell diagrams of
Fig.~\ref{f:emdiagram}(b). Due to the presence of the electromagnetic
vertices they contain, there is only a contribution 
in the charged channels from these diagrams.  Let us start with the diagram
where a pion is exchanged in the intermediate state. Label this diagram
$\Pi^{(\pi)}_{EM}$. Moving immediately to the chiral limit for the external pions,
one obtains
\begin{eqnarray}
\frac{1}{i}\Pi^{(\pi)}_{EM}(0)=\int \frac{d^4 q}{(2\pi)^4q^2}V^{(em)}_\mu(0,q)
\lambda_\pi(q^2)g^{\mu\nu}D_{PP}(q^2) V^{(em)}_{\nu}(q,0).
\label{e:pidumbbell}
\end{eqnarray}
The $\gamma \pi\pi$ electromagnetic vertex is given
by Eq.~(\ref{e:fullemvertex}), and the pion eigenmode propagator can be identified
from Eq.~(\ref{e:pimode}).  This expression does not yet contain the contributions
from the three remaining diagrams of Fig.~\ref{f:emdiagram}(b) that involve
the $\rho^0$ meson to photon conversion in the intermediate state. However,
none of these diagrams contribute in the chiral limit as we now show. From
Eq.~(\ref{e:fullemvertex}), the half off-shell behavior of the electromagnetic
vertex is
\begin{eqnarray}
V^{(em)}_\mu(0,q)=V^{(em)}_\mu(q,0)=-e_\pi\frac{f^2_p}{m^2}F_P(q^2)
(\cos\theta-\frac{2m}{\sqrt{q^2}}\sin\theta )q_\mu
\label{e:emvertexchrl}
\end{eqnarray}
in the chiral limit, which is purely longitudinal. Consequently the 
additional $\rho^0$ meson diagrams,
which are automatically included together with the direct photon exchange
by writing
\begin{eqnarray}
V^{(em)}_\mu(0,q)\rightarrow
 [L_{\mu\nu}+F_V(q^2)T_{\mu\nu}]V^{(em);\nu}(0,q),
\end{eqnarray}
as in Eq.~(\ref{e:modgam}), fail to contribute in this case due to the
longitudinal nature  of  $V^{(em)}_\mu(0,q)$; the transverse piece has been
projected out completely. The corresponding diagram for
the $a_L$ exchange has exactly the same form and properties.
We simply must substitute the eigenvalue
$\lambda_{a_L}(q^2)$ for $\lambda_\pi(q^2)$ in Eq.~(\ref{e:pidumbbell}).
Then the $\pi$ and $a_L$ exchanges taken together contribute an amount 

\begin{eqnarray}
\frac{1}{i}\Pi^{(\pi+a_L)}_{EM}(0)=e^2\big(\frac{f^2_p}{m^2}\big)^2
\int \frac{d^4 q}{(2\pi)^4}\Big\{
\lambda_\pi(q^2)\big[(\cos^2\theta+\frac{4m^2}{q^2}
\sin^2\theta)-\frac{4m}{\sqrt{q^2}}\sin\theta\cos\theta\big]
\nonumber\\
+\lambda_{a_L}(q^2)\big[(\sin^2\theta+\frac{4m^2}{q^2}
\cos^2\theta)+\frac{4m}{\sqrt{q^2}}\sin\theta\cos\theta\big]\Big\}
D^{PP}(q^2)F^2_P(q^2),
\label{e:piplusal}
\end{eqnarray}
the factor $1/q^2$ from the photon propagation having cancelled against a similar
factor coming from the contraction of the two electromagnetic vertices.
The combinations appearing in the integrand have the values
\begin{eqnarray}
\lambda_\pi\cos^2\theta+\lambda_{a_L}\sin^2\theta
=\frac{\lambda_\pi+\lambda_{a_L}}{2}+\frac{\lambda_\pi-\lambda_{a_L}}{2}\cos 2\theta=A
\nonumber\\
(\lambda_\pi-\lambda_{a_L})\sin\theta\cos\theta
=\frac{\lambda_\pi-\lambda_{a_L}}{2}\sin 2\theta =B
\nonumber\\
\lambda_\pi\sin^2\theta+\lambda_{a_L}\cos^2\theta
=\frac{\lambda_\pi+\lambda_{a_L}}{2}-\frac{\lambda_\pi-\lambda_{a_L}}{2}\cos 2\theta=C
\end{eqnarray}
from the properties of the eigenvalues and eigenvectors given by
Eqs.~(\ref{e:eigen}) and (\ref{e:theta}). The value of the contents of the
curly brackets in the integrand of Eq.~(\ref{e:piplusal}) is thus
\begin{eqnarray}
\Big\{\cdots\Big\}=\Big\{A+\frac{4m^2}{q^2}C-\frac{4m}{\sqrt{q^2}}B\Big\}
=1-8G_2f^2_\pi(q^2)=\frac{1}{1+8G_2f^2_p(q^2)},
\label{e:curly}
\end{eqnarray}
 after inserting the values of $A$, $B$ and $C$ from Eqs.~(\ref{e:A}) to (\ref{e:C})
 and using the definition of  $f^2_\pi(q^2)$  from Eq.~(\ref{e:fpisc}).
 Since the renormalized pion  propagator of Eq.~(\ref{e:DDren})
 contains the inverse of the same factor,
 \begin{eqnarray}
 D^{PP}(q^2)=g^2_p[1+8G_2f^2_p(q^2)]\frac{1}{q^2F_P(q^2)}
 \end{eqnarray}
 this factor cancels out in the final expression for the integrand,
 together with one of the $F_P(q^2)$'s. Hence we obtain the simple expression
 \begin{eqnarray}
 \frac{1}{i}\Pi^{(\pi+a_L)}_{EM}(0)=e^2 \frac{f^2_p}{m^2}\int \frac{d^4 q}{(2\pi)^4q^2}
 F_P(q^2)
 \label{e:pial}
 \end{eqnarray}
 as the final result after using the Goldberger-Treiman relation
 $f^2_p g^2_p=m^2$  for the mean field quantities 
 once again.  The final answer in fact coincides with the NJL result where no
 vector mesons are present \cite{dstl95}.

 The calculations of the dumbbell diagrams involving the $a_1$ and $\rho$
 mesons
 follow an identical pattern. The basic
 electromagnetic vertex for these two cases is obtained by replacing one of the
 incoming (or outgoing) $\pi$'s in Fig.~\ref{f:gammavert} by an $a_1$ or a $\rho$
 respectively.  In the first case this leads to
 \begin{eqnarray}
 V^{(\gamma \pi a_1)}_{\mu\nu}(k^\prime,k)=V^{PA_\nu}_\mu(k^\prime,k)
 (\cos\theta^\prime-\frac{2m}{\sqrt{k^{\prime 2}}}\sin\theta^\prime)+
 ie_\pi\big[\Pi^{AA}_{\mu\nu}(k^2)-\Pi^{VV}_{\mu\nu}(q^2)\big]
 \frac{\sin\theta^\prime}{\sqrt{k^{\prime 2}}}
\end{eqnarray}
 for an incoming $a_1(k)$ and outgoing $\pi(k^\prime)$; here
 $q_\mu=(k^\prime-k)_\mu$ is the momentum transferred by the photon.
 The required half off-shell
 value in the chiral limit is
 \begin{eqnarray}
 V^{(\gamma \pi a_1)}_{\mu\nu}(0,q)=
 V^{PA_\nu}_\mu (0,q)[1-8G_2f^2_\pi]+i\frac{e_\pi}{m}8G_2f^2_\pi
 \big[\Pi^{AA}_{\mu\nu}(q^2)-\Pi^{VV}_{\mu\nu}(q^2)\big],
 \end{eqnarray}
 due to the limiting values $\cos\theta^\prime\rightarrow 1$
 and $2m\sin\theta^\prime/\sqrt{k^{\prime 2}}
 \rightarrow 8G_2f^2_\pi$ for the emitted chiral pion. But the
 axial vector minus vector polarization equals $4f^2_pF_P(q^2)$ according
 to the Ward identity (\ref{e:ward1}), and
 \begin{eqnarray}
 V^{PA_\nu}_\mu (0,q)=2ime_\pi\frac{f^2_p}{m^2}F_P(q^2)g_{\mu\nu},
 \end{eqnarray}
 that follows immediately from the relation (\ref{e:VPA}) and the expression
 (\ref{e:VPP}) for $V^{PP}(k^\prime,k)$.
 The terms involving $G_2$ cancel, and one is left with
$V^{(\gamma \pi a_1)}_{\mu\nu}(0,q)=
 V^{PA_\nu}_\mu (0,q)$.
 The additional three diagrams that include the $\rho^0$ degrees of freedom
 can be incorporated into this expression by replacing $g_{\mu\nu}$ with
 $[L_{\mu\nu}+F_V(q^2)T_{\mu\nu}]$ as before.  The final expression for the
 $\gamma \pi a_1$ vertex including $\rho^0$ contributions is thus
 \begin{eqnarray}
 V^{(\gamma \pi a_1)}_{\mu\nu}(0,q)=
 2ime_\pi\frac{f^2_p}{m^2}F_P(q^2)[L_{\mu\nu}+F_V(q^2)T_{\mu\nu}]=
 -V^{(\gamma \pi a_1)}_{\nu\mu}(q,0).
\label{e:gammapia1}
\end{eqnarray}
The structure of the $a_1$ dumbbell diagram is thus
\begin{eqnarray}
\frac{1}{i}\Pi^{(a_1)}_{EM}(0)=\int \frac{d^4 q}{(2\pi)^4q^2}
V^{(\gamma \pi a_1)}_{\mu\nu}(0,q)g^{\mu\mu^\prime}
D^{AA}_T(q^2)T^{\nu\nu^\prime}
V^{(\gamma \pi a_1)}_{\mu^\prime\nu^\prime}(q,0).
\end{eqnarray}
In contrast with the case of $\pi+a_L$ exchange in Eq.~(\ref{e:piplusal}) 
where
the transverse part of the vertex involving $F_V(q^2)$ fell away, 
here only the
transverse part of $V^{(\gamma \pi a_1)}_{\mu\nu}(0,q)$ survives due
 to the transverse nature of the $a_1$
mode propagator; now the longitudinal part falls  away completely.  One has
\begin{eqnarray}
[L_{\mu\nu}+F_V(q^2)T_{\mu\nu}]T^{\nu\nu^\prime}
[L_{\mu^\prime\nu^\prime}+F_V(q^2)T_{\mu^\prime\nu^\prime}]
=F^2_V(q^2)T_{\mu\mu^\prime}.
\nonumber
\end{eqnarray}
Using $g^{\mu\mu^\prime}T_{\mu\mu^\prime}=3$, the electromagnetic polarization
contribution due to
the $a_1$ exchange becomes
\begin{eqnarray}
\frac{1}{i}\Pi^{(a_1)}_{EM}(0)=12e^2m^2\big(\frac{f^2_p}{m^2}\big)^2
\int \frac{d^4 q}{(2\pi)^4q^2}F^2_P(q^2)F^2_V(q^2)D^{AA}_T(q^2),
\label{e:a1}
\end{eqnarray}
after inserting Eq.~(\ref{e:gammapia1}).

The basic $\gamma \pi\rho$ vertex is obtained by replacing the spinor
$A^a_\nu=\gamma_\nu\gamma_5\tau^a$ by
$V_\nu=\gamma_\nu\tau^a$ in Eq.~(\ref{e:PAnu}). Then this diagram 
assumes the
same structure as the anomalous $\pi^0\rightarrow 2\gamma$ vertex, see for example
Ref.\cite{iz80}, p 552. After attaching the photon to
both the quark and antiquark lines, and adding, one finds
\begin{eqnarray}
V^{PV_\nu}_{\mu}(k^\prime,k)&=&(-eN_c) tr(\{T_f^\dagger,T_i\}e_q)
\int\frac{d^4 p}{(2\pi)^4}
tr[\gamma_5 S(p+k^\prime)\gamma_\mu S(p+k)\gamma_\nu S(p)]
\nonumber\\
&=&(-\frac{2}{3}eN_c)(4im\varepsilon_{\mu\nu\rho\sigma} k^{\prime \rho}
 k^{\sigma})\times J(k^\prime,k),
\end{eqnarray}
since the trace that arises in the numerator of the integrand after inserting the quark propagators
has the value $4im\varepsilon_{\mu\nu\rho\sigma} k^{\prime \rho} k^{\sigma}$.
The factor $J(k^\prime,k)$ is the integral
\begin{eqnarray}
J(k^\prime,k)=\int\frac{d^4p}{(2\pi)^4}
\frac{1}{[(p+k^\prime)^2-m^2][(p+k)^2-m^2][p^2-m^2]}.
\end{eqnarray}
The three additional diagrams containing the $\rho^0$ can be incorporated
by including the factor $[L_{\mu\mu^\prime} +F_V(q^2)T_{\mu\mu^\prime}]$
as before, and we obtain the final result
\begin{eqnarray}
V^{(\gamma\pi\rho)}_{\mu\nu}(k^\prime,k)=-\frac{8}{3}ieN_c m
J(k^\prime,k)[L_{\mu\mu^\prime} +F_V(q^2)T_{\mu\mu^\prime}]
\varepsilon^{\mu^\prime}_{\;\;\nu\rho\sigma} k^{\prime \rho} k^{\sigma}.
\end{eqnarray}
The half off-shell value of this vertex is zero in the chiral limit,
$ V^{(\gamma\pi\rho)}_{\mu\nu}(0,q)=0$, so the dumbbell diagram with a
$\rho$ replacing the $a_1$ vanishes.

\subsection{Dashen's theorem}
The complete set of diagrams that determine the corrections to the pion polarization to
${\cal O}(\alpha N_c)$ are shown in Figs.~\ref{f:selfenergy}
and \ref{f:emdiagram}.  We have already calculated the scattering and
meson pole, or dumbbell contributions of Fig.~\ref{f:emdiagram}
in the previous section. The remaining diagrams in
Fig.~\ref{f:selfenergy} dress either the quark or antiquark line, i.e. they
change the
quark mass circulating in the polarization loop. Since the
quark and the antiquark lines are dressed identically by the same interaction
to ${\cal O}(\alpha)$, both  diagrams have the same structure
apart from their isospin weights.
One finds for Fig.~\ref{f:selfenergy} (a)+(b) that
\begin{eqnarray}
\Pi^{(a+b)}_{EM}(0)=-e^2N_c\int\frac{d^4q}{(2\pi)^4q^2}\Big\{2tr(e_q^2)g^{\mu\nu}+
(F^2_V(q^2)-1)T^{\mu\nu}\Big\}M_{\mu\nu},
\end{eqnarray}
where
\begin{eqnarray}
M_{\mu\nu}&=& \int\frac{d^4p}{(2\pi)^4}tr\Big[\gamma_5S(p)\gamma_\mu
S(p+q)\gamma_\nu S(p)\gamma_5S(p)\Big].
\end{eqnarray}
But
\begin{eqnarray}
g^{\mu\nu}M_{\mu\nu}=4[I(q^2)+I(0)-(q^2+2m^2)K(q^2)]
\quad {\rm and}\quad L^{\mu\nu}M_{\mu\nu}=2I(q^2),
\end{eqnarray}
where we have defined
\begin{eqnarray}
K(q^2)= \int\frac{d^4p}{(2\pi)^4}\frac{1}{[(p+q)^2-m^2][p^2-m^2]^2}.
\end{eqnarray}
Hence
\begin{eqnarray}
\Pi^{(a+b)}_{EM}(0)=
e^2N_c\int\frac{d^4q}{(2\pi)^4q^2}\Big\{8tr(e^2_q)[-I(q^2)-I(0)+
(q^2+2m^2)K(q^2)].
\nonumber\\
-(F^2_V(q^2)-1)[2I(q^2)+4I(0)-4(q^2+2m^2)K(q^2)]\Big\}.
\end{eqnarray}
Diagrams 18(c) and 18(d) involve the dressing of the quark and
antiquark lines via the
exchange of a $\sigma$ mode with the condensate. The sum of these diagrams
can be expressed in terms of the quark electromagnetic self-energy $\Sigma_{EM}$,
since the $\sigma$ propagator terminates on an electromagnetic self-energy
insertion of the full quark self-energy:
\begin{eqnarray}
\Pi^{(c+d)}_{EM}(0)=-\frac{\Sigma_{EM}}{2m},
\end{eqnarray}
One has
\begin{eqnarray}
\Sigma_{EM}&=&(e^2G_1N_c)\int\frac{d^4q}{(2\pi)^4q^2}
\Big\{[2tr(e^2_q)+(F^2_V(q^2)-1)]J_\mu^{\;\;\mu}(q^2)\Big\}
\nonumber\\
&=&(8e^2mG_1N_c)\int\frac{d^4q}{(2\pi)^4q^2}\Big \{[2tr(e^2_q)+(F^2_V(q^2)-1)] 
\nonumber\\
&&\qquad\qquad\qquad\times [I(q^2)-I(0)+(q^2+2m^2)K(q^2)]\Big\}
\end{eqnarray}
using
\begin{eqnarray}
J_\mu^{\;\;\mu}(q^2)&=&\int\frac{d^4p}{(2\pi)^4}tr\big[\gamma_\mu S(p+q)\gamma^\mu S(p)S(p)\big]
\nonumber\\
&=&8m[I(q^2)-I(0)+(q^2+2m^2)K(q^2)]
\end{eqnarray}
The total contribution to the pion polarization from
Fig.~\ref{f:selfenergy} is then given by the sum
\begin{eqnarray}
\Big[\Pi^{(a+b)}_{EM}(0)+
 \Pi^{(c+d)}_{EM}(0)\Big]=
  -e^2N_c \int\frac{d^4q}{(2\pi)^4q^2}\Big\{16tr(e^2_q)+6(F^2_V(q^2)-1)\Big\}I(q^2)
\label{e:selfsum}
 \end{eqnarray}
On the other hand the polarization contribution due to the interaction of like-flavored
$\bar qq$ pairs in the $T_3=0$ channel can be found from
Eq.~(\ref{e:scatappendix}). It is
\begin{eqnarray}
\big\{\Pi^0_{EM}(0)\big\}_{scatt}= e^2N_c\int\frac{d^4q}{(2\pi)^4q^2}\Big\{
16tr(e^2_q)+6(F^2_V(q^2)-1)\Big\}I(q^2).
\end{eqnarray}
This is {\it precisely} the negative of the sum in Eq.~(\ref{e:selfsum}).
If we convert these two expressions into pion self-energies by supplying the missing
coupling constant $(ig_{\pi qq})^2$ and using  $F_P(q^2)=I(q^2)/I(0)$ plus
the GT relation $f^2_{\pi qq}g^2_{\pi qq}=m^2$ of the ENJL model, one reaches
the final conclusion that
\begin{eqnarray}
\Sigma_{self}=-\Sigma^0_{scatt}=ie^2g^{-1}_A \int\frac{d^4q}{(2\pi)^4q^2}
\Big\{4tr(e^2_q)+\frac{3}{2}(F^2_V(q^2)-1)\Big\}F_P(q^2).
\label{e:cancel}
\end{eqnarray}
This shows the exact cancellation between the self-energy of the neutral pion that
arises from the electromagnetic
self-energies of the quarks it contains, and the self-energy due to their
electromagnetic interaction, and illustrates how  Dashen's theorem is
realized in the ENJL model. A direct comparison of the various self-energies
is given in Fig.~\ref{f:dashenoper}. One sees there that
$\Sigma_{self}>0$, and  that the scattering contribution is the
negative of this, in agreement with the attractive electromagnetic interaction between like-flavored
$\bar qq$ pairs.

\widetext
%
%
\begin{figure}
\caption[]{Modification of the quark axial isospin current vertex due to the
exchange of the pion plus longitudinal axial eigenmode, and the
transverse $a_1$ mode. Quark propagation is shown as solid lines, mesons as broken lines.
The wavy line represents an isospin current
component of momentum $q$ entering the vertex.}
\label{f:isospin current}
\end{figure}

%
%
\begin{figure}
\caption[]{ Self-energy diagrams entering the non-chiral gap equation
to ${\cal O} (\hat m)$ for $m^*$. (a) The Hartree contribution $-i\Sigma_H$
in Eq.~(\ref{e:gap}). In this figure, (b)+(c) represent the 
non-chiral corrections arising from a finite
current quark mass, $-i\hat m$, denoted by the cross. The $\sigma$ 
mode exchange
is denoted by the double solid line.}
\label{f:currentmass}
\end{figure}

%
%
\begin{figure}
\caption[]{Behavior of the quark mass and condensate density as a function of the
ratio $x=\Lambda^2/m^2$.  The upper pair of curves labelled $m$ and
$-\langle\bar qq\rangle^{1/3}$ refer to the ENJL calculation with $f_\pi=93$MeV and
$g_A=0.75$. The lower pair of curves refer to the results for the NJL model.
They were obtained by scaling the upper pair by $\protect\sqrt{g_A}=0.866$.}
\label{f:nchrlm}
\end{figure}

%
%
\begin{figure}
\caption[str]{The vector and axial vector
 spectral densities of the chiral ENJL model versus
the momentum transfer squared $s=q^2$. The interaction strengths and regulating
cutoff are those given in the second line of Table~\ref{table1}.
The $\bar qq$ threshold then lies at $4m^2=
(2\times 0.264\,GeV)^2=0.279\,$GeV$^2$. The peak
 values of the vector and axial
strengths lie at $m^2_\rho=(0.713 \,$GeV$)^2$ and $m^2_{a_1}=(1.027\,$GeV$)^2$
 respectively
that are taken to identify the $\rho$ and $a_1$ masses of the ENJL model. 
Their ratio 
$m_{a_1}/m_\rho =\protect\sqrt{2.07}$ lies close to the original 
Weinberg estimate
\cite{wein67} of $\protect\sqrt{2}$.}
\label{f:strengths}
\end{figure}

%
%
\begin{figure} 
\caption[]{Comparison of the PV-regulated density functions of the ENJL and
NJL model respectively.    The parameters are the same
as for Fig.\ \protect\ref{f:strengths}. The position and spacings
of the $\bar q q$ threshold at $4m^2=0.279$ GeV$^2$, and the two
thresholds introduced by the Pauli-Villars regularization
at $4m^2(1+\Lambda^2/m^2)=4.8$ GeV$^2$ and
$4m^2(1+2\Lambda^2/m^2)=9.2$ GeV$^2$ for this parameter choice are also
shown.
Note (a) the strong redistribution of strength to lower energies of the 
vector mode, and (b) the strong suppression of the unphysical negative
density at large $s$ for the ENJL case.}
\label{f:neg}
\end{figure}

%
%
\begin{figure}
\caption[diffstr]{Comparison of the PV-regulated ENJL and NJL
strength function differences $\rho^V_1-\rho^A_1$ and
$\rho^{(0)V}_1-\rho^{(0)A}_1$ as obtained from
Eqs.~(\ref{e:rhov}) and (\ref{e:rhoa}) and
Eqs.~(\ref{e:vecdens}) and (\ref{e:axvecdens}). The parameters are the same
as for Fig.\ \protect\ref{f:strengths}.  The PV induced thresholds are as
described in the previous figure.
The strong concentration of strength at much lower energies
 in the ENJL case is due to the presence of
the vector and axial mode
form factors that essentially damp out
the PV-induced unphysical behavior of the NJL spectral density at energies
considerably below the first PV-generated threshold.}
\label{f:diffstr}
\end{figure}

%
%
\begin{figure}
\caption[]{The behavior of the real part of the function $J(z)$
along the real axis on the first
Riemann sheet, and its analytic continuation  $\tilde J(z)$ through the
cut $z=1$ to $\infty$, onto the second sheet.}
\label{f:ReJ}
\end{figure}

%
%
\begin{figure}
\caption[]{The behavior of the real part of the vector
 polarization $\Pi^{VV}_T(s)$
as a function of the variable $z=s/4m^2$ on the first and second sheets
of the cut $s$-plane with the cut running from $z=1$ to $\infty$ along the real
axis. The values of the polarization along the upper lip of the cut on the
first sheet join smoothly with their analytic continuation onto the lower lip of the cut on the
second sheet. At the branch point these values bifurcate into two branches
of the function that assume different values along the sector $0\le z\le 1$ of the two sheets.
As explained in more detail in connection with Eq.~(\ref{e:Jtildc}), the values on the first and second sheets are reached by allowing the
variable $z$ to pass either infinitesimally above or
through the cut infinitesimally
below the branch point
to reach this sector of the real axis. The solid circles indicate the values
of $s/4m^2$   where the polarization function equals  $-1/(8G_2\,m^2)$
for the
indicated interaction strengths $G_2$. These intercepts determine
the position of the real poles of vector mode propagator lying in the
interval $0\le s/4m^2\le 1$.}
\label{f:branch}
\end{figure}

%
%
\begin{figure}
\caption[]{The switch-over of the real pole of
the $\rho$ propagator $D^{VV}_T$
of Eq.~(\protect\ref{e:vecprot}) from the second to the first Riemann sheet
as the interaction strength $G_2$ increases through the values
$2\,(1)\,8,10$ and $16$ GeV$^{-2}$ (solid circles). The critical value of $G_2$ at which the switch-over
from the second to the first sheet occurs is $G^{(c)}_2=7.40$ GeV$^{-2}$
in this illustration.}
\label{f:vpole}
\end{figure}

%
%
\begin{figure}
\caption[]{The positions of the complex
poles on the second Riemann sheet of the $\rho$ and $a_1$ propagators
in Eqs.~(\ref{e:vecprot}) and (\ref{e:axvecpro}) in units of $4m^2$.
 The value of the
interaction strength $G_2 $ is recorded in GeV$^{-2}$ next to each pole.}
\label{f:cmplxpoles}
\end{figure}

%
%
\begin{figure}
\caption[]{Top figure (a): Comparison of the behavior on the
 first Riemann sheet of 
$Re\bar \Pi^{JJ}_T(s)$ for $J=V,A$ 
as obtained from an unsubtracted 
dispersion relation, plotted as a function of $s= q^2/4m^2$. The bottom
figure, (b),
gives the same information for $Re \Pi^{JJ}_T(s)$, now obtained from
a once subtracted dispersion relation.   Note the occurrence of a 
space-like zero in both functions in this case.}
\label{f:compv}
\end{figure}

%
%
\begin{figure}
\caption[landau]{ Landau ghost  poles in the $\rho$,
$a_1$, and $\pi$ channels.
These poles correspond to the space-like roots
of the combination $1+2G\Pi(s)$
where the  polarization function has been calculated via a
once-subtracted dispersion relation. 
For the vector and axial vector modes, $G=G_2$ and
$\protect\Pi(s)$ comes from
Eqs.~(\ref{e:vecpolbartxt}) or (\ref{e:axvecpolbartxt}) with the bars removed.
The roots corresponding to the resulting $\rho$ and $a_1$ ghost modes are
almost degenerate at $s/4m^2=-18.30$ and $-18.04$
respectively for the standard parameter set of Table~\protect\ref{table1}.
For the pion mode, $\Pi=\hat\Pi^{PP}$ of Eq.~(\ref{e:polren}) and $G=-G_1$.
Apart from the expected Goldstone mode for the pion at zero, a second,  ghost pole
appears at $s_{\pi g}/4m^2=-13.48$. The extensions of the curves beyond 
the cusp
at $s/4m^2=1$ refer to the real part of the corresponding polarization function.}
\label{f:ghosts}
\end{figure}

%
%
\begin{figure}
\caption[]{Comparison of the modulus squared of the exact vector form factor
with the single pole approximation.
The $\rho$ pole is located at
$s_\rho =4m^2(0.522-1.326\,i)=(0.146-0.370\,i)$GeV$^2$
on the second Riemann sheet,
 as illustrated in detail in
 Fig.~\protect\ref{f:contour}. Only that part of the 
approximate curve that lies to the right of the threshold
at $4m^2=0.279$GeV$^2$, marked by the
cusp in the exact curve, is
accessible on the real axis of the physical sheet.}
\label{f:veccomp}
\end{figure}

%
%
\begin{figure}
\caption[]{The same comparison as in Fig.~\protect\ref{f:veccomp} between the
exact and single pole approximation, but for the axial form factor 
$|G_A(s)|^2$. In this
figure, the $a_1$ pole lies at $s_{a_1}=4m^2(2.123-2.141\,i)
=(0.592-0.597\,i)$GeV$^2$.}
\label{f:aveccomp}
\end{figure}

%
%
\begin{figure} 
\caption[]{Closed integration contour on the first, or physical, Riemann
sheet
for evaluating the sum rules.
The physical cut starts at $s=4m^2$.  Additional poles 
 can appear on
the real axis of this sheet  at bound states  of the $\bar q q$ system
in $0<s<4m^2$ if $G_2$ is large enough, or as
ghost poles in $-\infty<s<0$. As discussed in the main text, such ghosts
only appear
if once-subtracted dispersion relations are used in the construction of
the propagators.  The crosses indicate the pole positions
of the virtual $\rho$ mode at $0.1658$, and complex $\rho $ and $a_1$
modes at $0.5221-1.326i$ and $2.1227-2.1405i$
respectively in units of $4m^2$ for the standard parameter set of
Table~\ref{table1}.
As these singularities all lie on the second sheet, they do not fall within
the integration contour.}
\label{f:contour}
\end{figure}

%
%
\begin{figure}
\caption[]{The $\gamma \pi\pi$ electromagnetic vertex for absorbing a
photon of four momentum $q=k^\prime-k$, where $k$ and $k^\prime$ are the
initial and final pion four momenta. Quark propagators are shown as solid lines,
pions as broken lines and photons as wavy lines.
In this figure, the small open circle
represents the $qq$ vertex of Fig.~\ref{f:isospin current} with the isospin current replaced
by the electromagnetic current. Thus the pseudoscalar and pseudovector
exchange contributions in that diagram are to be replaced
by $\rho^0$ exchange, indicated by the heavy solid line.}
\label{f:gammavert} 
\end{figure}

%
%
\begin{figure}
\caption[]{The space- and time-like behavior of the electromagnetic form factor
$|F_\pi(q^2))|^2$ of the pion. The experimental data (open circles) have
been compiled from
the  various sources listed under Ref.~\cite{piexp}. The calculated value of this
quantity from the ENJL model is
given by Eq.~(\ref{e:myhope}). The modulus squared of this expression has
been plotted as a solid line. Although the
ENJL model is restricted to low energies as an effective theory, where the agreement
is quantitatively significant, the plot has been extended to cover the full
range of available data. The possibility of $\rho-\omega$ mixing as evidenced by the
interference structure around $q^2\sim 3$GeV$^2$ has not been considered in the
ENJL model employed in this paper.}
\label{f:pionf2}
\end{figure}

%
%
\begin{figure}
\caption[]{The calculated electromagnetic mass squared splitting of the pion as
a function of the photon cutoff squared, $\lambda^2_{photon}$
for the standard parameter set of Table~\protect\ref{table1}.
The observed value
of the splitting, $\Delta m_\pi^2=1261$MeV$^2$ is
 indicated by the horizontal line. This
value of the splitting is obtained for $\lambda^2_{photon}=(1.27\rm MeV)^2$.}
\label{f:deltapi2}
\end{figure}

%
%
\begin{figure}
\caption[]{The complete set of pion polarization  diagrams to ${\cal
O}(N_c)$,
where the mass of either the quark or the antiquark line is renormalized
to ${\cal O}(\alpha)$. The conventions are the same as in
Fig.~\ref{f:gammavert}. The double solid line indicates the exchange of
the $\sigma$ mode of the model.}
\label{f:selfenergy}
\end{figure}

%
%
\begin{figure}
\caption[]{The two classes of irreducible pion polarization
diagrams involving single photon exchange that determine the electromagnetic
mass splitting of the charged to neutral pions to ${\cal O}(\alpha)$: (a) diagrams that describe
the scattering of $\bar qq$ pairs in the intermediate state,
and (b) meson pole, or ``dumbbell''
diagrams where the exchange of the $\pi\,+a_L$ and $a_1$ meson eigenmodes
together with the photon between identical vertices occurs. The conventions are the same
as in Fig.~\ref{f:gammavert}.
There is no contribution from the dumbbell diagram with
a $\rho$  substituted for an $a_1$ in the chiral limit.
The small circle at a photon vertex indicates that $\rho^0$ exchange has to be
included as in Fig.~\ref{f:gammavert}.  As a consequence, both diagrams are actually 
 a sum of four diagrams. In (c) we have written out this sum
 explicitly for the dumbbell diagram.}
\label{f:emdiagram}
\end{figure}

%
%
\begin{figure}
\caption[]{Breakdown of the $U_V(1)\times SU_L(2)\times SU_R(2)$ chiral symmetry
of the original ENJL Lagrangian into $U_A(1)\times U_V(1)$ by the
photon gauge field. This the origin of the electromagnetic mass
 of the  Goldstone pions of the ENJL model. The physical mechanism
that keeps the  neutral chiral
pion massless in agreement with Dashen's theorem is
illustrated explicitly, with the mass shift $\Sigma_{self}$ being exactly
cancelled out by the internal
electromagnetic interaction $\Sigma_{scatt}$ of the neutral pion.}
\label{f:dashenoper}
\end{figure}


\mediumtext

\begin{table}
\caption{Comparison of the NJL and ENJL model coupling constants,
regulating cutoff, as well as the quark properties they give  at $x=16$.
The input parameters are  $f_p=93$MeV for the NJL case (first row),
and $f_\pi=93$MeV together with a form factor $g_A=0.75$ in the
ENJL case (second row). We refer to the latter set of parameters as the
standard set in the text.}
\label{table1}
\begin{tabular}{ccccc}
$G_1($GeV$^{-2}$)&$G_2($GeV$^{-2}$)&$\Lambda($GeV$)$&$m({\rm MeV})$ & 
$-{\langle \bar qq\rangle}^{1\over 3}({\rm MeV})$\\          
\tableline
3.29 & 0 & 0.916 & 229 & 259 \\
2.47 & 3.61 & 1.058 & 264 & 299  \\
\end{tabular}
\end{table}

\begin{table}
\caption[]{The combination $\bar l_i+\ln(m^2_\pi/\mu^2)$ and
$l_7$ of the chiral perturbation theory coupling constants as
given by the ENJL or NJL model.}
\label{table2a}
\begin{tabular}{ccc}
${\it i}$&ENJL&NJL\\
\tableline
 & & \\
3&$\big\{8\pi^2\frac{f^2_\pi}{m^2}(2g_A-1)-4g_A(3g_A-1)\big\}g_A$&$8\pi^2
\frac{f^2_p}{m^2}-8$\\
 & &\\
4&$\big\{4\pi^2\frac{f^2_\pi}{m^2}-3g_A\big\}g_A^2$&$4\pi^2\frac{f^2_p}
{m^2}-3$\\
 & &\\
5&$1+6g_A^2+8\pi^2\frac{f^2_\pi}{m^2}(\frac{1-g^2_A}{g_A})$&7\\
 & &\\
6&$1+12g_A^2+8\pi^2\frac{f^2_\pi}{m^2}(\frac{1-g^2_A}{g_A})$&13\\
 & &\\
7&$\frac{2\hat{m}^2}{g_A}\frac{f^2_\pi}{m^2_\pi}$&
$2\hat{m}^2\frac{f^2_p}{m^2_p}$\\
 & &
\end{tabular}
\end{table}

\begin{table}
\caption[]{Comparison of the Gasser-Leutwyler chiral perturbation theory
parameters with the predictions of both the NJL and ENJL model calculations.
The relevant input parameters are: $f_\pi=93$MeV, $m_\pi=135$MeV, $x=16$, and
quark masses $m=229$MeV and $264$MeV for the NJL or ENJL cases as
listed in Table \protect\ref{table1}. The contribution from
$-\ln(m^2_\pi/\mu^2)$ with $\mu=2m$ is displayed as an intermediate step in
both cases.}
\label{table2b}
\begin{tabular}{cccc}
Parameter & Empirical value, Ref.~\cite{gl84} & NJL & ENJL \\
\tableline
$\bar l_3$ & $2.9\pm 2.4$ & $5.0+2.4=7.4$ & $4.5+2.7=7.2$\\
$\bar l_4$ & $4.3\pm 0.9$ & $3.5+2.4=5.9$ & $1.5+2.7=4.2$\\
$\bar l_5$&$13.9\pm 1.3$&$7.0+2.4=9.4$&$10.1+2.7=12.8$\\
$\bar l_6$&$16.5\pm 1.1$&$13+2.4=15.4$&$13.5+2.7=16.2$\\
$l_7$&$\sim 5\times 10^{-3}$&$1.4\times 10^{-3}$&$0.5\times 10^{-3}$\\
\end{tabular}
\end{table}

\begin{table}
\caption{Comparison of second sheet real and complex poles of the 
$\rho$ and $a_1$
 propagators in
units of $4m^2$. The type of dispersion relation used to generate the
 polarization
amplitude employed in
Eqs.~(\protect\ref{e:vecrts}) and (\protect\ref{e:axvecrts}) is indicated
on the left. The ghost poles only occur if subtracted dispersion relations are
used, and lie on the first, or physical sheet.
The input parameters are the standard set as given in the second line
of Table~\protect\ref{table1}.}
\label{table3}
\begin{tabular}{lccc}
& $\rho$ pole & $\rho$ pole & $a_1 $ pole\\
\tableline
Unsubtr. & 0.1658 & $0.5221-1.3260 i$&$2.1227-2.1405i$\\
Subtr.   &0.1659 &$0.5557-1.3473i$&$2.2340-2.1728i$\\
Ghost poles&-&$-18.3038$&$-18.0418$\\
\end{tabular}
\end{table}

\begin{table}
\caption{Comparison of the second sheet  complex poles of the $\rho$
and $a_1$ propagators and the peak positions in their spectral
densities.
The complex poles
are recorded in the usual form $s/M=M-i\Gamma$ 
in order to display a mass and a width contribution. We have used 
a capital letter
for the mass that appears as the real part of the complex pole in 
order to distinguish it
from the lower case $m_\rho$ and $ m_{a_1}$ that we associate with 
the position of the
peak value of the corresponding spectral function. Parameters are as for
 Table~\protect\ref{table3}. The experimental $\rho(770)$ and
 $a_1(1260)$ meson masses are given in brackets. The energy unit is MeV.}
\label{table5}
\begin{tabular}{ccc}
 & $\rho$-mode  &$a_1$-mode \\          
\tableline
Pole position  & $382-970i$&$770-777i$ \\
Peak position  &713 (768$\pm$0.6)&1027 (1230$\pm$40)\\
\end{tabular}
\end{table}

\begin{table}
\caption{Comparison of the asymptotic behavior as $|s|\rightarrow \infty $ 
in the
complex $s$-plane of Pauli-Villars regulated Feynman loop diagrams, 
employing either
unsubtracted or once-subtracted
dispersion relations. The amplitudes that are generated by the first 
method carry
a bar in the main text.}
\label{table4}
\begin{tabular}{lcc}
Amplitude&Unsubtr(with bar)&Subtr(without bar)\\          
\tableline
$iI(s)$ & $1/s$ & $\ln(-s)$ \\
$\Pi^{PP}(s)$   &$(1/s)\ln(-s)$ &$-s\ln(-s)$\\
$\Pi^{VV}_T(s)\sim \Pi^{AA}_T(s)$ &$-(1/s)\ln(-s)$ & $s\ln(-s)$\\
$F_P(s)$&$-1/s$&$-\ln(-s)$\\
$F_V(s)\sim G_A(s)$&$1$&$1/(s\,\ln(-s))$\\
$F_P(s)F_V(s)G_A(s)$ &$-1/s$&$-1/(s^2\ln(-s))$\\
$\rho^{(0)V}_1(s)\sim\rho^{(0)A}_1(s)$&$-1/s$&$s$\\
$\rho^{V}_1(s)\sim \rho^{A}_1(s)$&$-1/s$&$1/s(\ln|s|)^2$\\
$\rho^{(0)V}_1(s)-\rho^{(0)A}_1(s)$&$-1/s^2$&${\rm positive\: constant}$\\
$\rho^{V}_1(s)-\rho^{A}_1(s)$&$-1/s^2$&$-1/(s\,\ln|s|)^2$\\
\end{tabular}
\end{table}

\begin{table}
\caption[]{Values of isospin traces}
\label{table of trace}
\begin{tabular}{crrc}
Trace of&$\pi^+$&$\pi^-$&$\pi^0$\\
\tableline
$T^\dagger_f e_q T_ie_q$&$-\frac{4}{9}$& $-\frac{4}{9}$&$\frac{5}{9}$\\
$T^\dagger_f\frac{\tau_3}{2}T_i\frac{\tau_3}{2}$&$-\frac{1}{2}$&
$-\frac{1}{2}$&$\frac{1}{2}$\\
$T^\dagger_f e_qT_i\frac{\tau_3}{2}$&$-\frac{1}{3}$&$-\frac{2}{3}$&$
\frac{1}{2}$\\
$T^\dagger_f\frac{\tau_3}{2}T_ie_q$&$-\frac{2}{3}$&$-\frac{1}{3}$&$
\frac{1}{2}$\\
$T^\dagger _fe^2_qT_i$&$\frac{8}{9}$&$\frac{2}{9}$&$\frac{5}{9}$\\
$T^\dagger_fT_ie^2_q$&$\frac{2}{9}$&$\frac{8}{9}$&$\frac{5}{9}$\\
$T^\dagger_fe_q\frac{\tau_3}{2}T_i$&$\frac{2}{3}$&$\frac{1}{3}$&$
\frac{1}{2}$\\
$T^\dagger_fT_ie_q\frac{\tau_3}{2}$&$\frac{1}{3}$&$\frac{2}{3}$&$
\frac{1}{2}$\\
$T^\dagger_f(\frac{\tau_3}{2})^2T_i$&$\frac{1}{2}$&$\frac{1}{2}$&$
\frac{1}{2}$\\
\end{tabular}
\end{table}
\end{document}